\newif\ifOneCol

\ifOneCol
	\documentclass[preprint,letterpaper]{elsarticle}
\else
	\documentclass[5p,letterpaper]{elsarticle}
\fi
\pdfoutput=1
\usepackage[cmex10]{amsmath}
\usepackage{amsfonts,amssymb}
\usepackage{verbatim}
\usepackage{color}
\usepackage{hyperref}
\usepackage{multirow} 
\usepackage{caption} 
\newcommand{\specialcell}[2][l]{\begin{tabular}[#1]{@{}l@{}}#2\end{tabular}}

  \graphicspath{{graphics/}}
  \DeclareGraphicsExtensions{.pdf,.jpeg,.png}

\usepackage{algorithm}
\usepackage{algpseudocode}

\usepackage{array}

\usepackage{mdwmath}
\usepackage{mdwtab}
\DeclareMathOperator{\erf}{erf}

\newcommand{\x}{x}
\newcommand{\y}{y}
\newcommand{\z}{z}
\newcommand{\rad}[1]{\vec{r}_{#1}}

\newcommand{\setMolTypes}{\Omega_\mathrm{Mol}}
\newcommand{\indMolType}{m}
\newcommand{\setChemRxn}{\Omega_\mathrm{C}}
\newcommand{\indChemRxn}{c}
\newcommand{\indChemRxnRev}{c^\ast}
\newcommand{\reactant}[1]{A_{#1}}
\newcommand{\product}[1]{B_{#1}}
\newcommand{\kth}[1]{k_{#1}}
\newcommand{\rBind}{r_\mathrm{B}}
\newcommand{\rUnbind}{r_\mathrm{U}}
\newcommand{\prob}[1]{P_\mathrm{#1}}
\newcommand{\rxn}{\mathrm{rxn}}

\newcommand{\Dx}[1]{D_{#1}}
\newcommand{\dist}{d}
\newcommand{\Nx}[1]{N_{#1}}
\newcommand{\Cx}[1]{C_{#1}}

\newcommand{\rx}[1]{r_{#1}}
\newcommand{\rrx}{r}

\newcommand{\EXP}[1]{\exp\left(#1\right)}
\newcommand{\ERF}[1]{\erf\left(#1\right)}
\newcommand{\ERFC}[1]{\mathrm{erfc}\left(#1\right)}
\newcommand{\ERFCX}[1]{\mathrm{erfcx}\left(#1\right)}

\newcommand{\numTrials}{n}
\newcommand{\numSuccess}{k}
\newcommand{\numSuccessThresh}{i}
\newcommand{\trialProb}{p}
\newcommand{\varOne}{X}
\newcommand{\varOneVal}{x}
\newcommand{\varTwo}{Y}
\newcommand{\varTwoVal}{y}

\renewcommand{\O}[1]{\mathcal{O}\left(#1\right)}
\newcommand{\subBase}{h_\mathrm{base}}
\newcommand{\subLength}[1]{h_{#1}}
\newcommand{\Vol}[1]{V_{#1}}
\newcommand{\setRegion}{\Omega_\mathrm{R}}
\newcommand{\setRegionNeigh}{\Omega_{\mathrm{neigh},\indRegion}}
\newcommand{\indRegion}{r}
\newcommand{\indRegionNeigh}{w}

\newcommand{\indActor}{a}
\newcommand{\normRV}[1]{n_{#1}}
\newcommand{\uniRV}[1]{u_{#1}}
\newcommand{\pdf}[1]{f_{#1}}
\newcommand{\tCur}{t_\mathrm{cur}}
\newcommand{\tEnd}{t_\mathrm{end}}
\newcommand{\seed}{\psi}
\newcommand{\config}{\mathrm{CONFIG}}
\newcommand{\numMol}[1]{N_{#1}}
\newcommand{\numNewMol}{\numMol{\mathrm{new}}}
\newcommand{\complexity}[1]{C_{#1}}
\newcommand{\complexityMicro}{\complexity{\mathrm{micro}}}
\newcommand{\complexityMeso}{\complexity{\mathrm{meso}}}

\newcommand{\dtMicro}{\Delta t}

\newcommand{\numMolMicroRegion}{\Nx{\mathrm{micro},\indRegion}}
\newcommand{\numMolMicroRegionX}[1]{\Nx{\mathrm{micro},#1}}
\newcommand{\regimeMicro}{\Omega_\mathrm{micro}}
\newcommand{\timeRxn}[1]{t_{#1}}
\newcommand{\timeX}[1]{t_{#1}}

\newcommand{\numSub}{\Nx{\mathrm{Sub}}}
\newcommand{\numEvent}{\Nx{\mathrm{Event}}}
\newcommand{\prop}[1]{\alpha_{#1}}
\newcommand{\propTotal}{\alpha_\mathrm{T}}

\newcommand{\dtInf}{\delta t}

\newcommand{\setSub}{\Omega_\mathrm{Sub}}
\newcommand{\indSub}{s}
\newcommand{\indSubNeigh}{q}
\newcommand{\regimeMeso}{\Omega_\mathrm{meso}}
\newcommand{\numMolSub}{U_{\indSub,\indMolType}}
\newcommand{\numMolSubX}[1]{U_{\indSub,{#1}}}
\newcommand{\areaOverlap}{A_\mathrm{o}}

\newcommand{\hybridDist}[1]{l_{#1}}
\newcommand{\ProbIntra}{P_\mathrm{intra}}

\newcommand{\meter}{\mathrm{m}}
\newcommand{\second}{\mathrm{s}}

\newcommand{\mol}{\mathrm{mol}}
\newcommand{\TX}{\mathrm{TX}}
\newcommand{\RX}{\mathrm{RX}}

\newcommand{\varX}[1]{\gamma_{#1}}

\begin{document}

%
\title{Simulating with AcCoRD: Actor-Based Communication via Reaction-Diffusion}

	\begin{frontmatter}
		
	\author[uOttawa]{Adam~Noel}
	\ead{anoel2@uottawa.ca}
	
	\author[UBC]{Karen~C.~Cheung}
	\ead{kcheung@ece.ubc.ca}
	
	\author[IDC]{Robert~Schober}
	\ead{robert.schober@fau.de}
	
	\author[uOttawa]{Dimitrios~Makrakis}
	\ead{dimitris@eecs.uottawa.ca}
	
	\author[DIRO]{Abdelhakim~Hafid}
	\ead{ahafid@iro.umontreal.ca}
	
	\address[uOttawa]{School of Electrical Engineering and Computer Science, University of Ottawa, Ottawa, ON, Canada}
	\address[UBC]{Department of Electrical and Computer Engineering, University of British Columbia, Vancouver, BC, Canada}
	\address[IDC]{Institute for Digital Communication, Friedrich-Alexander-Universit\"{a}t Erlangen-N\"{u}rnberg (FAU), Erlangen, Germany}
	\address[DIRO]{Department of Computer Science and Operations Research, Universit\'{e} de Montr\'{e}al, Montr\'{e}al, QC, Canada}

\begin{abstract}
This paper introduces AcCoRD (Actor-based Communication via Reaction-Diffusion) version 1.0. AcCoRD is a sandbox reaction-diffusion solver designed for the study of molecular communication systems. It uses a hybrid of microscopic and mesoscopic simulation models that enables scalability via user control of local accuracy. AcCoRD is developed in C as an open source command line tool and includes utilities to process simulation output in MATLAB. The latest code and links to user documentation can be found at \url{https://github.com/adamjgnoel/AcCoRD/}. This paper provides an overview of AcCoRD's design, including the motivation for developing a specialized reaction-diffusion solver. The corresponding algorithms are presented in detail, including the computational complexity of the microscopic and mesoscopic models. Other novel derivations include the transition rates between adjacent mesoscopic subvolumes of different sizes. Simulation results demonstrate the use of AcCoRD as both an accurate reaction-diffusion solver and one that is catered to the analysis of molecular communication systems. A link is included to videos that demonstrate many of the simulated scenarios. Additional insights from the simulation results include the selection of suitable hybrid model parameters, the impact of reactive surfaces that are in the proximity of a hybrid interface, and the size of a bounded environment that is necessary to assume that it is unbounded. The development of AcCoRD is ongoing, so its future direction is also discussed in order to highlight improvements that will expand its potential areas of application. New features that are being planned at the time of writing include a fluid flow model and more complex actor behavior.
\end{abstract}


\end{frontmatter}

\section{Introduction}
\label{sec_intro}

\subsection{Background and Motivation}

There has been recent interest in designing synthetic wireless communication networks for environments where conventional (i.e., radio frequency) wireless technologies are unsafe, infeasible, or impractical, such as in biological systems. This interest has inspired the idea of adapting natural communication strategies from these biological systems. One such strategy is molecular communication (MC), initially proposed for synthetic networks in \cite{Hiyama2005}, where transmitters use physical molecules as information carriers. MC is used for signaling in nature over a wide range of physical scales, from quorum sensing in bacterial communities to communication via pheromones over a kilometer or more, as described in \cite{Antunes2009} and \cite[Ch.~53]{Sadava2014}, respectively. In particular, MC is ubiquitous in communication within and between cells; see \cite[Ch.~16]{Alberts}.

Natural MC systems are typically designed for the transmission of limited quantities of information, such as a time-varying ON/OFF control signal for some process. However, synthetic MC networks are envisioned for transmitting arbitrarily large amounts of information. These networks could enable new applications in fields including biological engineering, medicine, manufacturing, and environmental modeling; see \cite{Nakano2013c}.

Commonly-studied forms of MC include molecular \emph{diffusion}, where molecules passively propagate via collisions with other molecules in a fluid environment. The interest in diffusion for communication engineering, as demonstrated in \cite{Farsad2016}, can be attributed to its speed over very short distances (particularly on the scale of a micron or less), its simplicity (requiring no active propagation mechanism), and the availability of mathematical models to facilitate analysis. In particular, mathematical models are needed to determine a channel's impulse response (i.e., the time-varying signal observed at a receiver due to the release of an instantaneous signal by a transmitter). Knowledge of the channel impulse response is essential for meaningful transceiver system design and performance analysis.

There are many seminal texts on the analysis of diffusion, e.g., \cite{Crank1979,Cussler1984,Berg1993}. However, closed-form expressions for the impulse response of a diffusive channel generally require simplifying assumptions and specific system geometries. This reality is an obstacle to the development of MC networks. Most existing research has considered a variation of a common system model. Typically, authors will consider a one- or three-dimensional (i.e., 1D or 3D) unbounded environment, possibly with a uniform fluid flow, with a point source and a receiver that is either an absorbing surface (i.e., molecules are ``consumed'') or a passive observer\footnote{A (perhaps surprisingly) large fraction of papers reviewed in \cite{Farsad2016}, including the current authors' own work on MC, can be classified as using such a model.}.

While the resulting analysis of simplified models is convenient, and uniformity is helpful for comparing transceiver designs, there are a couple of issues with this trend. First, realistic diffusive environments are generally bounded. Approximating an environment as unbounded is only appropriate if it is symmetric and in the absence of other local obstacles, but environments such as cells and cellular tissues are full of obstacles; see \cite[Ch.~20]{Alberts}. Second, diffusion is not the only phenomenon that can affect the behavior of a diffusing particle. In addition to fluid flow (which is in general not uniform; see \cite{Truskey2009}), molecules can undergo chemical reactions (besides absorption at a surface) that convert them into other molecular species or transport them across boundaries. For example, see \cite[Ch.~9]{Chang2005} for elementary analysis of chemical reactions.

Flow and reactions can significantly modify the channel response, and could even be deliberately introduced to improve communication performance. This was observed by using an electric fan in the macroscale testbench developed in \cite{Farsad2013a}, and by adding enzymes to the propagation environment in our own work in \cite{Noel2014f}. Neither of these strategies could be accurately described using the commonly-studied channel models; the experiments in \cite{Farsad2013a} were fitted to a corrected 1D model in \cite{Farsad2013}, and the enzyme kinetics in \cite{Noel2014f} were simplified to a first order degradation reaction so that the impulse response could be derived.

The notion that we simplified the analytical model in \cite{Noel2014f} immediately begs the question of how we verified its accuracy. Furthermore, it was insufficient to determine the \emph{expected} channel behavior; for communications analysis, we are interested in the \emph{probability distribution} of the channel behavior. We addressed both issues in \cite{Noel2014f} by simulating the detailed model. Unfortunately (and to the best of our knowledge), no existing simulation platform would accommodate the detailed model in \cite{Noel2014f}. Even though we could have used a \emph{generic reaction-diffusion solver} such as Smoldyn (see \cite{Andrews2004}) to evaluate the expected channel behavior, it was not suitable for assessing the time-varying channel statistics or for configuring a source to release molecules based on modulating a sequence of random binary data. Our solution in \cite{Noel2014f} was to internally develop a simulator in MATLAB. However, this simulator was specific to the environment of the model in \cite{Noel2014f} and not easily portable to other system models. We claim that existing publicly-available MC simulators have similar limitations, e.g. \cite{Wei2013a,Felicetti2013,Llatser2014,Yilmaz2014a,Jian2016} (which we will discuss in further detail in Section~\ref{sec_mc_sim}).

\subsection{The AcCoRD Simulator}

With these limitations in mind, we have developed the AcCoRD simulator (Actor-based Communication via Reaction-Diffusion). AcCoRD is a generic reaction-diffusion solver that is developed in C and designed for communications analysis. It is an open source project in active development on Github; see \cite{Noel2016}. As of the time of writing (November 2016), the latest release is version 1.0. It has the following primary features:
\begin{itemize}
	\item AcCoRD is a hybrid solver that integrates two simulation models to define 3D environments with flexible local accuracy. Each local region is classified as \emph{microscopic} or \emph{mesoscopic}. Microscopic regions define each molecule individually and evolve over discrete time steps. Mesoscopic regions count the number of molecules in disjoint virtual bins (called \emph{subvolumes}) and evolve over time steps with continuous granularity. We increase the scalability by accommodating the placement of adjacent mesoscopic regions that have subvolumes of different sizes. This feature is based on an extension of the 2D system that we proposed in \cite{Noel2015a}.
	\item Actors can be distributed throughout the environment as \emph{active} molecule sources (i.e., transmitters) or \emph{passive} observers (i.e., receivers). Transmitters release molecules according to the modulation of a random binary data sequence. The precise number of molecules released and the release times for a given symbol interval can be deterministic or randomized. Receivers can record the number of molecules and (optionally) their positions at any specified interval. Future development will couple these two actor classes to enable transceivers that behave according to their observations.
	\item Zeroth, first, and second order chemical reactions can be defined locally or globally, i.e., over the entire propagation environment or in a particular set of regions. This general framework can accommodate reactions such as molecule degradation, enzyme kinetics, reversible or irreversible surface binding, ligand-receptor binding, transitions across boundary membranes, and simplified molecular crowding. Surface binding reactions include absorption, i.e., \emph{consumption}, adsorption, i.e., \emph{sticking}, and desorption, i.e., \emph{release} from a surface. Generally, we will refer to adsorption as reversible absorption.
	\item AcCoRD implements some microscopic behavior in continuous time. Specifically, the release of molecules by active actors, zeroth order reactions, and most first order reactions can occur at any time. Thus, a molecule can undergo multiple reactions in a single microscopic time step, and the accuracy of these phenomena are independent of the chosen time step.
	\item Independent realizations of a simulation can be repeated an arbitrary number of times (and on different computers) and then aggregated to determine the average behavior and channel statistics.
	\item The online documentation includes installation and usage instructions for the latest version, descriptions of all configuration options, and many sample configuration files; see \cite{Noel2016}. The sample configurations are provided to demonstrate all of AcCoRD's functionality.
	\item AcCoRD's interface has been designed to be helpful to novice users by providing descriptive output messages. AcCoRD also includes post-processing tools developed in MATLAB. These tools enable the aggregation of simulation output files, plotting receiver observations (either the time-varying behavior or empirical distributions at specified times), and visualizing the physical environment (either as still images or compiled into a video\footnote{A series of eight videos are discussed throughout this paper and can be found at \url{https://www.youtube.com/playlist?list=PLZ7uYXG-7XF8UyhFrIuQIiZig1XA89e3i}; see \cite{Noel2016c}.}).
\end{itemize}

\begin{figure}[!tb]
	\centering
	\includegraphics[width=3in]{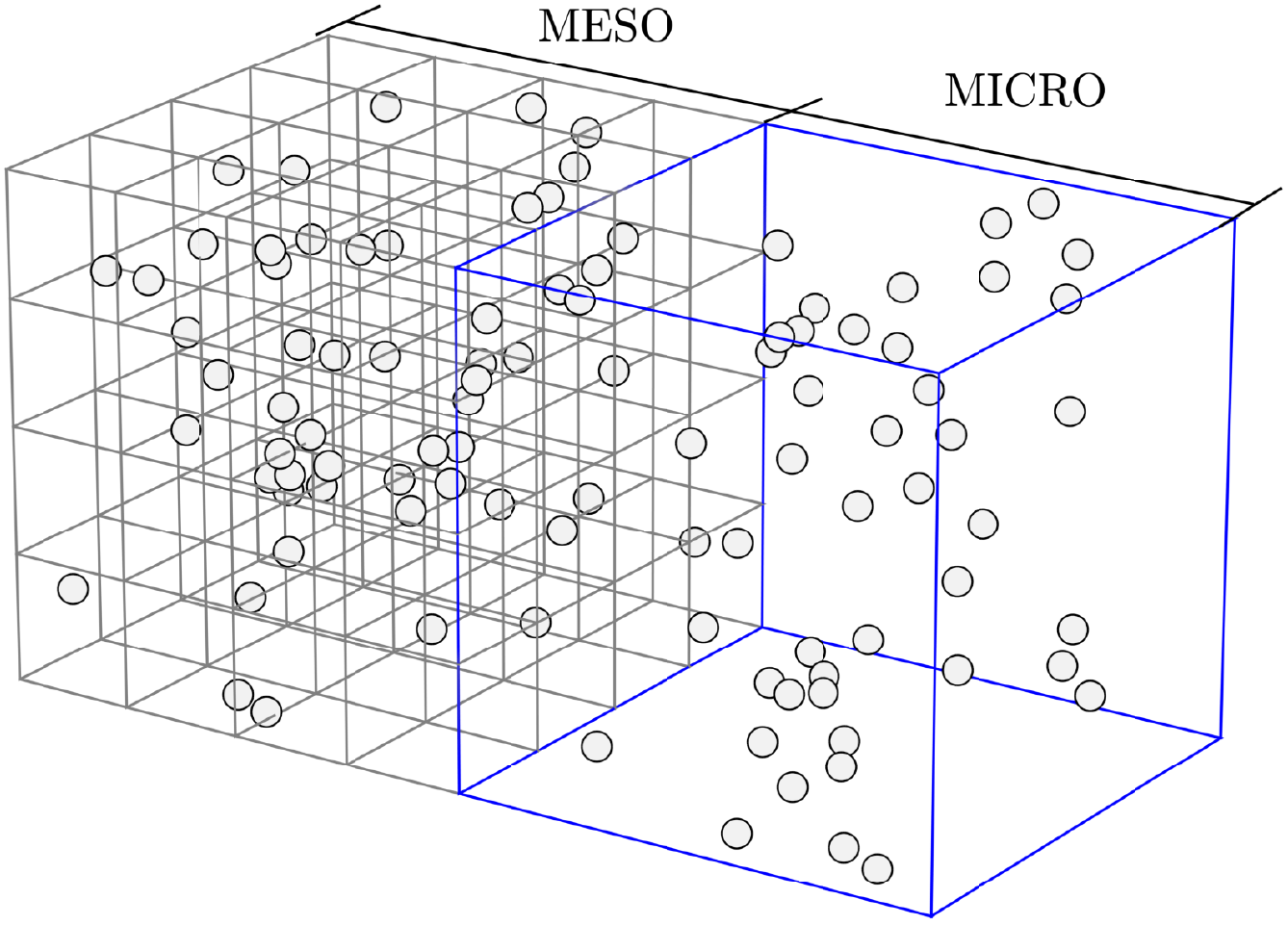}
	\caption{Simulation snapshot of a hybrid environment where the diffusing molecules are initially uniformly distributed. The left half is \emph{mesoscopic}, where molecules are counted in one of 64 cubes (i.e., \emph{subvolumes}), and the right half is \emph{microscopic}, where molecules are tracked individually. 100 diffusing molecules are displayed. This environment is labeled System 1 and variations of it are simulated in Section~\ref{sec_results_hybrid}. A sample video of this simulation is available as Video 2 in \cite{Noel2016c}.}
	\label{fig_hybrid_env_mol}
\end{figure}

\begin{figure}[!tb]
	\centering
	\includegraphics[width=3.25in]{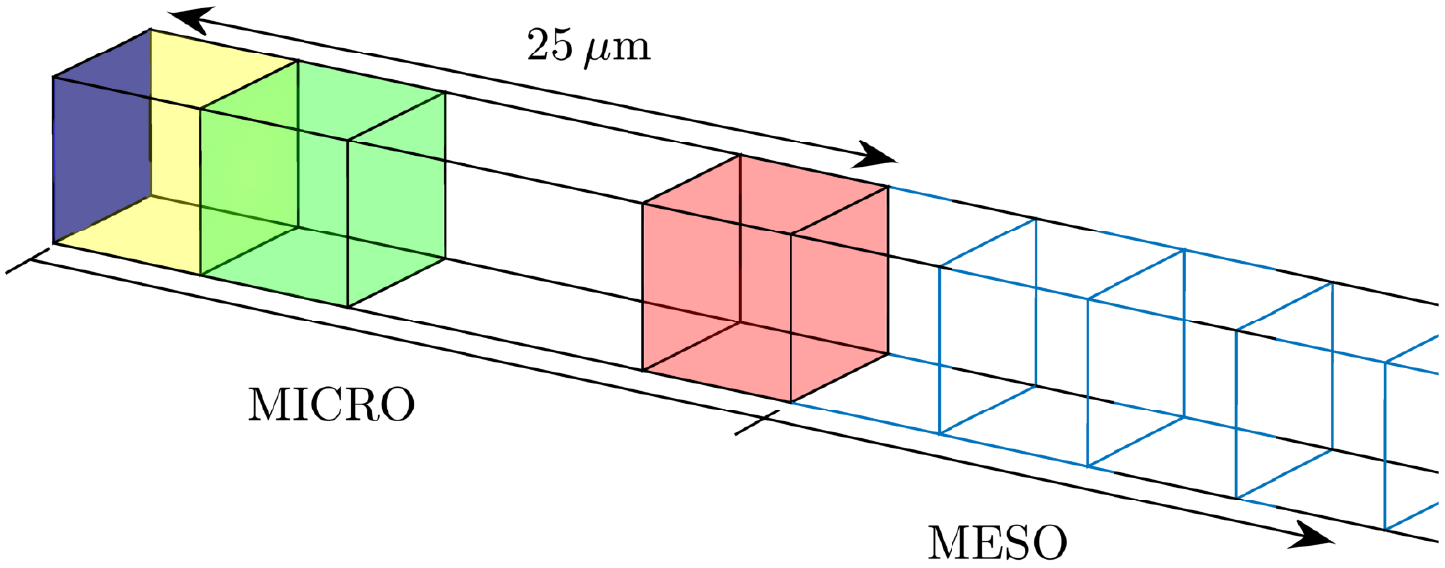}
	\caption{A portion of a hybrid environment with a reactive surface. The environment is a $0.5\,\meter\meter$ long rod. At one end (shown in dark blue) is an absorbing surface. Three observers watch the molecules present over $5\,\mu\meter$ sections of the rod (shown in yellow, green, and red). This environment is labeled System 2 and variations of it are simulated in Section~\ref{sec_results_hybrid_rxn_diff}. The variation shown has a hybrid interface between microscopic and mesoscopic regions at a distance of $25\,\mu\meter$ from the absorbing end, such that the cubes with blue outlines are mesoscopic subvolumes. A sample video of the simulation of this environment is available as Video 3 in \cite{Noel2016c}.}
	\label{fig_hybrid2_env}
\end{figure}

\begin{figure}[!tb]
	\centering
	\includegraphics[width=2in]{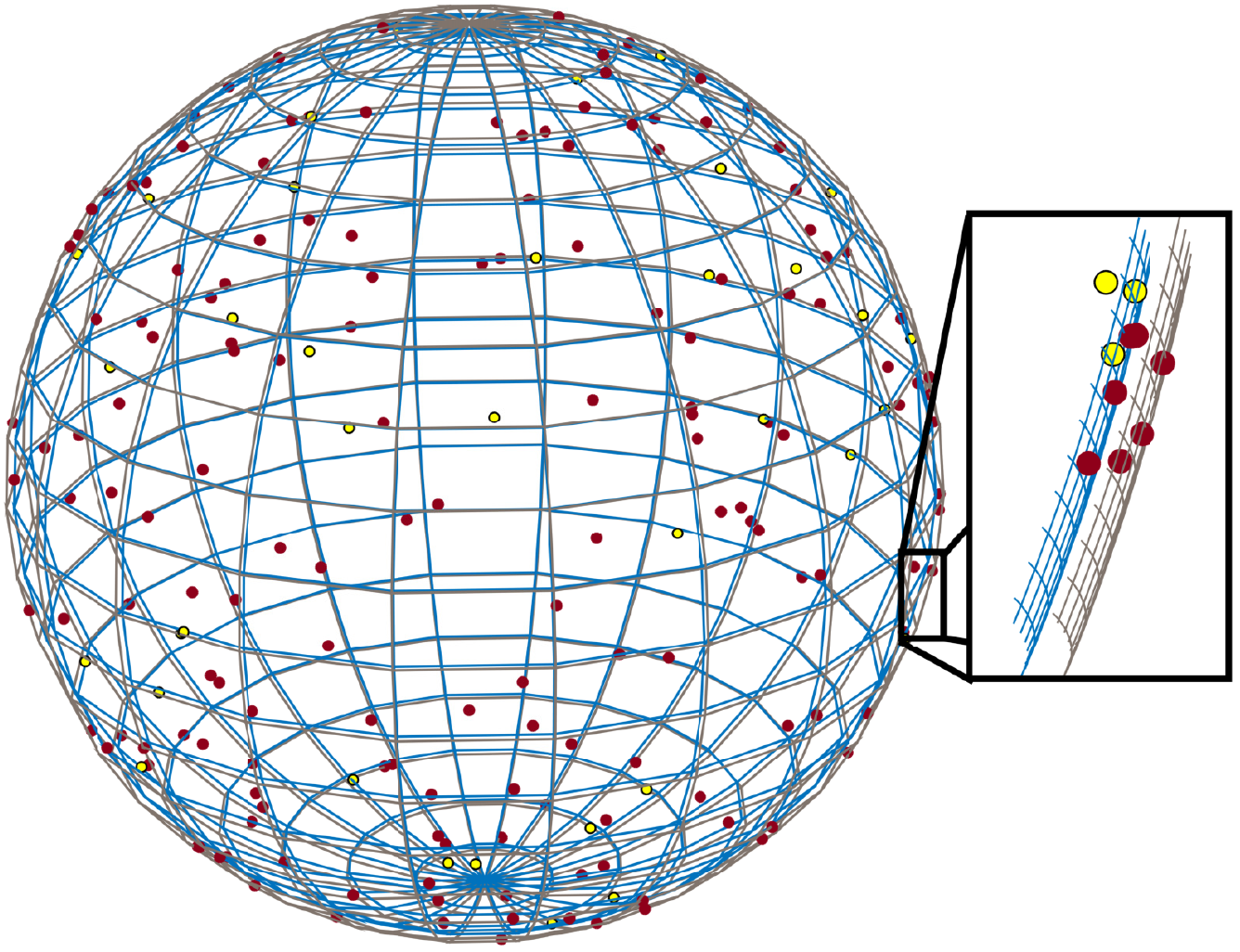}
	\caption{Simulation snapshot of an environment with two large concentric spheres with radii $120\,\mu\meter$ and $122\,\mu\meter$. Molecules are initialized in the space between the two spheres. This environment is a spherical analog to the 1D system studied in \cite[Fig.~6a]{Andrews2009}. The inner and outer spherical surfaces are shown with blue and grey outlines, respectively. This environment is labeled System 3 and variations of it are simulated in Section~\ref{sec_results_surface}. In the variation shown, molecules can probabilistically diffuse through the inner surface, i.e., the inner surface acts as a membrane. The molecules turn from red to yellow when they transition inside the inner sphere (see inset). Sample videos of this simulation are available as Video 4 (for absorption to the outer sphere) and Video 5 (for transitions through the membrane) in \cite{Noel2016c}.}
	\label{fig_surface_env}
\end{figure}

\begin{figure}[!tb]
	\centering
	\includegraphics[width=3in]{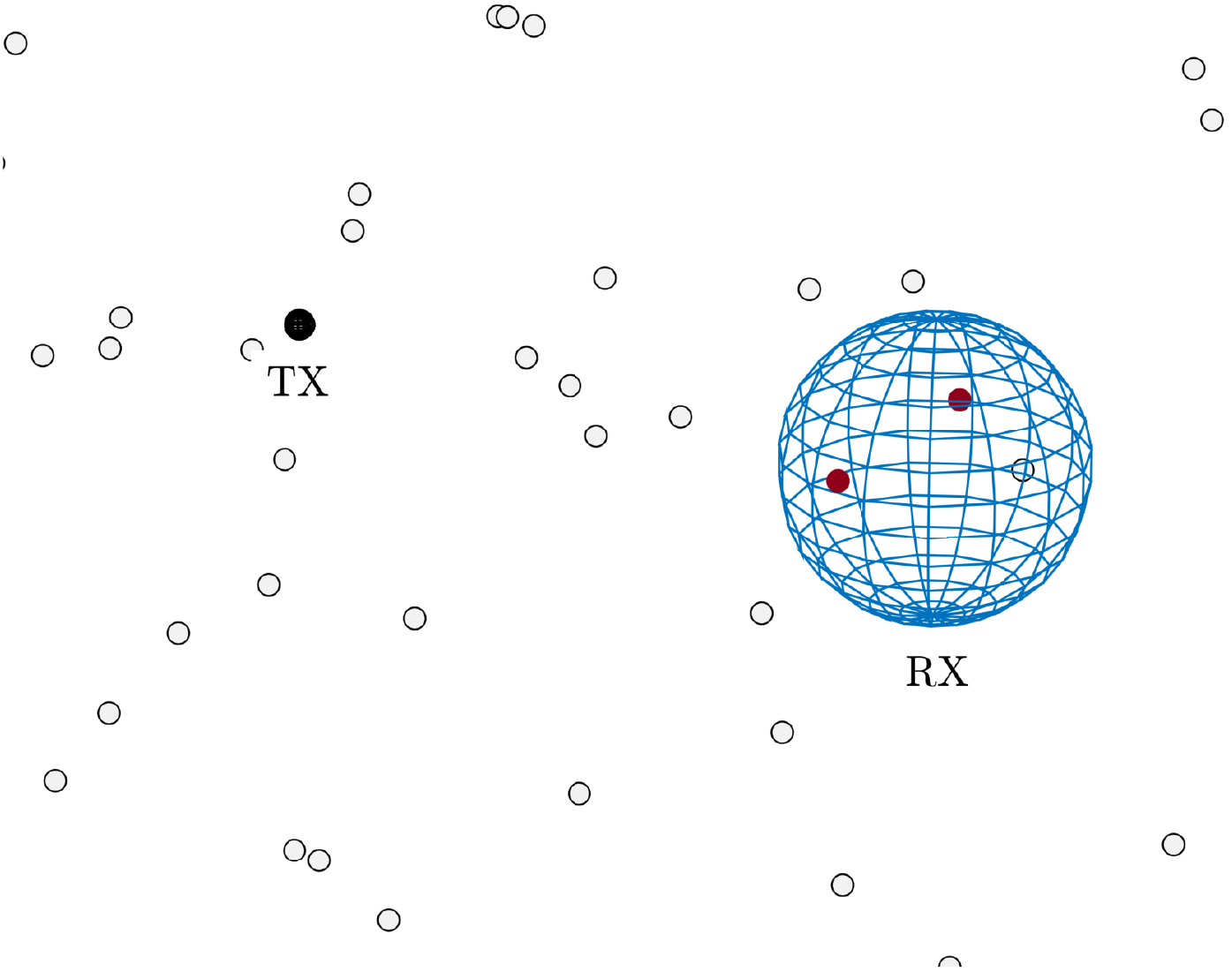}
	\caption{Simulation snapshot of a simple communication environment. Molecules are released by a point transmitter and observed at a spherical receiver (radius $1\,\mu\meter$) centered $5\,\mu\meter$ from the transmitter. This environment is labeled System 4 and variations of it are simulated in Section~\ref{sec_results_comm}. In the variation shown, the released molecules (light grey) can irreversibly bind to the receiver's absorbing surface (and turn red when bound). One molecule that appears to be inside the receiver is actually behind it. A sample video of this simulation is available as Video 8 in \cite{Noel2016c}. Other variations of this system are shown in Videos 6 and 7 in \cite{Noel2016c}.}
	\label{fig_comm_env}
\end{figure}

Four sample environments that illustrate AcCoRD's primary features are shown in Figs.~\ref{fig_hybrid_env_mol} to~\ref{fig_comm_env}. Figs.~\ref{fig_hybrid_env_mol} and~\ref{fig_hybrid2_env} show hybrid environments with both microscopic and mesoscopic regions. Fig.~\ref{fig_surface_env} has reversible surface reactions. Fig.~\ref{fig_comm_env} has a distinct transmitter and receiver. More details about these environments, including their simulation results, are discussed in Section~\ref{sec_results}.

AcCoRD provides significant flexibility for a user to define the physical environment, specify chemical reactions, and place sources and observers of molecules. We describe AcCoRD as a reaction-diffusion ``sandbox'', i.e., this flexibility enables users to explore (``play'') and create their own system. As a communications analysis tool, it can model data modulation at transmitters and generate channel statistics at observers by repeating a simulation an arbitrary number of times. By developing AcCoRD as a ``sandbox'' reaction-diffusion solver from the perspective of molecular communication, we anticipate that it will contribute the following for the MC community:
\begin{itemize}
	\item Encourage the use of simulations. Most authors in this domain have thus far verified their work via numerical evaluations of their analysis or via Monte Carlo simulations. In the latter case, the time-varying probability distribution of the channel behavior is assumed to be known. By omitting simulations, assumptions about the model and their accuracy for different parameter values might be untested.
	\item Improve the understanding and visualization of known reaction-diffusion environments and their channel responses. This is especially important for improving the accessibility of this multi-disciplinary field and encouraging new researchers to contribute without needing to develop their own simulation tools.
	\item Provide a platform to verify new analysis and test transceiver designs. For example, we used AcCoRD in \cite{Noel2016a} to study the accuracy of the common assumption that the molecule source is a point transmitter.
	\item Enable the exploration of new channels that have not or cannot be precisely examined analytically, such as the diffusion model with enzyme kinetics that we considered in \cite{Noel2014f}.
\end{itemize}

\subsection{Contributions}

This paper introduces the AcCoRD simulator, and also makes the following novel technical contributions:
\begin{enumerate}
	\item We derive the transition rate between adjacent 3D mesoscopic subvolumes of different sizes that have partially-overlapping faces. The 2D version of this rate was presented in \cite{Noel2015a}. Accommodating mesoscopic subvolumes with different sizes increases the flexibility of local accuracy.
	\item We derive the continuous event time for a first order microscopic chemical reaction event that is ``known'' to have occurred in the current microscopic time step. This derivation is inspired by how chemical reactions occur in both the microscopic and mesoscopic models, and enables a molecule to react multiple times within a single microscopic time step (albeit with an increased computational cost if multiple reactions do occur).
	\item We verify AcCoRD's accuracy by comparing its simulation output with analytical results for known reaction-diffusion environments. Examples include surface reactions, enzyme kinetics, and common molecular communication channels. Early versions of some of these results in 2D environments were presented in \cite{Noel2015}, but no specific simulations are repeated here.
	\item We demonstrate scalability by investigating the trade-off between computational efficiency and local accuracy in hybrid and multi-scale environments. We provide insights into the appropriate size and placement of subvolumes in the mesoscopic regime, the impact of a reactive surface near the proximity of the hybrid interface, and make general comments on when a hybrid model is appropriate. Preliminary versions of some of these results with simpler hybrid transition rules and in 2D environments were presented in \cite{Noel2015a,Noel2015}.
	\item We demonstrate that a bounded environment can be treated with an unbounded model if the distance from the area of interest to the nearest boundary is at least three times the average diffusion distance for the time scale of interest.
\end{enumerate}

\subsection{Organization}
    
The rest of this paper is organized as follows. We review related simulation tools, including generic reaction-diffusion solvers and diffusive MC simulators, in Section~\ref{sec_related}. In Section~\ref{sec_components}, we describe the components of an AcCoRD simulation. Section~\ref{sec_theory} presents the underlying theory for reaction and diffusion behavior. It includes our derivation for the transition rate between mesoscopic subvolumes of different sizes and for the reaction times of first order chemical reactions that have occurred within a microscopic time step. We present the overall AcCoRD algorithm and discuss computational complexity in Section~\ref{sec_algorithm}. We summarize the interface and work flow for using AcCoRD in Section~\ref{sec_use}. In Section~\ref{sec_results}, we present detailed simulation results to demonstrate AcCoRD's functionality and to verify its accuracy by comparing with analytical results. In Section~\ref{sec_future}, we identify features for AcCoRD's future development. Conclusions are drawn in Section~\ref{sec_concl}.

The Appendices focus on additional technical details that we include for completeness.~\ref{app_constraints} lists constraints and limitations to defining simulation environments.~\ref{app_surf} describes the surface reaction probabilities that were mostly derived in \cite{Andrews2009} but are also implemented in AcCoRD. In~\ref{app_algorithms}, we present the detailed algorithms of the main simulation steps. We review probability distributions and the calculation of mutual information in~\ref{app_prob}. Furthermore, sample simulation videos of many of the environments studied in this paper are included in \cite{Noel2016c} and summarized in~\ref{app_video}.

A reader who is only interested in using the AcCoRD software can focus on Sections~\ref{sec_components} and~\ref{sec_use}. A reader who is interested in the details of the implementation should refer to Sections~\ref{sec_components}, \ref{sec_theory}, \ref{sec_algorithm}, \ref{app_surf}, and~\ref{app_algorithms}.

\section{Related Work}
\label{sec_related}

In this section, we summarize existing simulation tools that implement features comparable to those developed in AcCoRD. First, we present a general discussion of reaction-diffusion solvers, including the simulation models that are commonly used over various physical scales. We also discuss efforts to develop simulations that integrate multiple models for improved scalability. Then, we focus on describing the simulation tools that have been developed specifically for the study of diffusive molecular communication systems. Finally, we draw comparisons between AcCoRD and existing tools.

\subsection{Reaction-Diffusion Solvers}
\label{sec_related_solvers}

Generic reaction-diffusion solvers make trade-offs between simulation accuracy and computational efficiency to model molecule behavior. As we discussed in \cite{Noel2015}, solver models can usually be placed into one of four categories that correspond to the degree of local detail of molecular behavior. In order of increasing computational efficiency (and therefore decreasing accuracy), these categories are \emph{molecular dynamics} models, \emph{microscopic} models, \emph{mesoscopic} models, and \emph{continuum} models. Here, we discuss each of these categories in sequence, and then describe efforts to implement simulation models over multiple scales.

\subsubsection{Molecular Dynamics}

The most detailed solvers use \emph{molecular dynamics} and account for all interactions between all individual molecules, including fluid solvents (e.g., water). A classical molecular dynamics simulator is the Large-scale Atomic/Molecular Massively Parallel Simulator (LAMMPS); see \cite{Plimpton1995} and details of ongoing development in \cite{StevePl}. For example, LAMMPS can model rigid body particles, charge potentials, and electron force fields, where the orientations of individual particles are accounted for.

\subsubsection{Microscopic Models}

The next most-detailed solvers use \emph{microscopic} models which treat solvent molecules as a continuum but all other molecules are individually tracked. Modeling the solvent as a continuum enables the introduction of a diffusion coefficient to describe the motion of an individual molecule as it moves within the solvent; see \cite{Cussler1984}. One microscopic solver is the Smoluchowski Dynamics simulator (Smoldyn); see \cite{Andrews2004,Andrews}. Smoldyn uses a constant time step and tracks the number and state of molecules over subsequent time steps. An alternative to this approach is Green's-function reaction dynamics (GFRD), which uses a continuous time model and tracks changes in molecules' states (e.g., when two molecules collide and react); see \cite{VanZon2005}.

\subsubsection{Mesoscopic Models}

More computationally efficient solvers do not model individual molecules but partition the simulation environment into virtual containers. This approach becomes more efficient as the environmental complexity grows (e.g., with the presence of more molecules). The third category of solvers uses \emph{mesoscopic} models, which refer to the containers as ``subvolumes'' and track discrete numbers of molecules in each container. Reactions involving two or more molecules can only occur if all reactants are in the same subvolume. Chemical reactions occur as ``events'', and diffusion is modeled as an event where a molecule moves from one subvolume to another. In their most accurate form, mesoscopic models execute one reaction at a time, and this approach can capture system behavior \emph{exactly} (in a statistical sense) if molecule populations throughout each subvolume are homogeneous and the size of each subvolume is chosen in consideration of the time scales of potential events; see \cite{Ramaswamy2011}. 

There are a few ways to increase computational efficiency via scalability in mesoscopic models. One method is by adjusting the size and placement of the subvolumes themselves. Examples of this method use subvolumes that are tetrahedrons (as in the URDME simulator in \cite{Drawert2012}) or subvolumes whose size can change in both time and space (as considered for square subvolumes in \cite{Bayati2011}). Another method is to evolve the system according to time steps and execute multiple events in each step. This approach is known as ``tau-leaping''; see \cite{Gillespie2001} for its introduction in reaction-only systems and \cite{Iyengar2010} for an implementation in reaction-diffusion systems. Depending on the size of the molecule populations within a given subvolume and the likelihood of reaction events, the number of events in one step could be found as a Poisson random variable, a Gaussian random variable (i.e., the Langevin method), or a deterministic value (i.e., a continuum model).

\subsubsection{Continuum Models}

In the limit of large molecule populations, mesoscopic models become \emph{continuum} models, where local molecule concentrations have continuous values and the system is in effect evolving via the solution of a set of partial differential equations with finite element analysis. Such a system is no longer stochastic but has deterministic behavior. COMSOL Multiphysics (see \cite{COMSOL}) and ANSYS (see \cite{ANSYS}) are commercially-available continuum solvers.

\subsubsection{Hybrid Models}

There have been multiple attempts to integrate multiple simulation models into a single reaction-diffusion solver. Doing such an integration improves flexibility in the trade-off between accuracy and computational complexity by using a more accurate model only where (or when) it is needed. For example, approaches have been proposed that combine microscopic and mesoscopic models, as in \cite{Klann2012,Flegg2014,Hellander2012}. The hybrid model in \cite{Hellander2012} has been implemented in the (primarily mesoscopic) solver URDME (see \cite{Drawert2012}) and the model in \cite{Flegg2014} was recently implemented in the (primarily microscopic) solver Smoldyn (see \cite{Robinson2015}). LAMMPS has also coupled its molecular dynamics model to a continuum model; see \cite{Wagner2008}. Virtual Cell is a continuum solver that added Smoldyn for microscopic behavior; see \cite{Resasco2012}.

\subsection{Diffusive Molecular Communication Simulators}
\label{sec_mc_sim}

The existing generic reaction-diffusion solvers described in Section~\ref{sec_related_solvers} are, in general, useful ``sandbox'' tools for exploring the dynamics of molecular behavior. However, there are several characteristics that limit their applicability to the study of molecular communication networks. We identify these characteristics as follows:
\begin{itemize}
	\item Existing generic solvers are not designed to accommodate the behavior of a transmitter that is releasing molecules according to the modulation of a data sequence. It can be possible to release finite pulses of molecules at specific times, for example via commands in Smoldyn, but other arbitrary molecule signal waveforms are not as easily accommodated (e.g., frequency shift keying). Furthermore, the molecule release times would need to be hard-coded into the configuration. Such an approach would be inconvenient for measuring the average bit error probability at a receiver over a large number of randomly-generated transmitter bit sequences, since every bit sequence would need its own corresponding configuration.
	\item In communications analysis, we are not necessarily interested in accurately observing system behavior \emph{everywhere}. Instead, we are ultimately interested in the behavior at the \emph{receiver(s)} of information. Selecting a solver model typically imposes a particular range of accuracy over the entire simulation environment. Thus, computational resources might be ``wasted'' to maintain accuracy in regions of the environment that have minimal impact on the receiver(s) of interest. However, this limitation can be mitigated by choosing an appropriate hybrid simulation model.
	\item Existing generic solvers are not designed for generating channel statistics over a large number of independent realizations. The reliability of a receiver is often measured as the probability that it correctly detects a message sent from a transmitter, which implies that a particular observation (or series of observations) was made by the receiver. If we want to simulate the time-varying probability distributions of receiver observations, i.e., the non-stationary channel statistics, then we may need to simulate one scenario many thousands of times and aggregate the results. Such a functionality is not native to existing generic solvers.
\end{itemize}

Given the aforementioned limitations in existing generic reaction-diffusion solvers, communications researchers have developed a number of simulators specifically for simulating diffusive molecular communication systems. A recent discussion comparing most molecular communication simulators is in \cite{Farsad2016}. Here, we highlight the features of simulators whose source code or executables are publicly available. Thus, we exclude simulators that have been described but not released, including NanoNS (Nanoscale Network Simulator) in \cite{Gul2010}, dMCS (Distributed Molecular Communication Simulator) in \cite{Akkaya2014}, and other tools developed internally by authors for their own research.


BNSim (Bacterial Network Simulator) was presented in \cite{Wei2013a} and is a multi-scale mesoscopic reaction-diffusion solver for simulating interactions between mobile bacteria populations. The bacteria undergo chemotaxis (i.e., run-and-tumble motion) and collisions between them are accounted for. Key chemical reactions are simulated with the complete Stochastic Simulation Algorithm (SSA; see \cite{Gillespie1976}) whereas other reactions are simulated over larger time scales via tau-leaping or the Langevin method; see a review of this and other methods in \cite{Gillespie2007}.

BiNS2 (Biological Nanoscale Simulator) was described in \cite{Felicetti2013} and is a microscopic reaction-diffusion simulator for flowing cylindrical environments (e.g., blood vessels). Both transmitters and receivers can be mobile and molecules are detected via receptor reactions. The simulation environment is customizable, can include local obstacles, and can be visualized at run time.


N3Sim was presented in \cite{Llatser2011,Llatser2014} and is a microscopic simulator for a square or unbounded 3D environment. In addition to Brownian motion, it can account for molecule inertia and collisions between solvent molecules. Circular or spherical emitters can release molecules according to predefined waveforms or a user-defined pattern. Receivers can be squares or circles in 2D and spheres in 3D, and can be either passive or fully absorbing. Notably, N3Sim is the only publicly-available molecular communication simulator (excluding AcCoRD) with a detailed user guide and instructions for use (see \cite{Llatser}).

MUCIN (MolecUlar CommunicatIoN) was presented in \cite{Yilmaz2014a} and is a microscopic simulator for unbounded 3D environments developed in MATLAB. It models a point or spherical transmitter that releases molecules according to one of several modulation schemes. The spherical receiver can be passive, partially absorbing (i.e., molecules are reflected with some probability), or fully absorbing.

The IEEE P1906.1/Draft 1.0 Recommended Practice for Nanoscale and Molecular Communication Framework, summarized in \cite{Bush2015}, uses a simulation tool implemented on top of ns-3 as a reference simulator. It is used to compare molecular and electromagnetic communication schemes. However, this simulator uses analytical results from a specific physical environment and does not simulate the environment directly.

Finally, nanoNS3 was introduced in \cite{Jian2016} and is a continuum simulator for bacteria-based molecular communication. It is implemented on top of ns-3 and has models implemented for the signal observed at a receiver bacterium, microfluidic channel loss, ON/OFF transmitter modulation, and addressing via signal amplitudes.

\subsection{Comparing with AcCoRD}

Some of the existing simulators for diffusive MC, namely BNSim and BiNS2, are very detailed for their intended environments and how they can be configured. However, these tools and the other MC simulators were not designed as generic reaction-diffusion solvers, so they are not as flexible as AcCoRD for studying new and different environments. Instead of making exhaustive comparisons between AcCoRD and all aforementioned simulation tools listed in this section, we make a simplified comparison in Table~\ref{table_simulators}. The first ``model'' listed for each simulator is its primary model, but we emphasize that only AcCoRD was initially designed as a hybrid of simulation models. Furthermore, AcCoRD is the only generic reaction-diffusion solver (i.e., ``sandbox'' simulator) that can accommodate molecule sources that modulate a sequence of randomly-generated data. We believe that this combination facilitates studying the performance of a variety of diffusion-based molecular communication systems.

\begin{table}[!tb]
	\centering
	\caption{Simplified comparison of simulation tools with AcCoRD. Under ``Model'', ``MD'' refers to molecular dynamics, ``Micro'' refers to microscopic, ''Meso'' refers to mesoscopic, and ``Cont'' refers to continuum. Simulators that use more than one model list the primary model first. For each simulator, we also indicate whether each is a generic reaction-diffusion solver (i.e., ``sandbox'') and whether they can release molecules according to the modulation of a data sequence. The IEEE P1906.1 reference simulator described in \cite{Bush2015} is omitted because it does not simulate the environment directly.}
	
	{\renewcommand{\arraystretch}{1.2}\footnotesize
		\begin{tabular}{l||l|l|l}
			\hline
			\bfseries Simulator & \bfseries Model & \bfseries Sandbox & \bfseries \specialcell{Modulate\\Data}  \\ \hline \hline
			LAMMPS \cite{Plimpton1995} & MD/Cont & Yes & No \\
			Smoldyn \cite{Andrews2004} & Micro/Meso & Yes & No \\
			URDME \cite{Drawert2012} & Meso/Micro & Yes & No \\
			COMSOL \cite{COMSOL} & Cont & Yes & No \\
			ANSYS \cite{ANSYS} & Cont & Yes & No \\
			Virtual Cell \cite{Resasco2012} & Cont/Micro & Yes & No \\
			BNSim \cite{Wei2013a} & Meso & No & No \\
			BiNS2 \cite{Felicetti2013} & Micro & No & Yes \\
			N3Sim \cite{Llatser2014} & Micro & No & Yes \\
			MUCIN \cite{Yilmaz2014a} & Micro & No & Yes \\
			nanoNS3 \cite{Jian2016} & Cont & No & Yes \\
			AcCoRD & \bfseries Micro/Meso & \bfseries Yes & \bfseries Yes \\ \hline
		\end{tabular}
	}
	\label{table_simulators}
\end{table}

We will find it most insightful to make a direct comparison between AcCoRD and the microscopic reaction-diffusion solver Smoldyn. AcCoRD and Smoldyn have a number of similar underlying algorithms and features, such as for chemical reactions and hybrid microscopic-mesoscopic interfaces, which will be discussed in greater detail in Section~\ref{sec_theory}, \ref{app_surf}, and~\ref{app_algorithms}. Smoldyn, which has been in active development for over a decade, has a more mature code base, more features such as more options for defining physical environments and surface interactions, and the flexibility to change parameters while a simulation is in progress. However, we identify the following primary advantages for AcCoRD:
\begin{itemize}
	\item AcCoRD addresses the aforementioned limitations of generic reaction-diffusion solvers to facilitate adoption for communications analysis. In particular, AcCoRD can accommodate the release of molecules according to the modulation of a random binary sequence. Also, simulation realizations can be easily aggregated (even if they were run on different computers) to compile the simulation statistics.
	\item AcCoRD implements some microscopic behavior in continuous time, whereas microscopic behavior in Smoldyn is completely discrete. Therefore, the accuracy of some phenomena in AcCoRD (namely, zeroth order and some first order reactions) is independent of the chosen time step.
	\item A primary motivation for AcCoRD was the integration of microscopic and mesoscopic models to enable flexibility in local accuracy, as we initially proposed in \cite{Noel2015a}. Smoldyn is primarily a microscopic simulator that recently added a mesoscopic ``module'' that is not as well documented as the rest of the platform. Thus, even though AcCoRD and Smoldyn both implement transitions between the two models using the hybrid model in \cite{Flegg2014}, the integration in AcCoRD is central to the simulator's design. In AcCoRD, it is relatively seamless for a user to apply both models. Furthermore, we accommodate adjacent mesoscopic regions that have subvolumes of different sizes, whereas Smoldyn does not.
	\item AcCoRD includes visualization tools that were developed in MATLAB, which make them accessible and familiar to a wide research audience. Smoldyn can display animations of simulation progress and save images for future use, but the images must be displayed and captured \emph{online}; i.e., while the environment is being simulated. AcCoRD performs its visualization \emph{offline}; a user can preview the physical environment without running a simulation, and simulation output can be processed to either generate images or make a video with the freedom to choose what environment features to display and at what times.
	\item Smoldyn uses the Mersenne Twister (see \cite{Matsumoto1998}) as its random number generator (RNG). The Mersenne Twister is a common RNG and the default RNG in MATLAB. However, the permuted congruential generator (PCG) family of RNGs, recently proposed in \cite{ONeill}, is claimed to have improved statistical quality. A detailed discussion of the PCG's advantages is outside the scope of this work, but we note that it has a faster generation speed and a smaller code footprint. AcCoRD uses a PCG RNG that we have confirmed in internal testing to be faster than an efficient implementation of the Mersenne Twister.
\end{itemize}

\section{System Model Components}
\label{sec_components}

\begin{figure}[!tb]
	\centering
	\includegraphics[width=3.25in]{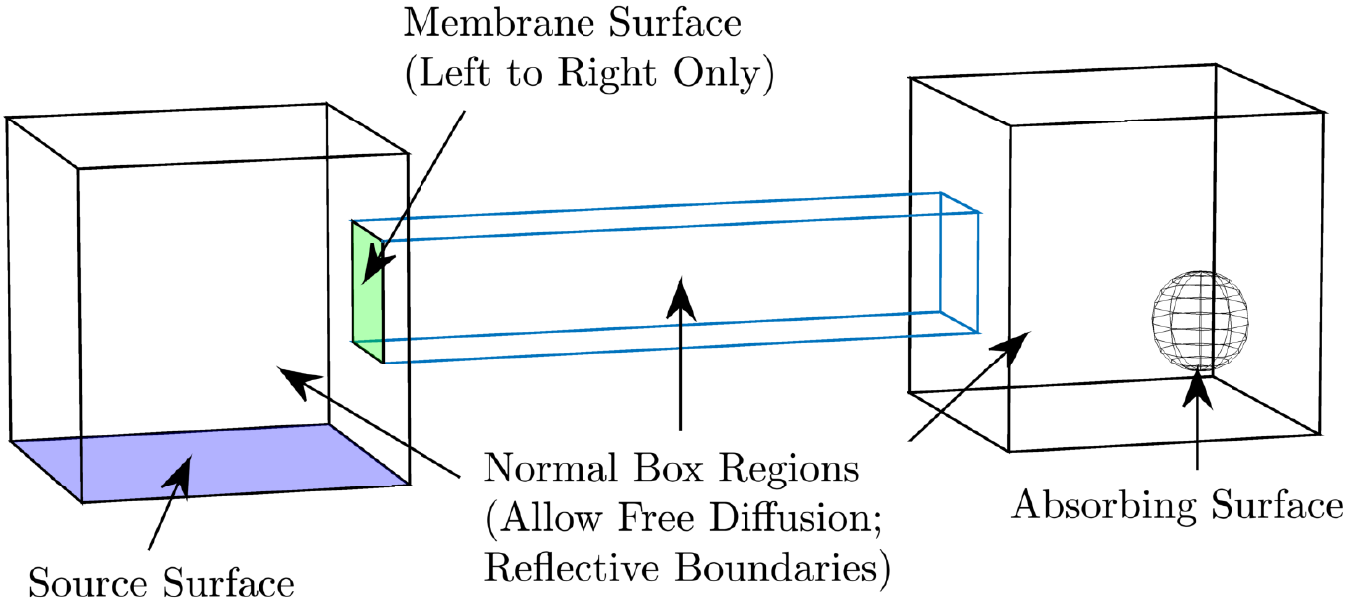}
	\caption{Example of a complex simulation environment with many of the system model components described in Section~\ref{sec_components}. There are normal and surface regions, actors, and multiple chemical reactions. The entire system is microscopic. Two normal cubic regions are joined by a normal rod region with a membrane surface region at one end (green), such that the membrane's neighbors are the left cube and the rod, and the rod's neighbors are the membrane and the right cube. The membrane enables molecules to pass from left to right only. The left cube has a neighboring surface region (blue) that is a source of molecules. The only \emph{parent/child} region relationship is between the right cube and the absorbing spherical surface inside of it. Actors observe the entire environment so that molecule coordinates are generated for the sample video, which is available as Video 1 in \cite{Noel2016c}.}
	\label{fig_fancy_env}
\end{figure}

We have introduced AcCoRD as a ``sandbox'' reaction-diffusion solver for molecular communications design and analysis. Therefore, before we discuss the underlying theory (in Section~\ref{sec_theory}) or the implementation of AcCoRD's algorithms (in Section~\ref{sec_algorithm} and~\ref{app_algorithms}), we describe the system model components that are available to the user. In this section, we present these components, which are categorized into \emph{regions}, \emph{actors}, \emph{chemical parameters}, and \emph{global settings}. We also summarize how these components are included in an AcCoRD configuration file. A sample environment demonstrating many of these components is shown in Fig.~\ref{fig_fancy_env} (which is also shown in Video 1 in \cite{Noel2016c}). Throughout this section, most parameters that can be defined by the configuration are \emph{italicized}. A listing of constraints and limitations on model components are included in~\ref{app_constraints}.

\subsection{Regions}

Regions are the literal ``building blocks'' of a simulation environment. They define the physical space where molecules can move or be created. Other components, i.e., actors and chemical reactions, can be defined for all or some subset of an environment's regions. As long as some (minor) constraints are satisfied, regions can be placed in solitude, adjacent to other regions, or nested inside of other regions. A \emph{parent} region is one that has another region (i.e., its \emph{child}) nested inside of it. Generally, if a region's face is adjacent to another region, then those regions are automatically classified as neighbors. As long as neighboring regions are not separated by a surface region that restricts molecule transitions, then molecules can move freely between them.

There are three ways to classify a region:
\begin{enumerate}
	\item \emph{Shape}. AcCoRD is primarily 3D and the implemented 3D shapes are spheres and rectangular boxes. 2D rectangles are also implemented.
	\item \emph{Type}. A region can be \emph{normal} or a \emph{surface}. Normal regions occupy all of the space that they are defined in, except where they are a \emph{parent} to other regions nested inside. A 3D surface region can have either a 3D shape, such as a box or sphere, or it can also have a 2D shape (i.e., a rectangle). 2D rectangle regions are implemented but have only been tested as surfaces to normal 3D regions. Surfaces also have an associated ``direction'' as defined by the \emph{surface type}. A one-sided surface means that molecules can only approach and interact with one of the two sides and not both. A two-sided surface, defined as a \emph{membrane}, can have molecules interact from either side. However, all surfaces are reflecting by default, unless the user defines chemical reactions at a surface to enable other interactions (as described later in this section).
	\item \emph{Model}. Each region uses either a \emph{microscopic} or \emph{mesoscopic} model. All molecules in microscopic regions are tracked individually and evolve according to a global \emph{microscopic time step}. Mesoscopic regions keep track of how many of every type of molecule are in each subvolume. In a given simulation, we define the union of all microscopic regions as the \emph{microscopic regime}, and the union of all mesoscopic regions as the \emph{mesoscopic regime}. \emph{All} regions, whether they are microscopic or mesoscopic, are partitioned into \emph{subvolumes}. Subvolumes are the foundation of a mesoscopic model, but they are also useful for the implementation of microscopic regions when determining where regions are located relative to each other. Spherical regions are always microscopic and have only one subvolume.
\end{enumerate}

The location of a region is defined by its \emph{anchor coordinate}, which for example is the center of a sphere or the ``lower corner'' $\{\x,\y,\z\}$ coordinate of a box. The size of a sphere is defined by its \emph{radius}, which can in principal be any positive real number (including $\infty$ for an unbounded environment). The sizes of boxes are constrained by a global \emph{subvolume base size} $\subBase$. The length of a box along each dimension $\{\x,\y,\z\}$ is defined as an integer multiple of $\subBase$, and each subvolume in a box is a cube whose length, the \emph{region subvolume length} $\subLength{}$, is also an integer multiple of $\subBase$. We discuss some of the physical and computational considerations for choosing an appropriate subvolume length in Section~\ref{sec_complexity}. 2D rectangle regions have a length of 0 along one of the three dimensions. Defining $\subBase$ facilitates the placement of regions (and also actors) by avoiding issues due to floating point operations (e.g., when $2\times1\neq2$ due to different data types or floating point rounding errors).

Given the above definitions, regions can be placed almost arbitrarily. Generally, regions can be placed adjacent to each other or inside of each other (i.e., nested), such that the \emph{parent} outer region is identified by the \emph{child} inner region. Any two regions that are adjacent or has one nested directly inside the other are detected as \emph{neighbors}. Surface regions are usually only defined along the outer boundaries of normal regions, unless the surface is a 3D shape nested inside of a normal 3D region. For clarity of exposition, we list more details of the constraints on region placement in~\ref{app_constraints}.

There are two additional properties that are not specific to any one region but are needed when a simulation has both a microscopic regime and a mesoscopic regime. These properties control how molecules can transition from one regime to the other (i.e., at the hybrid interface), and are described in further detail in Section~\ref{sec_diff_hybrid}. The first property selects whether the mesoscopic subvolumes at the hybrid interface are assumed to be small relative to the average displacement in the microscopic regime. The second property defines how far away a microscopic molecule should be from the boundary, either before or after it diffuses, to ignore the possibility that it entered the mesoscopic regime \emph{during} the diffusion step.

\subsection{Actors}

Actors in AcCoRD provide the interface to a simulation by enabling input of molecules or observing molecules as output. The two classes of actors are \emph{active} and \emph{passive}. These classes share some common properties but otherwise have unique behavior. Active actors (e.g., transmitters) enable input by creating molecules according to the modulation of a binary data sequence. Passive actors (e.g., receivers) enable output by observing the number of molecules that are present at some location. Both classes of actors are currently immobile and \emph{independent}, i.e., their behavior is established by the user and does not depend on that of the system or other actors. However, the generic actor design facilitates the future implementation of \emph{dependent} actors which will have coupled behavior. We discuss this further in Section~\ref{sec_future}.

All actors can be placed using one of two methods. In the first method, an actor is defined as the union of a subset of the regions. This method is preferred when an actor is intended to exist over an entire region or regions, and enables an actor to be disjoint or to have non-standard shapes (e.g., if an actor is defined either over disjoint regions or over some region and not that region's nested children). In the second method, an actor is defined as a single virtual shape. Currently, all region shapes are valid actor shapes, and it is also possible to define an active actor as a \emph{point}. Actors defined with the second method are always defined over the entire specified shape (i.e., and not just its surface).

It is important to clarify that actors are virtual and \emph{do not} have a tangible physical presence in a simulation environment. They only define where molecules can be added or observed. Regions must be used to impose physical boundaries. For example, if a user wants a receiver that is an absorbing sphere, then the configuration needs to define a spherical surface region, the absorbing chemical reaction at the surface (see the following subsection), and a passive actor to observe the molecules that are absorbed. However, if the user wants a receiver that is a passive sphere within some larger environment, then it is most computationally efficient to just define a spherical actor without an accompanying region (due to the significant overhead to identify transitions between microscopic regions).

Every actor has its own \emph{start time} to begin its first action (i.e., releasing or observing molecules). The actions are repeated according to an \emph{action interval}, until either a \emph{maximum number of actions} has been reached or the simulation time is complete.

The release pattern of an active actor is based on its binary \emph{bit sequence}, which can be defined by the user or randomly generated according to the specified independent probability of bit-1. Multiple bits can be used in each action interval (i.e., symbol interval). Currently, \emph{concentration shift keying} (CSK) is implemented, where one type of molecule is released. Other types of modulation can be added in future development; see \cite{Kuran2011} for examples. Molecules for each symbol are released over a \emph{release interval}, which can be as short as 0 seconds or as long as desired (even longer than the action interval). An actor can be configured to behave as a noise source and release molecules continuously by setting the probability of bit-1 to 1 and the release interval to be equal to the action interval. The precise number of molecules created and their release times within the release interval can be random (i.e., generated via a Poisson arrival process, such that the time between consecutive molecule releases is an exponential random variable) or deterministic (by dividing the release interval into discrete \emph{slots} and releasing molecules at the start of each slot). Every added molecule is placed at a uniformly-distributed random location within the actor. If an active actor is being recorded, then its bit sequence is written to the simulation output file.

The behavior of passive actors is simpler. Each passive actor acts by counting the number of molecules of the \emph{specified types} that are within their defined space. It is also possible to \emph{record molecule positions}, which are copied for molecules in the microscopic regime and randomly generated within the corresponding subvolume in the mesoscopic regime. Another option is to \emph{save the environment time} associated with each observation.

\subsection{Chemical Properties}

The chemical properties describe the molecules that can exist in the environment and how they can interact via reactions. Even if there are no chemical reactions, the number of molecule types and their \emph{diffusion coefficients} must be defined. There are two methods to represent different states of the same molecule species, for example to specify that a molecule should not diffuse once it has been absorbed by a surface. In the first method, the user can define two types of molecules, set one of them to have a diffusion coefficient of zero, and make that molecule the product of the other molecule's absorption reaction. In the second method, the user can define one type of molecule and set that molecule's local diffusion coefficient to zero at the surface region (via an optional region property).

\emph{Chemical reactions} define processes for molecules to be created, destroyed, or transformed; refer to \cite[Ch.~9]{Chang2005} for an elementary discussion. Every chemical reaction has a set of \emph{reactants} (which the reaction consumes), a set of \emph{products} (which the reaction creates), and some way to indicate the likelihood of the reaction occurring (typically a \emph{reaction rate}). A molecule that is a reaction catalyst and facilitates the reaction without being consumed (i.e., an enzyme) can be represented as both a reactant and a product. Reactions with no reactants, one reactant, and two reactants are referred to as \emph{zeroth order}, \emph{first order}, and \emph{second order}, respectively. Reactions with more reactants are typically decomposed into a sequence of elementary reactions that are zeroth, first, or second order. First and second order reactions are also known as unimolecular and bimolecular reactions, respectively.

We classify each chemical reaction in AcCoRD as either a \emph{surface reaction} or a \emph{non-surface reaction}. Surface reactions can be either 1) an interaction between one molecule and a surface \emph{region}, e.g., \emph{absorption} (i.e., consumption) by a surface, \emph{desorption} (i.e., release) from a surface, and transition across a \emph{membrane} surface, \emph{or} simply 2) a reaction where the reactants \emph{must} be on the surface. We consider adsorption (i.e., sticking to a surface) to be reversible absorption. Non-surface reactions do not require a surface region, but can also occur at a surface. Ligand-receptor surface binding is \emph{not} considered a surface reaction under our definition, because one of the reactants (i.e., the ligand) is initially not on the surface and it reacts with the receptor and \emph{not} the surface\footnote{Our classification of reactions is similar but not identical to classifying reactions as heterogeneous or homogeneous; see \cite[Ch.~15]{Petrucci2016}. Heterogeneous reactions have reactants, products, or catalysts in different states (e.g., liquid and solid), whereas homogeneous reactions occur entirely in a single state. Surface reactions in AcCoRD are heterogeneous when a molecule interacts with a surface region, but non-surface reactions can also be heterogeneous, such as in the case of ligand-receptor surface binding.}. Every chemical reaction is associated with a default list of regions where it can occur, and this can be set as \emph{everywhere} (i.e., in all surface regions for a surface reaction or in all normal regions for a non-surface reaction) or nowhere. Regions that are exceptions to the default placement can also be listed.

Reaction rates for zeroth, first, and second order reactions have units $\mol\cdot\second^{-1}\meter^{-3}$, $\second^{-1}$, and $\mol^{-1}\second^{-1}\meter^{3}$, respectively, where ``$\mol$'' is number of molecules. Second order reactions that can occur in the microscopic regime also need a \emph{binding radius} $\rBind$ which is used instead of the reaction rate (in general, the binding radius can be derived from the reaction rate, but this is not currently part of AcCoRD's implementation; see \cite{Andrews2004} for further details). Bimolecular reactions in the microscopic regime with more than one product also have an \emph{unbinding radius} $\rUnbind$ to define how far apart products should be placed. We note that the random generation of molecules by an active actor is also modeled as a zeroth order reaction but is distinct from other zeroth order reactions because the generation rate for the actor is modulating a data sequence.

The named surface interaction reactions, i.e., absorption, desorption, and membrane reactions, are assumed to be first order (since they have one molecule interacting with a surface) and have additional properties. These reactions can be classified as \emph{reversible} if they define a corresponding \emph{reverse reaction}. They also need a \emph{surface transition probability}, which determines the calculation in \cite{Andrews2009} (and summarized in~\ref{app_surf}) that we use to calculate the probability of the reaction occurring. The three options for this probability are as follows:
\begin{enumerate}
	\item \emph{``Normal''}: The reaction rate is treated as if the reaction is a \emph{non-surface} first order reaction. This option can accommodate ``perfect'' absorption (i.e., when the absorption rate is infinite), but otherwise its application is limited.
	\item \emph{``Mixed''}: We assume that the distribution of molecules near the surface is in the well-mixed state, such that the concentration is uniform. This option is fast to calculate and accurate for systems that are actually well-mixed.
	\item \emph{``Steady state''}: We assume that the distribution of molecules near the surface is in the steady state, such that the concentration is constant but can vary locally. This option was shown in \cite{Andrews2009} to have excellent accuracy, even when the system is transient. Also from \cite{Andrews2009}, the reverse reaction rate is accounted for when determining this reaction probability.
\end{enumerate}

Membrane reactions should not have any product molecules defined (since we assume that the product molecule is the same type as the reactant) and they need to be classified as \emph{inner} or \emph{outer} to indicate the side of the membrane that the reactant can approach from. Non-membrane surface reactions need to explicitly indicate whether each product is automatically detached from the surface (although this is generally intended only for desorption reactions) and if so then how the products are placed. The options for placing detached product molecules are as follows:
\begin{enumerate}
	\item \emph{``Leave''}: Molecules are placed in the normal region at the point where they were bound to the surface. This option is the simplest but is only accurate if the molecule detaches at the end of the microscopic time step.
	\item \emph{``Full diffusion''}: Molecules diffuse perpendicular to the surface for the elapsed time since the detachment occurred. This option should be the most accurate in cases of irreversible desorption.
	\item \emph{``Steady state diffusion''}: Molecules diffuse perpendicular to the surface for an unknown period of time that is no greater than the microscopic time step. This option is the most appropriate when the detachment reaction has a reverse adsorption reaction whose surface transition probability is steady state, because then the precise adsorption time is also unknown. The placement distribution for this method was derived in \cite{Andrews2009} in consideration of the steady state transition probability.
\end{enumerate}

The methods for placing molecules according to the ``full diffusion'' and ``steady state diffusion'' options are discussed in~\ref{app_surf_desorption}.

\subsection{Other Settings}

Every simulation starts at time $t=0$ but needs a specified \emph{final time} and \emph{number of times to be repeated}. The \emph{random number seed} is used to initialize the PCG RNG (see \cite{ONeill}), although it can be over-ridden with a different seed via an extra command line input argument. The \emph{prefix of the simulation output} file is specified such that the random number seed (that is actually used) is appended to it. Finally, the user can also choose to \emph{override the display of warnings} from the loading of the configuration file and control the \emph{maximum number of progress updates} that will be printed to the console screen with an estimate of the remaining run time to complete the simulation. There will be no more than one progress update per independent realization, since it is difficult to estimate the time remaining while within a realization.

\subsection{Configuration Files}

AcCoRD's configuration files are written in JavaScript Object Notation (JSON) format. JSON is a lightweight data interchange format, easy to read, and has parsers available in many programming languages; see \cite{Crockford}. AcCoRD's configuration files accommodate all of the system model components introduced in this section with the following consistent format:
\begin{itemize}
	\item Output Filename
	\item Switch to over-ride configuration warnings
	\item Simulation Control
	\begin{itemize}
		\item Number of repeats and random number seed $\seed$
		\item Final simulation time $\tEnd$ and microscopic time step $\dtMicro$
		\item Hybrid interface parameters
	\end{itemize}
	\item Chemical Properties
	\begin{itemize}
		\item Number of types of molecules and their diffusion coefficients $\Dx{\indMolType}$
		\item One object describing each chemical reaction
	\end{itemize}
	\item Environment
	\begin{itemize}
		\item Subvolume Base Size $\subBase$
		\item Region Specification
		\begin{itemize}
			\item One object describing each region
		\end{itemize}
		\item Actor Specification
		\begin{itemize}
			\item One object describing each actor
		\end{itemize}
	\end{itemize}
\end{itemize}

As long as the correct nesting of parameters is maintained, JSON files can be arbitrarily reordered. Even though JSON does not support comments, extra fields can be created to include additional information. Many of the sample configuration files included with the AcCoRD source code have ``Notes'' fields for comments.

\section{Reaction-Diffusion Theory and Algorithms}
\label{sec_theory}

In this section, we describe the underlying theory for the microscopic and mesoscopic simulation models. We present the equations and algorithms for modeling diffusion, chemical reactions in solution, and surface reactions. This section includes our derivation of the transition rate between mesoscopic subvolumes of different sizes, and the continuous reaction event time for first order chemical reactions that are known to have occurred in the current microscopic time step.

Before we describe the analytical details of the individual phenomena, we establish some global model definitions and notation. The total simulation environment is partitioned into a set of regions $\setRegion$. Each region is either in the microscopic regime $\regimeMicro$ or the mesoscopic regime $\regimeMeso$, and $\setRegionNeigh$ is the set of regions that are neighbors of region $\indRegion$. The mesoscopic regime is partitioned into the set of subvolumes $\setSub$. The set of defined molecule types is $\setMolTypes$ and the set of chemical reactions is $\setChemRxn$. The number of molecules of the $\indMolType$th type that are in the $\indSub$th mesoscopic subvolume is $\numMolSub$.

\subsection{Model Evolution}

The most intuitive way to distinguish between the two simulation models is to summarize how they behave over time and space. The microscopic model describes molecule locations over continuous space but evolves in discrete time steps of constant size $\dtMicro$. For each time step, \emph{every} microscopic molecule can diffuse and react.

The mesoscopic model describes molecule locations over a discrete grid (of subvolumes) but evolves over continuous time. In its most accurate form (i.e., without tau-leaping, which we will consider in future development), only \emph{one} mesoscopic event occurs at a time. The potential events are every possible chemical reaction in each subvolume and every possible transition via diffusion between adjacent subvolumes. Each potential event is assigned a \emph{propensity}, $\prop{}$, such that the probability of the event occurring within the infinitesimal time step $\dtInf$ is $\prop{}\dtInf$.

The integration of the two models within the overall AcCoRD simulation algorithm is described in Section~\ref{sec_algorithm}, along with a discussion of their computational complexity and comments on the appropriate selection of time step $\dtMicro$ and the mesoscopic subvolume size. The implementation of both individual models is discussed in greater detail in~\ref{app_algorithms}.

\subsection{Diffusion}
\label{sec_diff}

Diffusion is modeled in both regimes with diffusion coefficients, which describe the variance of motion of individual molecules by assuming that the molecules are moving in a continuum of solvent (e.g., air, water, or blood). The diffusion coefficient of the $\indMolType$th molecule type is $\Dx{\indMolType}$, which we will occasionally write as $\Dx{}$ when referring to an arbitrary molecule.

\subsubsection{Diffusion in One Regime}

In the microscopic regime, every molecule tries to diffuse in every time step. In the absence of collisions with surfaces or other molecules, the displacement of a single molecule in one time step is $\normRV{}\sqrt{2\Dx{\indMolType}\dtMicro}$ along each dimension, where $\normRV{}$ is a normal random variable with mean 0 and variance 1.  $\normRV{}$ is independently drawn in each dimension for every molecule.

In the mesoscopic regime, it is assumed that all molecules within a given subvolume are uniformly distributed. The propensity that some molecule diffuses from a subvolume into a neighboring subvolume is directly proportional to the number of molecules in that subvolume that are of the same type. For a uniform grid of square or cubic subvolumes of length $\subLength{}$, where neighboring subvolumes share an \emph{entire} face, the diffusion propensity for the $\indMolType$th molecule type from the $\indSub$th subvolume to the $\indSubNeigh$th subvolume is given as \cite[Eq.~(1.6)]{Flegg2014}
\begin{equation}
\label{eq_meso_diff}
\prop{\indSub,\indSubNeigh,\indMolType} = \frac{\Dx{\indMolType}}{\subLength{}^2}\numMolSub,
\end{equation}
assuming that subvolumes $\indSub$ and $\indSubNeigh$ share a face (otherwise, $\prop{\indSub,\indSubNeigh,\indMolType} = 0$).

In AcCoRD, mesoscopic subvolumes can have different sizes and so they might have faces that are only partially shared. Thus, we need a more general expression to describe the diffusion propensity. As we considered for the 2D case in \cite{Noel2015a}, we start with the propensity from a 1D subvolume of length $\subLength{\indSub}$ to one of a different length $\subLength{\indSubNeigh}$. This was previously derived as \cite[Eq.~(15)]{Bernstein2005}
\begin{equation}
\label{eq_meso_diff_1d}
\prop{\indSub,\indSubNeigh,\indMolType} = \frac{2\Dx{\indMolType}}{\subLength{\indSub}(\subLength{\indSub} + \subLength{\indSubNeigh})}\numMolSub.
\end{equation}

The advantage of (\ref{eq_meso_diff_1d}) over (\ref{eq_meso_diff}) is that (\ref{eq_meso_diff_1d}) accounts for the size of the destination subvolume while satisfying Fick's law for the diffusion flux. The impact is that a molecule will be less likely to enter a larger neighbor than one that is the same size. While this might be an unintuitive result, it can be shown that (\ref{eq_meso_diff_1d}) will maintain a uniform distribution of molecules between the two subvolumes whereas (\ref{eq_meso_diff}) will not.

\begin{figure}[!tb]
	\centering
	\includegraphics[width=3in]{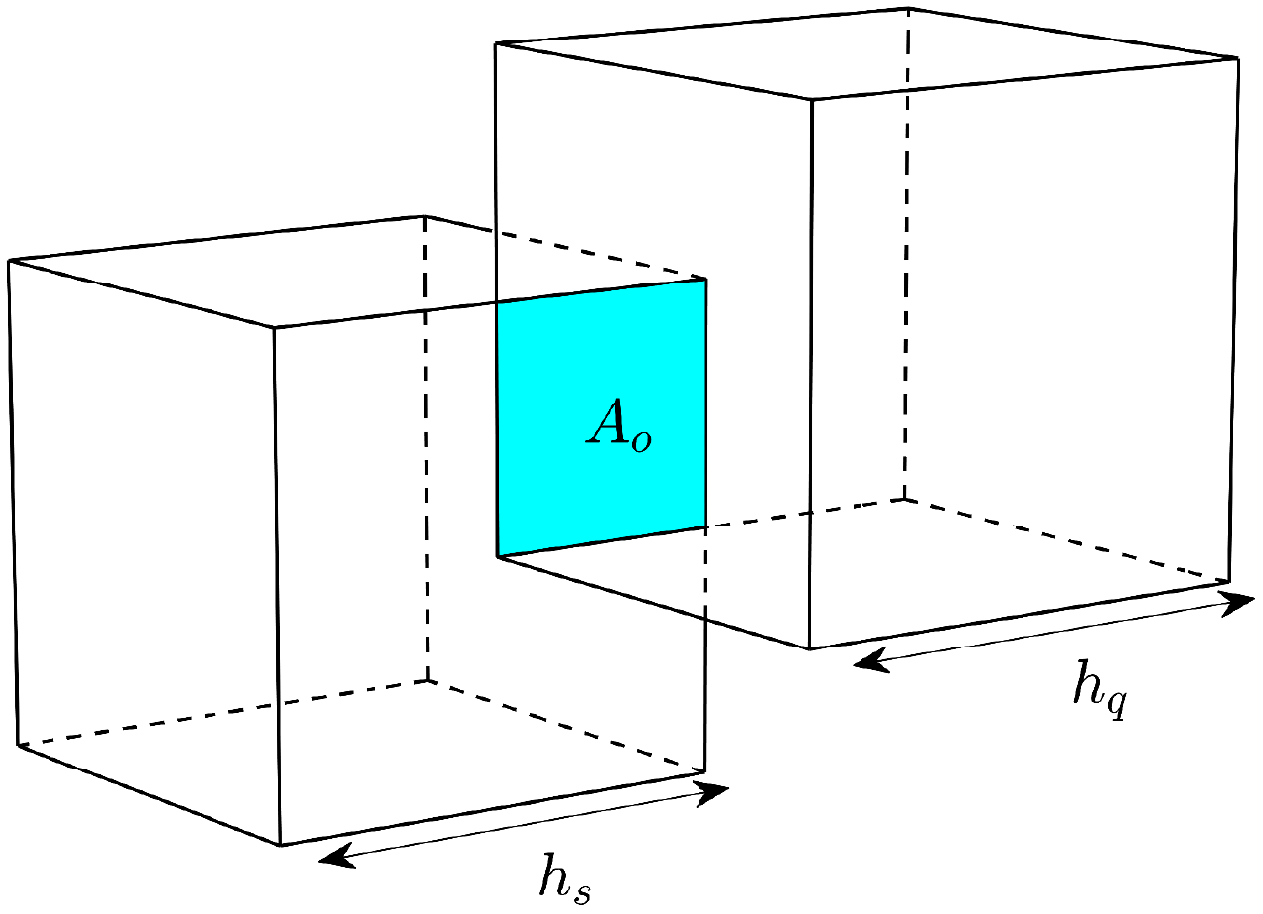}
	\caption{Two 3D mesoscopic subvolumes that partially overlap. The overlap area with size $\areaOverlap$ is shown in aqua blue. Such overlap areas can occur in order to accommodate regions that have subvolumes of different sizes or complex environment geometries. In this example, the two subvolumes actually have equal widths, i.e., $\subLength{\indSub} = \subLength{\indSubNeigh}$.}
	\label{fig_sub_overlap}
\end{figure}

Eq.~(\ref{eq_meso_diff_1d}) immediately applies to the 3D case \emph{if} the face shared by subvolume $\indSub$ with subvolume $\indSubNeigh$ is completely covered by subvolume $\indSubNeigh$. However, this does not always occur in the 3D case. In \cite{Noel2015a}, we scaled (\ref{eq_meso_diff_1d}) by the \emph{relative overlap length} for the 2D case, i.e., the fraction of the line segment that subvolume $\indSub$ shares with subvolume $\indSubNeigh$. Here, we consider the \emph{overlap area}, $\areaOverlap$, which is the size of the shared surface between subvolume $\indSub$ and subvolume $\indSubNeigh$ (see Fig.~\ref{fig_sub_overlap}). The likelihood of a molecule diffusing from subvolume $\indSub$ to subvolume $\indSubNeigh$ should be scaled by $\areaOverlap/\subLength{\indSub}^2$, which is the fraction of the face of subvolume $\indSub$ that is actually shared with subvolume $\indSubNeigh$. Thus, we can write the propensity for the $\indMolType$th molecule type to diffuse from subvolume $\indSub$ to subvolume $\indSubNeigh$ as
\begin{equation}
\label{eq_meso_diff_3d}
\prop{\indSub,\indSubNeigh,\indMolType} = \frac{2\Dx{\indMolType}\areaOverlap}{\subLength{\indSub}^3(\subLength{\indSub} + \subLength{\indSubNeigh})}\numMolSub.
\end{equation}
where $\areaOverlap \le \min\{\subLength{\indSub}^2,\subLength{\indSubNeigh}^2\}$.

\subsubsection{Hybrid Diffusion Between Microscopic and Mesoscopic Regimes}
\label{sec_diff_hybrid}

When both regimes exist, then we have hybrid diffusion and we need to account for diffusion across the interface between the two regimes. We adopt the transition rules described in \cite{Flegg2014}. The rules define how an individual molecule in the microscopic regime can be placed in a mesoscopic subvolume, and how a diffusion event in a mesoscopic subvolume can result in a new microscopic molecule. We summarize both processes here.

A molecule in the microscopic regime has two ways to enter the mesoscopic regime. First, a molecule is automatically placed in a mesoscopic subvolume if its diffusion trajectory crosses the subvolume. Otherwise, if the molecule's destination region (i.e., after diffusion) has mesoscopic neighbors, then we still consider the possibility that the molecule entered the mesoscopic regime \emph{during} the time step (since actual diffusion is not along a straight line). The probability that the molecule entered mesoscopic region $\indRegion$ \emph{within} arbitrary time step $\dtInf$, $\ProbIntra$, is given by \cite[Eq.~(1.9)]{Flegg2014}
\begin{equation}
\label{eq_micro_to_meso}
\ProbIntra = \EXP{-\frac{\hybridDist{\indRegion,i}\hybridDist{\indRegion,f}}{\Dx{\indMolType}\dtInf}},
\end{equation}
where $\{\hybridDist{\indRegion,i},\hybridDist{\indRegion,f}\}$ are the closest distances from the molecule to region $\indRegion$ at the start and end of the diffusion step, respectively. We introduce a maximum distance $\hybridDist{\mathrm{max}}$ to control computational requirements, i.e., (\ref{eq_micro_to_meso}) is ignored if either of $\{\hybridDist{\indRegion,i},\hybridDist{\indRegion,f}\}$ are greater than $\hybridDist{\mathrm{max}}$. Alternatively, defining relatively small microscopic regions that neighbor the mesoscopic regime will also limit the number of checks using (\ref{eq_micro_to_meso}), although this will also increase the number of checks for diffusion across microscopic region boundaries.

For molecules that originate in the mesoscopic regime, we need to consider the propensity for a molecule to enter the microscopic regime and where to place the molecule when a diffusion event occurs. For ease of implementation, we assume that if a mesoscopic subvolume is adjacent to microscopic region $\indRegion$ along some face, then the \emph{entire} face is shared with region $r$, i.e., the overlap area for subvolume $\indSub$ is $\areaOverlap = \subLength{\indSub}^2$. Thus, the propensity for the $\indMolType$th  molecule type to diffuse from subvolume $\indSub$ to region $\indRegion$ is \cite[Eqs.~(1.10), (1.11)]{Flegg2014}
\begin{equation}
\label{eq_prop_hybrid}
\prop{\indSub,\indRegion,\indMolType} = \frac{2}{\subLength{\indSub}}
\sqrt{\frac{\Dx{\indMolType}}{\pi\dtMicro}}\numMolSub,
\end{equation}
which replaces (\ref{eq_meso_diff_3d}) when the neighboring subvolume is in a microscopic region. We note that (\ref{eq_prop_hybrid}) depends on the microscopic time step $\dtMicro$, so a smaller time step will result in a higher propensity to leave subvolume $\indSub$.

\begin{figure}[!tb]
	\centering
	\includegraphics[width=3in]{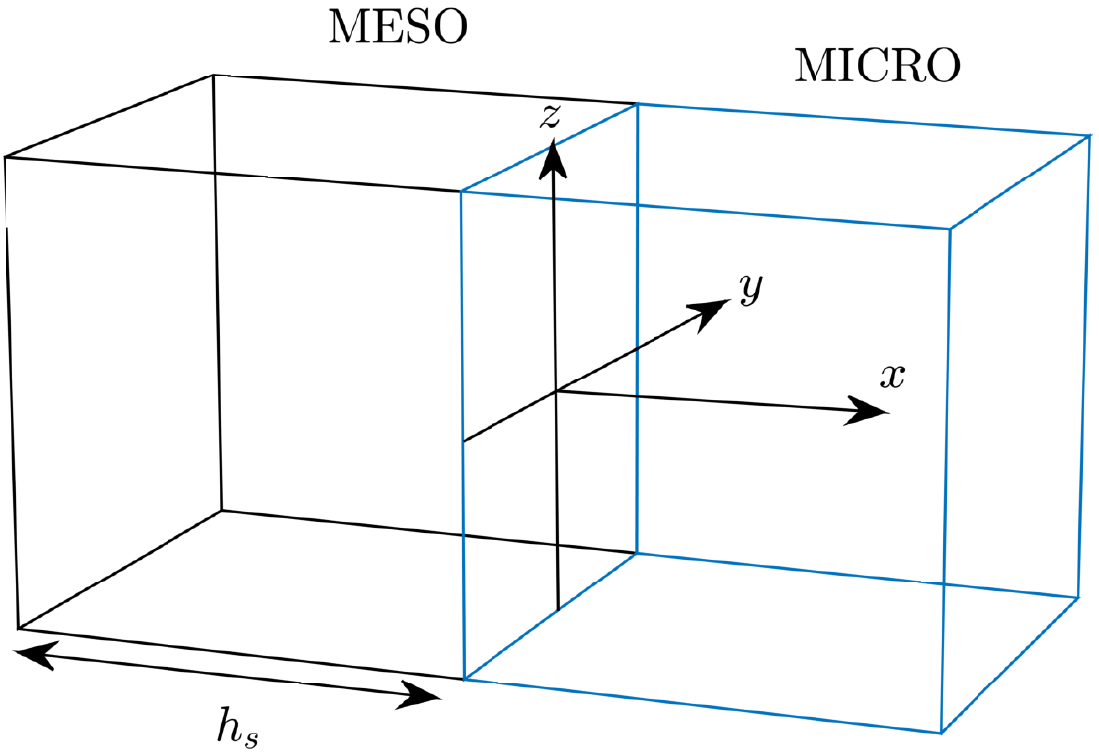}
	\caption{View of hybrid interface between a mesoscopic subvolume (``MESO'') and a microscopic region (``MICRO''). Coordinates are consistent with the microscopic molecule placement distributions in (\ref{eq_pdf_hybrid_normal}), (\ref{eq_hybrid_tangential}), and (\ref{eq_hybrid_tangential_large}).}
	\label{fig_hybrid_interface}
\end{figure}

If a diffusion event defined by the propensity in (\ref{eq_prop_hybrid}) occurs, then a microscopic molecule must be created. Specifically, tangential and normal coordinates must be chosen, since the molecule is able to diffuse away from the mesoscopic surface until the next microscopic time step. Without loss of generality in presentation (since any other orientation is analogous), we assume that the diffusion event occurs at a subvolume $\indSub$ whose shared face with the microscopic region is in the $\y\z$-plane and centered at the origin with the microscopic region in positive $\x$, as shown in Fig.~\ref{fig_hybrid_interface}. Then, the probability distribution to initialize the microscopic molecule in the $\x$-direction is \cite[Eq.~(1.12)]{Flegg2014}
\begin{equation}
\label{eq_pdf_hybrid_normal}
\pdf{\x}(\x) = \sqrt{\frac{\pi}{4\Dx{\indMolType}\dtMicro}}\ERFC{\frac{\x}{\sqrt{4\Dx{\indMolType}\dtMicro}}}, \quad \x > 0,
\end{equation}
where $\ERFC{\xi} = 1 - \ERF{\xi}$ is the complementary error function (from \cite[Eq.~(3.1.2)]{Ng1968}), and the error function is \cite[Eq.~(3.1.1)]{Ng1968}
\begin{equation}
\label{eq_erf}
\ERF{\xi} = \frac{2}{\sqrt{\pi}}\int\limits_0^\xi \EXP{-\gamma^2} \mathrm{d}\gamma.
\end{equation}

As presented in \cite[Eq.~(35)]{Andrews2009}, an efficient approximation to sample from the distribution in (\ref{eq_pdf_hybrid_normal}) is to generate a uniform random number $\uniRV{}$ between 0 and 1 and then $\x$ is
\begin{equation}
\label{eq_hybrid_normal}
\x = \frac{0.729614\uniRV{} - 0.70252\uniRV{}^2}
{1 - 1.47494\uniRV{} + 0.484371\uniRV{}^2} \sqrt{2\Dx{\indMolType}\dtMicro}.
\end{equation}

For the tangential directions $\{\y,\z\}$ there are two models presented in \cite{Flegg2014}. Both models make assumptions about the relative size of the mesoscopic subvolumes in order to simplify the infinite sum that appears when balancing the diffusion distributions at the interface. The models either assume that the mesoscopic subvolumes are relatively \emph{small} (i.e., that the spatial resolution of the mesoscopic regime is comparable to that of the microscopic regime) or that the subvolumes are relatively \emph{large} (i.e., that the spatial resolution of the mesoscopic regime is larger than that of the microscopic regime). In the first model, where the mesoscopic subvolume is assumed to be relatively \emph{small}, the model assumes that $\subLength{\indSub} \sim \sqrt{\Dx{\indMolType}\dtMicro}$ as $\subLength{\indSub}\to 0$ and $\dtMicro \to 0$. This assumption, which is adopted by the hybrid implementation in Smoldyn (see \cite{Robinson2015}), uses a triangular distribution to place the molecule along each tangential dimension, i.e., \cite[Eq.~(2.14)]{Flegg2014}
\begin{equation}
\label{eq_hybrid_tangential}
\pdf{\y}(\y) = \frac{1}{\subLength{\indSub}}\left(1 - \frac{|\y|}{\subLength{\indSub}}\right), \quad -\subLength{\indSub} < \y < \subLength{\indSub},
\end{equation}
for $\y$ (and analogously for $\z$). In the second model, where the subvolume is assumed to be relatively \emph{large}, the model assumes that $\subLength{\indSub} \gg \sqrt{\Dx{\indMolType}\dtMicro}$. This results in a uniform distribution along each tangential dimension, i.e., \cite[Eq.~(2.16)]{Flegg2014}
\begin{equation}
\label{eq_hybrid_tangential_large}
\pdf{\y}(\y) = \frac{1}{\subLength{\indSub}}, \quad -\frac{\subLength{\indSub}}{2} < \y < \frac{\subLength{\indSub}}{2},
\end{equation}
for $\y$ (and analogously for $\z$). We implement \emph{both} the large and small hybrid subvolume models in AcCoRD, so the user can choose the most appropriate model in consideration of the relative size of the subvolumes at the hybrid interface.

\subsection{Chemical Reaction Kinetics}
\label{sec_theory_rxn}

The accurate modeling of a chemical reaction relies on knowledge of the reaction's \emph{elementary} steps, which describe the actual reaction steps that occur at the molecular level; see \cite[Ch.~9]{Chang2005}. If a reaction's elementary steps are known, then the reaction kinetics can be determined and simulation is possible. The three common classes of elementary reactions are zeroth order, first order, and second order (higher order elementary reactions, which rely on the unlikely simultaneous collision of three or more molecules, are very rare). Here, we describe each of the three common orders and how they are simulated. We also discuss the special cases of surface reactions.

\subsubsection{Zeroth Order Reactions}

Zeroth order reactions are of the form
\begin{equation}
\label{eq_zeroth_order}
\emptyset \xrightarrow{\kth{0}} \mathrm{Products},
\end{equation}
such that product molecules are spontaneously created in the propagation environment. The zeroth order reaction rate constant, $\kth{0}$, with units $\mol\cdot\second^{-1}\meter^{-3}$, specifies the number of molecules created per second per cubic meter of solvent. When a zeroth order reaction is defined on a 2D surface, we define the reaction rate with units $\mol\cdot\second^{-1}\meter^{-2}$. We note that a true zeroth order chemical reaction is unphysical, but it models a source that adds molecules to the environment at some rate. In fact, we implement the addition of molecules by active actors as zeroth order reactions with a time-varying rate that is modulated by its data sequence (unless the actor is configured to release its molecules instantaneously).

A zeroth order reaction is a Poisson process, i.e., the times between reaction events are continuous, independent, and have a constant rate; see \cite[Ch.~5]{Ross2009}. Realizations of the time $\timeRxn{\indChemRxn}$ between consecutive events of zeroth order reaction $\indChemRxn$ with rate $\kth{0}=\kth{\indChemRxn}$ in region $\indRegion$ can be obtained by generating exponential random variables via the inverse transform method, i.e., \cite[Ch.~15]{Ross2009}
\begin{equation}
\label{eq_t_zeroth}
\timeRxn{\indChemRxn} = -\frac{\log{\uniRV{}}}{\kth{\indChemRxn}\Vol{\indRegion}},
\end{equation}
where $\uniRV{}$ is an independent random number uniformly distributed between 0 and 1, $\Vol{\indRegion}$ is the volume (or area if applicable) of region $\indRegion$, and $\log$ is the natural logarithm. In the microscopic regime, we create molecules on a continuous timescale, so zeroth order reactions are directly simulated using (\ref{eq_t_zeroth}). Molecule locations are created uniformly within the region. In the mesoscopic regime, we need to represent the reaction with a corresponding propensity. We will see in~\ref{sec_alg_meso} that \emph{all} events in the mesoscopic regime occur at times according to a sequence of exponential random variables. For zeroth order reaction $\indChemRxn$, the propensity in subvolume $\indSub$ is \cite[Eq.~(6)]{Bernstein2005}
\begin{equation}
\label{eq_prop_zeroth}
\prop{\indSub,\indChemRxn} = \kth{\indChemRxn}\Vol{\indSub},
\end{equation}
where $\Vol{\indSub}$ is the volume (or area) of subvolume $\indSub$.

In the case of actor $\indActor$ being active and releasing molecules at random times, we use (\ref{eq_t_zeroth}) in both the microscopic and mesoscopic regimes by replacing $\Vol{\indRegion}$ with the actor volume $\Vol{\indActor}$. This simplifies the implementation by minimizing changes to the mesoscopic algorithm when an actor starts or stops adding molecules. When an active actor creates a molecule that should be placed in the mesoscopic regime, then it is added to the corresponding subvolume.

\subsubsection{First Order Reactions}

First order (or unimolecular) reactions are of the form
\begin{equation}
\label{eq_unimolecular}
\reactant{} \xrightarrow{\kth{1}} \mathrm{Products},
\end{equation}
such that an $\reactant{}$ molecule needs to exist and it is transformed into one or more products at a rate $\kth{1}$, which has units $\second^{-1}$.  In the mesoscopic regime, the propensity of first order reaction $\indChemRxn$ with rate $\kth{1}=\kth{\indChemRxn}$ in subvolume $\indSub$ is \cite[Eq.~(6)]{Bernstein2005}
\begin{equation}
\label{eq_prop_first}
\prop{\indSub,\indChemRxn} = \kth{\indChemRxn}\numMolSub,
\end{equation}
where the reactant is a molecule of type $\indMolType$. Product molecules are placed in the same subvolume as the reactant.

Our implementation of first order reactions in the microscopic regime is more involved. If a non-surface reaction $\indChemRxn$ is the only first order reaction for which a molecule of type $\indMolType$ is the reactant, then the probability of a given molecule of type $\indMolType$ reacting within arbitrary time interval $\dtInf$ is \cite[Eq.~(13)]{Andrews2004}
\begin{equation}
\label{eq_prob_first}
\Pr\left(\mathrm{Reaction}~\indChemRxn\right) = 1 - \EXP{-\kth{\indChemRxn}\dtInf},
\end{equation}
but if the molecule can participate in multiple non-surface first order reactions, then the probability of reaction $\indChemRxn$ occurring is \cite[Eq.~(14)]{Andrews2004}
\begin{equation}
\label{eq_prob_first_mult}
\Pr\left(\mathrm{Reaction}~\indChemRxn\right) =
\frac{\kth{\indChemRxn}}{\sum \kth{\indMolType,1}}
\left[1 - \EXP{-\dtInf\sum \kth{\indMolType,1}}\right],
\end{equation}
where $\sum \kth{\indMolType,1}$ is the sum of the rate constants of all non-surface first order reactions for which a molecule of type $\indMolType$ can be a reactant. For both (\ref{eq_prob_first}) and (\ref{eq_prob_first_mult}), we can simulate whether reaction $\indChemRxn$ occurs by generating a uniform random variable $\uniRV{}$ and comparing it with the corresponding probability.

In AcCoRD, we implement non-surface first order reactions in the microscopic regime on a continuous time scale, which makes the accuracy of these reactions independent of the microscopic time step $\dtMicro$. So, when first order reaction $\indChemRxn$ occurs \emph{and} at least one product molecule is created, we need to determine the precise reaction time within the current microscopic time step. We consider the time interval $\left[0,\dtMicro\right]$, i.e., relative to the current microscopic step. Generally, we assume that the candidate reactant molecule was created within this interval at time $\timeX{i}$, i.e., $0 \le \timeX{i} < \dtMicro$ (and if the molecule previously existed, then $\timeX{i}=0$). The time remaining for this molecule to react is $\dtInf = \dtMicro - \timeX{i}$. From mesoscopic theory, we can generate the reaction time for a first order reaction as an exponential random variable via
\begin{equation}
\label{eq_t_first}
\timeX{1} = -\frac{\log{\uniRV{}}}{\sum \kth{\indMolType,1}},
\end{equation}
but this time is unconstrained. We need to generate a \emph{constrained} exponential random variable so that the reaction time $\timeX{1} \in \left[0,\dtInf\right)$. We can achieve this by constraining the uniform random variable $\uniRV{}$. If we set $\timeX{1} = \dtInf$ and re-arrange (\ref{eq_t_first}), then we can determine that the minimum value of $\uniRV{}$ is
\begin{equation}
\label{eq_umin}
\uniRV{\mathrm{min}} = \EXP{-\dtInf\sum \kth{\indMolType,1}}.
\end{equation}

Thus, the time of reaction $\indChemRxn$ can be simulated using (\ref{eq_t_first}), where $\uniRV{}$ is between $\uniRV{\mathrm{min}}$ and 1. The reaction time of reaction $\indChemRxn$ relative to the current microscopic time step is then $\timeRxn{\indChemRxn} = \timeX{i} + \timeX{1}$.

The absorption and membrane surface reactions, where a molecule in solution interacts with a surface, are special cases of first order reactions that can only occur \emph{if} the molecule attempts to diffuse across the surface. Furthermore, their corresponding reaction rates are more strictly referred to as reaction coefficients with units $\meter/\second$. Given that a molecule's diffusion trajectory intersects a reactive surface, then the surface reaction occurs with some probability. If the reaction does not occur, then the molecule is reflected off of the surface. We adopt the surface reaction probabilities derived for Smoldyn in \cite{Andrews2009}, which apply to surfaces in the microscopic regime. For clarity of exposition, the surface reaction probabilities and details for placing product molecules are described in~\ref{app_surf}. The primary difference between Smoldyn's implementation and AcCoRD's is that surface reactions in AcCoRD occur over the sub-microscopic time interval $\dtInf \le \dtMicro$, which is based on a molecule's \emph{creation time}. However, in order to adopt the analysis in \cite{Andrews2009}, we generally assume (unless otherwise noted) that the precise reaction times for surface reactions are \emph{unknown}. Thus, the accuracy of absorption and membrane reactions depends on the microscopic time step.

As noted in \cite{Andrews2009}, it is most accurate to also consider the possibility that a molecule reacts with a surface even if its diffusion trajectory does not intersect the surface. This is analogous to considering that a microscopic molecule might enter and exit the mesoscopic regime \emph{within} the time step, which occurs with probability (\ref{eq_micro_to_meso}) when the reactive surface is a flat plane. The author of \cite{Andrews2009} claimed that the corresponding potential for improving accuracy is minimal, so we also ignored this possibility in AcCoRD's implementation. However, given the small time steps that are needed for accurate surface reactions in Section~\ref{sec_results}, we can re-visit the benefits of this simplification in future work.

\subsubsection{Second Order Reactions}

Second order (or bimolecular) reactions are of the form
\begin{equation}
	\label{eq_bimolecular}
	\reactant{} + \product{} \xrightarrow{\kth{2}} \mathrm{Products},
\end{equation}
i.e., an $\reactant{}$ molecule collides with a $\product{}$ molecule and they are transformed into one or more products at a rate $\kth{2}$, which has units $\mol^{-1}\second^{-1}\meter^{3}$ (in 3D). In the mesoscopic regime, the propensity of the second order reaction $\indChemRxn$ with rate $\kth{2}=\kth{\indChemRxn}$ in subvolume $\indSub$ is \cite[Eq.~(6)]{Bernstein2005}
\begin{equation}
\label{eq_prop_second}
\prop{\indSub,\indChemRxn} = \frac{\kth{\indChemRxn}\numMolSubX{\reactant{}}\numMolSubX{\product{}}}{\Vol{\indSub}},
\end{equation}
where $\numMolSubX{\reactant{}}$ is the number of $\reactant{}$ molecules and $\numMolSubX{\product{}}$ is the number of $\product{}$ molecules in subvolume $\indSub$, respectively. The product $\numMolSubX{\reactant{}}\cdot\numMolSubX{\product{}}$ represents the number of potential reactive collisions between specific $\reactant{}$ and $\product{}$ molecules. A subvolume clearly needs both reactants present in order for a reaction to be possible. When reaction $\indChemRxn$ occurs, the product molecules are placed in the same subvolume as the reactants. If both reactants in reaction $\indChemRxn$ are the same type of molecule, i.e., if $\reactant{} \equiv \product{}$, then $\numMolSubX{\product{}}$ in (\ref{eq_prop_second}) is replaced with $(\numMolSubX{\reactant{}}-1)$ because an $\reactant{}$ molecule cannot collide with itself.

The implementation of second order reactions in the microscopic regime is distinct from zeroth order and first order reactions. The stochastic behavior of second order reactions is achieved via the randomness of diffusion. Instead of defining a reaction probability, a reaction is characterized by a binding radius $\rBind$; see \cite{Andrews2004}. Thus, reaction $\indChemRxn$ occurs if reactants $\reactant{}$ and $\product{}$ are separated by a distance of less than $\rBind$ after diffusing.

In general, determining the value of $\rBind$ from reaction rate $\kth{2}$ is non-trivial. In \cite{Andrews2004}, closed-form expressions for the binding radius are described in the limits of very small and very large time steps, i.e., when the mutual displacement of the reactants in one time step is much smaller and much larger than the binding radius, respectively. Otherwise, lookup tables are presented in the supplementary information of \cite{Andrews2004} to determine $\rBind$ for a given $\kth{2}$, time step, and \emph{mutual} diffusion coefficient (i.e., the sum of the diffusion coefficients of the reactants).

Another complication with bimolecular reactions is when they are \emph{reversible}. A bimolecular reaction's reverse reaction should also define an \emph{unbinding radius} $\rUnbind$ as an initial separation distance between the products. The unbinding radius is intended as a balance to ensure that the ``forward'' bimolecular reaction occurs at the correct frequency. Lookup tables to determine the binding radius in \cite{Andrews2004} include the unbinding radius for reversible bimolecular reactions.

In AcCoRD, the user must define the binding and unbinding radii explicitly. For a given candidate reactant molecule, we check the coordinates of every corresponding candidate reactant that is in the same region or in a neighboring region. Currently, we only apply the unbinding radius for the products of \emph{second order} reactions so that we can model collisions between spherically-shaped molecules with non-negligible volumes. Also, since second order reactions are determined after diffusion for the current time step, any given molecule can only participate in up to \emph{one} second order reaction in a single time step, either as a reactant or a product, so the accuracy of microscopic second order reactions is sensitive to the time step.

\section{AcCoRD Algorithms and Complexity}
\label{sec_algorithm}

In the previous section, we focused on modeling the individual reaction and diffusion phenomena. In this section, we describe the structure of AcCoRD's simulation algorithms. First, we discuss the overall algorithm. Then, we comment on the computational complexity of the microscopic and mesoscopic models, including how to select appropriate values for the microscopic time step and the mesoscopic subvolume size. Specific details of the primary simulation steps are described in detail in~\ref{app_algorithms}.

The most significant demands on memory usage are for tracking molecules in the microscopic regime and subvolumes in the mesoscopic regime. We will discuss these tasks in greater detail in~\ref{sec_alg_micro} and \ref{sec_alg_meso}, respectively, but we introduce them here so that we can discuss their use in other stages of the simulation. Microscopic molecules are stored in linked lists, which can be easily modified as needed when molecules are created or removed. We maintain \emph{two} lists for every type of molecule in every region. One list is for ``recent'' molecules that were created \emph{within} the current microscopic time step. Each recent molecule was created at some time $\timeX{i}$, where $0 \le \timeX{i} < \dtMicro$, and has its own \emph{individual} time step, i.e., $\dtMicro - \timeX{i}$. The other list is for all ``normal'' molecules that were created before the current microscopic time step, and which do not need the additional memory for an individual time step. Generally, we add molecules to the corresponding ``recent'' list when they are first created and then transfer them to the ``normal'' list after their first diffusion step.

In the mesoscopic regime, there is a constant number of subvolumes. Thus, we maintain arrays of subvolume structures where each structure describes a single subvolume. For each subvolume, we store the identities of its neighboring subvolumes, the number of each type of molecule present, and the propensities of all potential reactions as calculated in Section~\ref{sec_theory}.

\subsection{Overall Algorithm}
\label{sec_top_algorithm}

The overall AcCoRD algorithm is presented in Algorithm~\ref{alg_overall}. The two input arguments are the configuration filename $\config$ and the seed number $\seed$ for the random number generator (RNG). In line~\ref{step_load}, the configuration text file is parsed and read to determine the simulation parameters. The initialization step in line~\ref{step_initialization} performs routine tasks such as memory allocation and initializing the RNG, but also does the following:
\begin{itemize}
	\item Determine which regions are neighbors and which region subvolumes are neighbors.
	\item Determine where actors are placed relative to the regions.
	\item Define the network of chemical reactions.
	\item Initialize reaction and diffusion propensity calculations for mesoscopic subvolumes.
	\item Create the simulation output files.
\end{itemize}


\begin{algorithm}[!tb]
	\caption{Overall AcCoRD Algorithm}
	\label{alg_overall}
	\begin{algorithmic}[1]
		\Procedure{AcCoRD}{$\config, \seed$} 
		\State Load configuration specified by $\config$ \label{step_load}
		\State Initialize system \label{step_initialization}
			\For{Each realization} \label{step_sim_loop}
				\State Reset the environment \label{step_reset} 
				\State $\tCur \gets 0$
				\While{$\tCur \le \tEnd$}
					\State Find next simulation step \label{step_find_next}
					\If{Next step is active actor $\indActor$}
						\State \Call{Active\_Action}{$\indActor$} \label{step_active}
					\ElsIf{Next step is passive actor $\indActor$}
						\State \Call{Passive\_Action}{$\indActor$} \label{step_passive}
					\ElsIf{Next step is microscopic}
						\State \Call{Microscopic\_Step}{} \label{step_micro}
					\Else  \Comment{Next step is mesoscopic}
						\State \Call{Mesoscopic\_Step}{} \label{step_meso}
					\EndIf
					\State Update $\tCur$ \label{step_update_time}
				\EndWhile
				\State Write observations to file \label{step_write}
				\State Display progress \label{step_display}
			\EndFor
		\State Cleanup \label{step_cleanup}
		\State \textbf{return}
		\EndProcedure
	\end{algorithmic}	
\end{algorithm}

The outer simulation loop in the AcCoRD algorithm starts at line~\ref{step_sim_loop} and executes the number of independent realizations specified by the configuration. At the start of each realization, the physical environment is reset to its ``initial'' state, i.e., with no molecules present in any region. At the end of each realization, at line~\ref{step_write}, we write the observations of the specified actors to the \emph{primary simulation output file}. We then update the simulation progress to the console (as specified; the configuration file defines the frequency of progress updates). After the simulation loop, the cleanup step in line~\ref{step_cleanup} writes a summary of the simulation execution to a separate \emph{summary file} and releases all allocated memory.

The core of the overall algorithm is the evolution of the system in a single realization from time $\tCur = 0$ until time $\tCur > \tEnd$. The system evolves as a series of steps. The four possible steps are: 1) Action by an active actor; 2) Action by a passive actor; 3) Evolution of the microscopic regime; and 4) Evolution of the mesoscopic regime. These four steps are discussed in detail in~\ref{app_algorithms}. AcCoRD has an actor-based design with a hybrid of simulation regimes, so steps do \emph{not} occur in a predetermined order. Instead, every actor and each simulation regime has its own internal timer that defines when the corresponding step should be executed. All timers are sorted in a priority queue. The current step is determined in line~\ref{step_find_next} by the timer whose value equals $\tCur$. Once the corresponding behavior is executed, the timer is updated. Then, in line~\ref{step_update_time}, the queue is re-sorted and the new time $\tCur$ is determined from the new lowest timer value. This flexible design enables actors to behave at any frequency and accommodates the execution of both microscopic and mesoscopic simulation models.

\subsection{Computational Complexity}
\label{sec_complexity}

We focus our discussion of AcCoRD's computational complexity by discussing the factors that impact the execution speed. Specifically, we consider the running time of the microscopic and mesoscopic evolution steps. These two steps generally consume an overwhelming majority of a given simulation's total run time, such that the time required to execute actor behavior and the overall AcCoRD algorithm is usually negligible. With the environment configurations that we have tested, execution speed has been considerably more of a bottleneck than the required computer memory. However, we also comment briefly on considerations for memory usage.

\subsubsection{Mesoscopic Complexity}

The speed of simulation in the mesoscopic regime depends on the number of events, i.e., chemical reactions within individual subvolumes or diffusion between adjacent subvolumes, and the time to execute each event. Every mesoscopic simulation step executes a single event. Each simulation has a specified end time $\tEnd$, so the number of events is inversely proportional to the time between individual events. The time until the next mesoscopic event is inversely proportional to the total propensity $\propTotal$, which is the sum of all possible individual event propensities (we described the calculation of propensities in Section~\ref{sec_theory}, and we discuss the generation of event times from propensities in~\ref{sec_alg_meso}). Thus, by inspection of the propensity equations, we can infer the event frequency and therefore simulation run time complexity for different events.

The time required to simulate diffusion in the mesoscopic regime between subvolumes of the same size can be inferred from (\ref{eq_meso_diff}), i.e., the number of diffusion events of molecule type $\indMolType$ out of subvolume $\indSub$ grows with $\O{\Dx{\indMolType}\numMolSub/\subLength{\indSub}^2}$, where $\O{\cdot}$ refers to big O notation. Clearly, computation time will increase by increasing the diffusion rate, increasing the number of molecules, or decreasing the subvolume size. The same observation can be made by considering the diffusion between subvolumes of different sizes in (\ref{eq_meso_diff_1d}), but from (\ref{eq_prop_hybrid}) the number of diffusion events from the subvolume to an adjacent microscopic region grows with $\O{\sqrt{\Dx{\indMolType}}\numMolSub/\subLength{\indSub}}$. We emphasize that the number of molecules can be time-varying, so the overall diffusion simulation speed is also time-varying.

Unlike diffusion, which always adds a molecule in one subvolume after removing it from another, the number of chemical reaction events in the mesoscopic regime depends on the net change in the total number of molecules due to each reaction. The number of \emph{zeroth} order reactions is independent of the number of molecules, since they have no reactant. From (\ref{eq_prop_zeroth}), we can deduce that the number of instances of zeroth order chemical reaction $\indChemRxn$ in subvolume $\indSub$ grows with $\O{\kth{\indChemRxn}\Vol{\indSub}}$.

First and second order reactions do depend on the corresponding number of molecules. From (\ref{eq_prop_first}), we conclude that the number of instances of first order reaction $\indChemRxn$ in subvolume $\indSub$ with reactant $\indMolType$ grows with $\O{\kth{\indChemRxn}\numMolSub}$. From (\ref{eq_prop_second}), we find that the number of instances of second order reaction $\indChemRxn$ in subvolume $\indSub$ with reactants $\reactant{}$ and $\product{}$ grows with $\O{\kth{\indChemRxn}\numMolSubX{\reactant{}}\numMolSubX{\product{}}/\Vol{\indSub}}$.

Now that we have established the growth in the \emph{number} of mesoscopic simulation events, we comment on the speed of executing a \emph{single} event. This is described in further detail in \ref{sec_alg_meso}, where it is shown that a mesoscopic step takes $\O{\log\numSub + \numEvent}$ time, where $\numSub$ is the total number of subvolumes and $\numEvent$ is the total number of possible events in the current subvolume (i.e., the sum of all reaction and diffusion events).

Given the above discussion, and assuming that there are no hybrid interfaces, we can write the overall complexity of the mesoscopic model run time as
\ifOneCol
	\begin{align}
	\complexityMeso = &\, \mathcal{O}\Bigg(\tEnd\bigg[\sum_{\setSub}\bigg(\sum_{\indMolType}\frac{\Dx{\indMolType}\numMolSub}{\subLength{\indSub}^2} + \sum_{\kth{0}}\kth{0}\Vol{\indSub} + \sum_{\kth{1}}\kth{1}\numMolSubX{\kth{1}} \nonumber \\ &+ \sum_{\kth{2}}\frac{\kth{2}\numMolSubX{\kth{2,1}}\numMolSubX{\kth{2,2}}}{\Vol{\indSub}} \bigg)\bigg]\left(\log\numSub + \numEvent\right)\Bigg),
	\label{eq_meso_complexity}
	\end{align}
\else
	\begin{align}
	\complexityMeso = &\, \mathcal{O}\Bigg(\tEnd\bigg[\sum_{\setSub}\bigg(\sum_{\indMolType}\frac{\Dx{\indMolType}\numMolSub}{\subLength{\indSub}^2} \nonumber \\ 
	&+\sum_{\kth{0}}\kth{0}\Vol{\indSub} + \sum_{\kth{1}}\kth{1}\numMolSubX{\kth{1}}+ \sum_{\kth{2}}\frac{\kth{2}\numMolSubX{\kth{2,1}}\numMolSubX{\kth{2,2}}}{\Vol{\indSub}} \bigg)\bigg] \nonumber \\ &\times\left(\log\numSub + \numEvent\right)\Bigg),
	\label{eq_meso_complexity}
	\end{align}
\fi
where the summations are over all mesoscopic subvolumes, all types of molecules in subvolume $\indSub$, and all zeroth, first, and second order reactions in subvolume $\indSub$, respectively. $\numMolSubX{\kth{1}}$ is the number of molecules in subvolume $\indSub$ that are the reactant associated with reaction rate $\kth{1}$, and $\numMolSubX{\kth{2,1}}\numMolSubX{\kth{2,2}}$ is the product of the corresponding numbers of reactants associated with reaction rate $\kth{2}$. $\tEnd$ is the simulation end time, assuming that it starts at $t=0$. If hybrid interfaces are present, then we must include in (\ref{eq_meso_complexity}) a summation of $\sqrt{\Dx{\indMolType}}\numMolSub/\subLength{\indSub}$ over molecule types for the subvolumes that are adjacent to a microscopic region.

From (\ref{eq_meso_complexity}), using more subvolumes will increase both the number of simulation events and the time to execute each event. Thus, one factor in the selection of subvolume size is the simulation run time. However, it is not the only factor, and in fact there are practical bounds that constrain subvolume size for an accurate simulation, particularly if bimolecular reactions can occur. For example, as defined in \cite[Eq.~(7)]{Ramaswamy2011}, a 3D subvolume will only be ``well-mixed'' (such that the second order reaction propensity in (\ref{eq_prop_second}) is accurate) if the subvolume length $\subLength{\indSub}$ is much less than $\sqrt{6\Dx{}\timeX{\indChemRxn}}$, where $\timeX{\indChemRxn}$ is the characteristic time of the fastest bimolecular reaction. Alternatively, if subvolumes are too small, then bimolecular reaction propensities will depend on reactant molecules in neighboring subvolumes. A lower limit on 3D subvolume size for a given bimolecular reaction was calculated in \cite{Hellander2016} to be about $3.2\rBind$, where $\rBind$ is the binding radius of the two reactants. For subvolumes smaller than this limit, corrections are needed to modify the bimolecular reaction propensity, but such changes are outside the scope of this work.

\subsubsection{Microscopic Complexity}

Unlike the mesoscopic regime, the frequency of microscopic steps in simulation time does not vary over the course of the simulation. Instead, the number of microscopic simulation steps is explicitly defined by the constant simulation time step $\dtMicro$ as $\lceil\tEnd/\dtMicro\rceil+1$, where $\lceil\cdot\rceil$ is the ceiling function. Generally, a smaller time step increases the simulation accuracy, but as with subvolume size there are practical limits to consider. In order to assume Brownian motion, the time step must be large enough to ignore hydrodynamic memory effects, which occur over time scales less than $t=\varX{}^2\rho/\eta$, where $\varX{}$ is the diffusing molecule radius, $\rho$ is the fluid density, and $\eta$ is the fluid viscosity\footnote{Using typical values for water density ($10^3\,\mathrm{kg}/\meter^3$) and viscosity ($10^{-3}\,\mathrm{kg}/(\meter\cdot\second)$), and molecule radii on the order of less than $10\,\mathrm{n}\meter$, hydrodynamic memory effects would occur on a time scale of less than $1\,\mathrm{n}\second$, which is much less than the smallest microscopic time steps considered in this paper.}; see \cite{Huang2011}. For time steps that are too large, interactions between a molecule and environmental features (especially surfaces and other boundaries) will be modeled inaccurately, since we generally consider straight line diffusion trajectories. A simple bound is to consider a molecule's root mean square distance along each dimension in one time step, i.e., $\sqrt{2\Dx{}\dtMicro}$. This distance should be much smaller than the resolution of the smallest environmental features.

As we discuss in~\ref{sec_alg_micro}, a microscopic simulation step has separate stages for zeroth order reactions, non-surface first order reactions, diffusion (with surface reactions), and second order reactions. From the run time complexity results in~\ref{sec_alg_micro}, we can write the overall complexity of the microscopic model run time as
\ifOneCol
	\begin{align}
	\complexityMicro = &\, \mathcal{O}\Bigg(\frac{\tEnd}{\dtMicro}\sum_{\regimeMicro}\bigg[\sum_{\kth{0}}\kth{0}\Vol{\indRegion} + \sum_{\kth{1}}\kth{1}\Nx{\kth{1}, \indRegion} + \numMolMicroRegion|\setRegionNeigh| \nonumber \\ &+ \sum_{\kth{2}}\Nx{\kth{2,1},\indRegion}\Big(\Nx{\kth{2,2},\indRegion}+\sum_{\indRegionNeigh\in\setRegionNeigh}\Nx{\kth{2,2},\indRegionNeigh}\Big)\bigg]\Bigg),
	\label{eq_micro_complexity}
	\end{align}
\else
	\begin{align}
	\complexityMicro = &\, \mathcal{O}\Bigg(\frac{\tEnd}{\dtMicro}\sum_{\regimeMicro}\bigg[\sum_{\kth{0}}\kth{0}\Vol{\indRegion} + \sum_{\kth{1}}\kth{1}\Nx{\kth{1}, \indRegion} \nonumber\\ &+ \numMolMicroRegion|\setRegionNeigh| \nonumber \\ &+ \sum_{\kth{2}}\Nx{\kth{2,1},\indRegion}\Big(\Nx{\kth{2,2},\indRegion}+\sum_{\indRegionNeigh\in\setRegionNeigh}\Nx{\kth{2,2},\indRegionNeigh}\Big)\bigg]\Bigg),
	\label{eq_micro_complexity}
	\end{align}
\fi
where each summation is over the set of microscopic regions, the set of corresponding chemical reactions (i.e., zeroth, first, and second order with rates $\kth{0}, \kth{1}$, and $\kth{2}$, respectively), and the set of regions $\setRegionNeigh$ that neighbor region $\indRegion$. $\Vol{\indRegion}$ is the region's volume, $\Nx{\kth{1}, \indRegion}$ is the number of reactant molecules for the first order reaction associated with rate $\kth{1}$, $\numMolMicroRegion$ is the number of molecules in region $\indRegion$, and $|\setRegionNeigh|$ is the total number of regions that are neighbors to region $\indRegion$. $\Nx{\kth{2,1}, \indRegion}$ and $\Nx{\kth{2,2}, \indRegion}$ are the numbers of first and second reactant molecules in region $\indRegion$ for the second order reaction associated with rate $\kth{2}$, and $\Nx{\kth{2,2},\indRegionNeigh}$ is the corresponding number of second reactant molecules in neighboring microscopic region $\indRegionNeigh$. The term $\numMolMicroRegion|\setRegionNeigh|$ represents the cost for diffusing each molecule and verifying its trajectory, including testing for entering the mesoscopic regime.

We can compare the complexity in computational run time of the mesoscopic and microscopic models by comparing (\ref{eq_meso_complexity}) and (\ref{eq_micro_complexity}). Both rely on the characteristic parameters that define each regime, i.e., time step in the microscopic model and subvolumes in the mesoscopic model. However, they show similar sensitivity to all three orders of chemical reactions. We will gain more specific insights into the time for simulating diffusion by measuring simulation run times in Section~\ref{sec_results}.

\subsubsection{Memory Usage}

Finally, we comment on memory usage. AcCoRD's greatest memory demands are to store information about the individual subvolumes and the individual molecules. Subvolumes are used in both the mesoscopic and microscopic regimes, although more memory is used for each subvolume in the mesoscopic regime in order to: 1) count the number of each type of molecule in the subvolume, and 2) track the subvolume's place in the indexed priority queue to determine when its next event will occur. Memory for individual molecules is needed in microscopic regions, and also to store molecule locations (from both regimes) before a realization's observations are written to the output file. Generally, if too much memory is required, then we can either represent mesoscopic regions with fewer subvolumes or have fewer molecules in the microscopic regime.

\section{Using AcCoRD}
\label{sec_use}

In this section, we describe the general work flow for an AcCoRD user. This section does not provide step-by-step instructions; a user can refer to the online code documentation for specific details on the latest version; see \cite{Noel2016}. The online documentation includes installation and usage instructions, descriptions of all configuration options, and many sample configuration files. We described the format of a configuration file in Section~\ref{sec_components}. Here, we focus on AcCoRD's usability, and we intend for this section to accommodate future updates to the code.

\subsection{Preparing a Simulation}

Every simulation is based on the environment and behavior specified in a configuration file. The sample configuration files are intended to be copied and renamed for modification. To confirm that the environment is designed as intended (i.e., with the correct size and relative placement of objects), the user can draw the regions and actors via a configuration plotting function in MATLAB. For example, \emph{every} figure with a sketch in this paper (with the exception of Fig.~\ref{fig_sub_overlap}) was prepared using an AcCoRD configuration file and the plotting function.

The AcCoRD simulator is compiled as a single executable file. Executables and build scripts are currently provided for Windows and Linux operating systems. The executable can be called directly from a command line interface, and it is also possible to run the simulator remotely on a computing cluster (such as those operated by Compute/Calcul Canada, which were used for many of the simulations in Section~\ref{sec_results}). Providing a seed for the random number generator at run time will over-ride the seed defined in the configuration file. The intent is that a user can distribute the computing load of a simulation by running different realizations on different processors; providing a unique seed in each call will avoid repeated realizations.

\subsection{Running a Simulation}

To facilitate automated or remote execution, a simulation normally runs without any input from the user. The command line interface displays a brief summary of the configuration, including the number of regions, actors, and subvolumes. It also shows the relative directory where the simulation output is being written. We use a timer to measure the duration of every realization and provide updates on the estimated time remaining under the assumption that each realization will execute in comparable time. The total simulation time, i.e., the time required to execute the for-loop in Algorithm~\ref{alg_overall} on page~\pageref{alg_overall}, is also displayed at the end.

AcCoRD can return warnings or errors at run time. In general, we use warnings to indicate invalid or missing configuration information that can be replaced by hard-coded default values. We use errors to deal with memory allocation problems or identify invalid configuration information that is too late to automatically correct (for example, regions that are nested improperly). Warnings and errors are described in the command line with corresponding details, such as the name of the region, actor, or reaction that led to the warning or error. By default, the creation of any warnings during the loading of the configuration will pause the simulation and prompt the user to continue (especially since the ``corrected'' configuration may not behave as intended or may lead to other errors). Any errors will terminate the simulation.

\subsection{Post-Processing}

Running a simulation once, i.e., with one seed value, generates \emph{two} output files that are both labeled with the corresponding seed number. A \emph{summary} file is in JSON-format and it describes the information in the custom-formatted \emph{primary} output file. This information includes how many actors were being recorded and how much data is associated with each actor. All post-processing utilities were developed in MATLAB and the user can load simulation output with an import function. By specifying multiple seed values, multiple pairs of output files can be imported and aggregated simultaneously. For example, a simulation that was repeated 10 times (with 10 different seeds) and simulated 1000 realizations with each seed can be imported as a single simulation with $10^4$ realizations and saved to a MATLAB mat-file.

We have a number of visualization tools for AcCoRD's simulation output. Current options include the following:
\begin{enumerate}
	\item Visualizing individual molecules in an animation or video. We combine the configuration plotting tool, which displays the specified regions and actors, with plots of individual molecules at locations observed by passive actors within a \emph{single} realization. A user generally has full control over what simulation components are displayed, what time interval is viewed, how the camera is controlled (can be static or dynamic), and whether to create a series of figures or an exported video file. Custom overlay information can include a simulation progress timer or a display of the number of molecules of some type observed by one of the passive actors. A total of eight videos are included in \cite{Noel2016c}.
	\item Plotting the time-varying signal as a curve. Curves can correspond to individual realizations or be averaged over any subset of realizations in the imported simulation file. It is also possible with the same syntax to add curves of user-defined data, so a user can add analytical results or other information.
	\item Plotting the signal's statistical distribution as a probability mass function (PMF) or cumulative distribution function (CDF). These empirical distributions are determined from a specified subset of simulation realizations, and can be time-varying (i.e., drawn as a 3D surface) or averaged over multiple observation times (i.e., drawn as a 2D curve). Analogous to plotting the time-varying signal, we can also plot analytical distributions based on a user-defined trial probability (i.e., probability of observing a given molecule) and number of trials (i.e., number of molecules). A distribution can be Binomial or either the Poisson or Gaussian approximation of the Binomial distribution. We review these distributions in~\ref{app_prob}.
	\item Plotting the signal's time-varying mutual information relative to one or more reference observations in 2D and 3D, respectively. Mutual information can be used as an indicator of observation independence, as we considered in \cite{Noel2014d}. Consecutive observations that are not sufficiently separated in time will be identical and have mutual information equal to the entropy of the signal. Observations that are sufficiently separated in time should be independent and have zero mutual information. Since ``zero'' mutual information cannot be measured directly from a finite number of samples (see \cite{Goebel2005}), a user can also plot \emph{Monte Carlo mutual information}, which measures the mutual information between finite samples of \emph{independent} Binomial random variables. We review the calculation of mutual information in~\ref{app_prob}.
\end{enumerate}

Communications performance metrics such as the bit error rate (BER) depend on the specific design of the receiver's detector. AcCoRD does not impose a specific detector design. By making the receiver observations available for post-processing in MATLAB, a user can readily implement their own detector off-line and measure its performance. Utilities that implement common detector designs will be added in a future update. Also, the future implementation of dependent actors will include detectors as part of the online simulation.

\section{AcCoRD Simulation Results}
\label{sec_results}

In this section, we present a series of simulation results generated by AcCoRD. Our overall aim is to demonstrate the functionality and the accuracy of the software, including its design as a ``sandbox'' reaction-diffusion solver and its generation of output that is suitable for signal processing (including communications analysis). All results were generated using version 0.7 or 0.7.0.1 and can be equivalently generated by version 1.0. Some results have accompanying videos in \cite{Noel2016c} to show a sample of the evolution of the environment with a reduced number of molecules. These videos are also summarized in~\ref{app_video}. Generally, we have avoided focusing on results that we have already presented in preliminary work with earlier versions of AcCoRD. These earlier results include a study on the choice of statistical distribution to describe the number of observed molecules in \cite{Noel2015}, a demonstration of a system with a large number of molecule sources in \cite{Deng2016b}, a study on the accuracy of the assumption that a transmitter is a point source in \cite{Noel2016a}, and a direct comparison between passive and absorbing receivers in \cite{Noel2016b}. The reader may refer to these earlier works for results from those specific scenarios.

We consider four system environments in this section, each with a series of variations. These systems are representative of AcCoRD's features but in consideration of space they are not comprehensive\footnote{Specifically, we do not simulate a reversible first order reaction in solution or use bimolecular reactions to model molecule crowding. However, sample configurations for these cases are included with the source code in \cite{Noel2016}. We also only consider ``steady state'' reaction probabilities for surface reactions that have finite reaction rates.}. We now summarize each system and its contribution to our study. There is at least one video in \cite{Noel2016c} for each system.

\begin{figure}[!tb]
	\centering
	\includegraphics[width=3in]{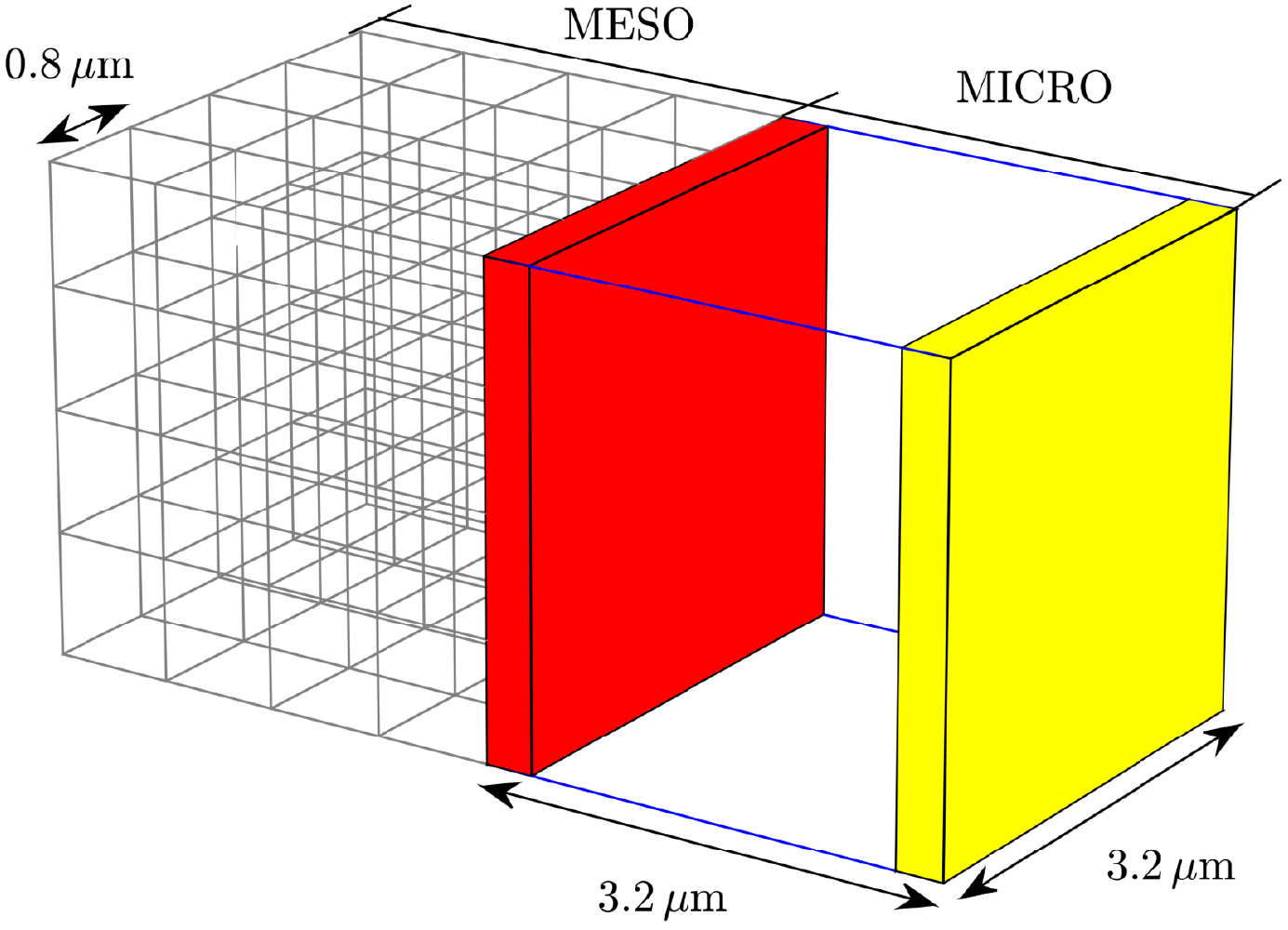}
	\caption{Layout of the hybrid environment in System 1. Both regions are cubes. The variation shown has a mesoscopic region (MESO) with subvolumes of width $\subLength{}=0.8\,\mu\meter$. The red and yellow areas of the microscopic (MICRO) region have a depth of $0.32\,\mu\meter$ and are observed by passive receivers. A sample video of the simulation of this environment is available as Video 2 in \cite{Noel2016c}.}
	\label{fig_hybrid_env}
\end{figure}

System 1 is a bounded rectangular environment, as shown in Fig.~\ref{fig_hybrid_env} and also in Fig.~\ref{fig_hybrid_env_mol} on page~\pageref{fig_hybrid_env_mol}. In Section~\ref{sec_results_hybrid}, we use System 1 to study the accuracy of using a hybrid of microscopic and mesoscopic simulation models for a diffusion-only environment. We also verify the accuracy of (\ref{eq_meso_diff_3d}) to describe the transition rate between mesoscopic subvolumes of different sizes. We measure the distribution of molecules observed in different parts of the system, the mutual information between consecutive observations at a given location, and the simulation run time.

System 2 is a long rectangular rod with a perfectly-absorbing surface at one end, as shown in Fig.~\ref{fig_hybrid2_env} on page~\pageref{fig_hybrid2_env}. System 2 is used as an example of a hybrid environment with both diffusion and chemical reactions. In Section~\ref{sec_results_hybrid_rxn_diff}, we compare the accuracy of the fully microscopic model with that of the hybrid model at different locations along the rod. We vary the location of the hybrid interface and also measure the simulation run time.

System 3 has two large concentric spherical surfaces and places molecules between them, as shown in Fig.~\ref{fig_surface_env} on page~\pageref{fig_surface_env}. Every molecule can either adsorb to one of the surfaces or transition through it (i.e., the surface is a membrane). The size of this system makes it locally 1D so that it is analogous to the environment considered in \cite[Fig.~6a]{Andrews2009}. Thus, this system is used in Section~\ref{sec_results_surface} to verify our implementation of the surface reaction probabilities that were derived in \cite{Andrews2009} (and summarized in~\ref{app_surf}), while also verifying the accuracy of the collision detection and reflection off of spherical surfaces (which are geometric problems whose details are outside the scope of this paper).

Finally, System 4 considers a point molecule source transmitter and a spherical receiver, as shown in Fig.~\ref{fig_comm_env} on page~\pageref{fig_comm_env}. This system and its variations have been commonly studied in the MC literature, cf. \cite{Noel2014f, Heren2015, Meng2014, Mahfuz2015}. In Section~\ref{sec_results_comm}, we consider the channel impulse response at a passive receiver when the system is unbounded versus having an absorbing or reflecting outer boundary. We also consider the impulse response with degradation while diffusing. To do so, we compare a first order degradation reaction with enzyme-mediated degradation via Michaelis-Menten kinetics (see \cite[Ch.~10]{Chang2005}). We measure the time-varying PMF at a passive receiver when the transmitter releases a series of finite pulses of molecules. We also verify the impulse response when the receiver has a fully- or partially-absorbing surface.

\subsection{System 1: Hybrid and Multi-Scale}
\label{sec_results_hybrid}

The layout of System 1 is a box as shown in Fig.~\ref{fig_hybrid_env}, with a width and depth of $3.2\,\mu\meter$ and a length of $6.4\,\mu\meter$. We first divide the environment into two cubic halves, where one half is microscopic with a time step of $\dtMicro=0.5\,\meter\second$ and the other half is mesoscopic with subvolumes of a uniform size that we vary from $\subLength{}=0.1\,\mu\meter$ to $\subLength{}=3.2\,\mu\meter$. We place two observers in the microscopic region (one at the hybrid interface and the other at the opposite end of the region; shown in red and yellow, respectively, in Fig.~\ref{fig_hybrid_env}), and each is a passive actor with a depth of $0.32\,\mu\meter$ so they observe $10\,\%$ of the microscopic region, respectively. An active actor uniformly initializes $1000$ molecules with a diffusion coefficient of $\Dx{}=10^{-10}\frac{\meter^2}{\second}$ throughout the entire system at time $t=0$, and the passive actors observe the system at intervals of $0.01\,\second$ for $1\,\second$ (i.e., 100 observations). With these parameters, the average diffusion distance of a microscopic molecule along one dimension in one time step is $\sqrt{2\Dx{}\dtMicro}\approx0.32\,\mu\meter$. Thus, the range of subvolume sizes include subvolumes that are relatively small (i.e., $\subLength{}\sim\sqrt{\Dx{}\dtMicro}$) and relatively large (i.e., $\subLength{}\gg\sqrt{\Dx{}\dtMicro}$), so we can compare rules (\ref{eq_hybrid_tangential}) and (\ref{eq_hybrid_tangential_large}) when determining the tangential placement of molecules that enter the microscopic region from a mesoscopic subvolume. We define $\hybridDist{\mathrm{max}}=10\,\mu\meter$, so no molecules in the microscopic region are excluded from the possibility of entering and exiting the mesoscopic region during a diffusion step.

We seek to get a sense of the trade-offs of the hybrid environment, including its speed and accuracy, when simulating diffusion. In this system, the hybrid interface between the microscopic and mesoscopic regions is relatively large compared to the total size of the environment. Thus, we expect that the accuracy will be very sensitive to that at the interface, and that the simulation run time will be very sensitive to the overhead introduced by having the interface. The total simulation time of $1\,\second$ is long enough a molecule's expected displacement along each dimension to be a few times larger than the entire system, so any long term bias in the diffusion between regions should be observable.

\begin{figure}[!tb]
	\centering
	\includegraphics[width=3.45in]{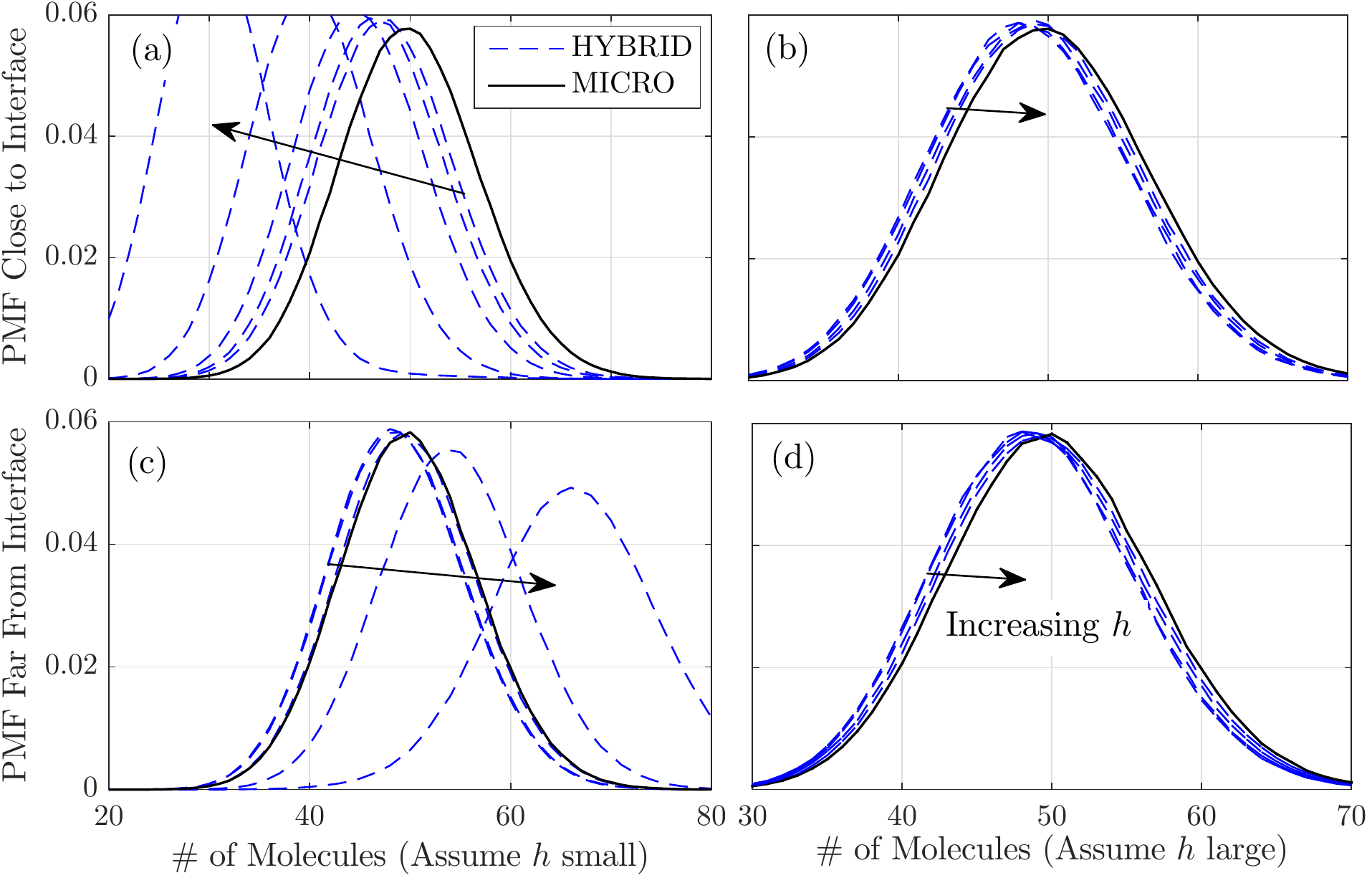}
	\caption{Probability mass functions of number of molecules observed in subregions of the hybrid environment in System 1. We observe $10\,\%$ of the microscopic region closest to the hybrid interface (in (a) and (b)) and $10\,\%$ of the microscopic region furthest from the hybrid interface (in (c) and (d)). The hybrid transitions are based on the assumption that the mesoscopic subvolumes are small (i.e., that $\subLength{} \sim \sqrt{\Dx{}\dtMicro}$, in (a) and (c)) or that they are large (i.e., that $\subLength{} \gg \sqrt{\Dx{}\dtMicro}$, in (b) and (d)) relative to the average diffusion distance ($\sqrt{2\Dx{}\dtMicro} \approx0.32\,\mu\meter$) in the microscopic region. The arrows are in the directions of the PMFs for increasing mesoscopic subvolume size $\subLength{}$. We plot $\subLength{}=\{0.2,0.4,0.8,1.6,3.2\}\,\mu\meter$, except we omit the $\subLength{}=3.2\,\mu\meter$ case in (b) and (d) because its PMFs are indistinguishable from those when $\subLength{}=1.6\,\mu\meter$. Simulations with $\subLength{}=0.1\,\mu\meter$ were also performed, but the resulting PMFs are indistinguishable from those when $\subLength{}=0.2\,\mu\meter$ so they are not shown.}
	\label{fig_hybrid_pmf}
\end{figure}

In Fig.~\ref{fig_hybrid_pmf}, we simulate each value of $\subLength{}$ $10^4$ times and plot the probability mass function of all 100 observations at each observer (as a single PMF for each observer constructed from $10^6$ observations). We separately consider the transition rules (\ref{eq_hybrid_tangential}) or (\ref{eq_hybrid_tangential_large}), i.e., assume that the mesoscopic subvolumes are small or large, respectively. The hybrid environments are compared with a variation where the entire system is microscopic (i.e., where there is no hybrid interface). Video 2 in \cite{Noel2016c} shows a sample realization when $\subLength{}=0.8\,\mu\meter$.

Given the uniform molecule distribution, we expect that each PMF in Fig.~\ref{fig_hybrid_pmf} should have a peak value at about $50$ molecules. This is the case for the benchmark microscopic system. We observe that all hybrid variations in Fig.~\ref{fig_hybrid_pmf} underestimate the actual number of molecules close to the hybrid interface (see subplots (a) and (b)). We see that accuracy in subplot (a), i.e., when we assume that $\subLength{}$ is small, improves with decreasing $\subLength{}$. The accuracy in subplot (b), i.e., when we assume that $\subLength{}$ is large, improves with increasing $\subLength{}$. However, none of the PMFs of the observer close to the interface match the microscopic benchmark well when we assume that $\subLength{}$ is small. This inaccuracy is most likely due to the hybrid transition rule for small $\subLength{}$, which does not account for the environment boundary when molecules leave the mesoscopic regime. This can be corrected in a future release of AcCoRD. The accuracy is better when we consider the observer far from the interface (subplot (c)), but is still not as accurate as the corresponding case when we assume that the subvolumes are large (subplot (d)). Again, this can be improved by accounting for molecules crossing the environment boundary when leaving the mesoscopic regime. Currently, the accuracy appears to be much less sensitive to the subvolume size when we assume that the subvolumes are large, so we will only consider this assumption and use (\ref{eq_hybrid_tangential_large}) for the remainder of our results in this paper.

\begin{figure}[!tb]
	\centering
	\includegraphics[width=3.45in]{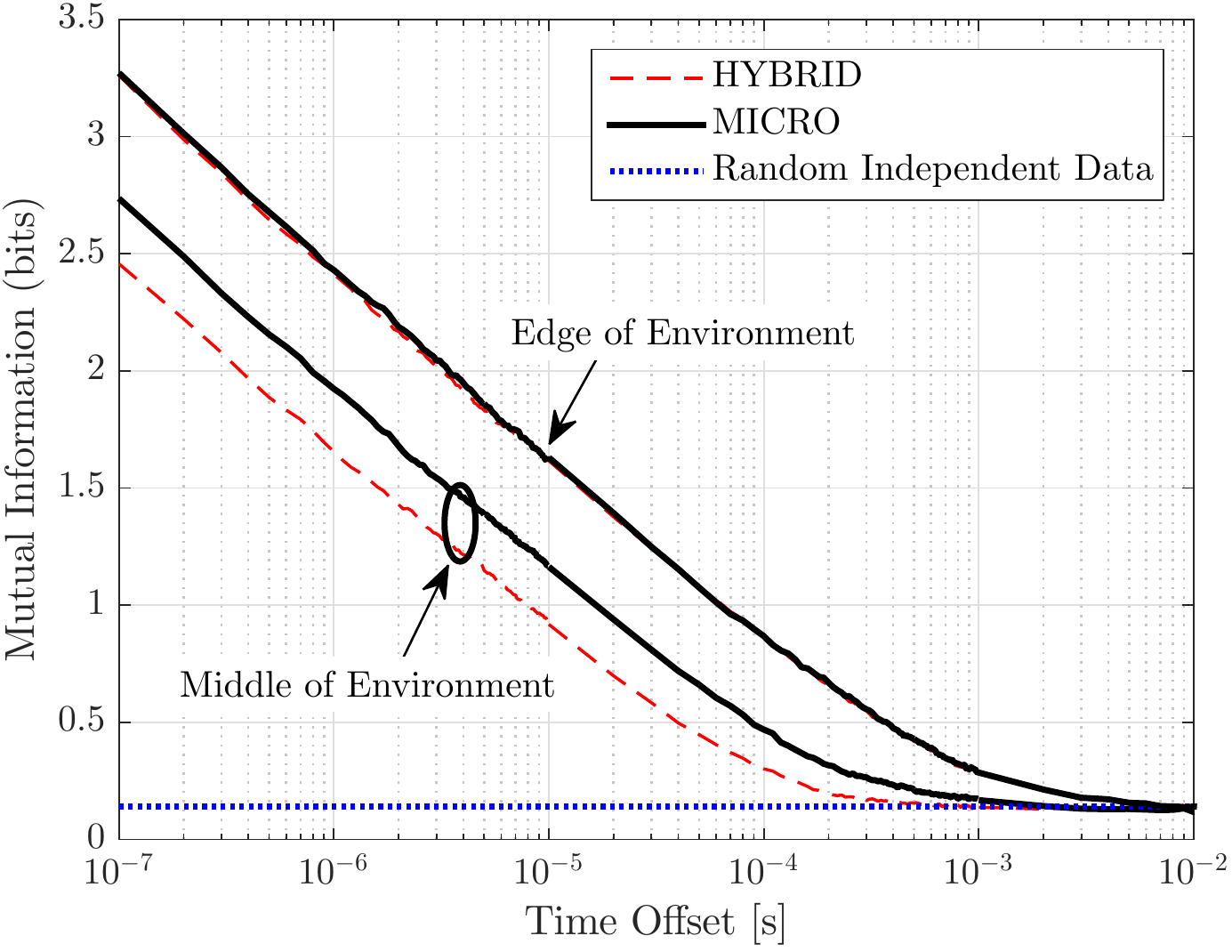}
	\caption{Mutual information between observations as a function of the time between the observations. The hybrid environment in System 1 with $\subLength{}=0.8\,\mu\meter$ is compared with the non-hybrid (i.e., microscopic) version. Mutual information between an equivalent number of randomly-generated independent observations is shown as a lower bound.}
	\label{fig_hybrid_mi}
\end{figure}

In Fig.~\ref{fig_hybrid_pmf}, we collected all observations by a passive actor into a single PMF and we did not consider the impact of the observation times. It is interesting to consider the impact of the hybrid interface on the joint observation statistics, since communications analysis and related signal processing often relies on assuming independent observations; see \cite{Noel2014d,ShahMohammadian2012}. We address this idea in Fig.~\ref{fig_hybrid_mi}, where we measure the mutual information using (\ref{eq_mutual_information}) between consecutive observations as a function of time between when the observations were taken. We consider the first observation as the reference time and vary the offset from $10^{-7}\,\second$ to $10^{-2}\,\second$, i.e., most of the offsets shown are smaller than the microscopic time step $\dtMicro=0.5\,\meter\second$ used in Fig.~\ref{fig_hybrid_pmf}. The mutual information for each offset is calculated from $10^4$ simulated observations, and we also compare with the mutual information measured from $10^4$ pairs of independently-generated random variables with the same mean (i.e., $50$) as a bound of ``true'' independence\footnote{As we discuss in~\ref{app_prob}, independent variables actually have zero mutual information, but calculating mutual information from a finite number of realizations generally results in a non-zero value; see \cite{Goebel2005}. We use such a value here as an independence bound.}. We use base 2 for the logarithm in (\ref{eq_mutual_information}) so that mutual information is measured in bits.

In Fig.~\ref{fig_hybrid_mi}, the only hybrid variation that we consider is when $\subLength{}=0.8\,\mu\meter$, but similar results can be observed with other subvolume sizes. When the observer is \emph{far} from the interface, i.e., at the edge of the environment, the mutual information between observations is identical to that when the entire system is microscopic. However, these observations are not truly independent until there is about $10\,\meter\second$ between them, i.e., more than an order of magnitude longer than the microscopic time step in Fig.~\ref{fig_hybrid_pmf}. In this case, one cannot automatically assume that observations taken more frequently than every $10\,\meter\second$ will be independent.

There are two interesting observations about the mutual information at the observer in the \emph{middle} of the environment, i.e., at the interface in the hybrid case. First, the mutual information in both system variations is \emph{less} than it is at the edge of the environment for the same offset. This is because more molecules at the edge reflect off of the system boundary and remain within the observer, so there are fewer molecule transitions into and out of the observer and hence less uncertainty about the values of consecutive observations. Second, the mutual information in the hybrid case is less than in the microscopic system; the region transitions at the hybrid interface introduce \emph{additional uncertainty} in the molecule locations. Determining the precise cause of this uncertainty at the interface is an interesting problem for future work. Nevertheless, we observe that the mutual information between observations in the middle of the environment tends towards that measured for randomly-generated independent observations for both simulation models.

\begin{figure}[!tb]
	\centering
	\includegraphics[width=3in]{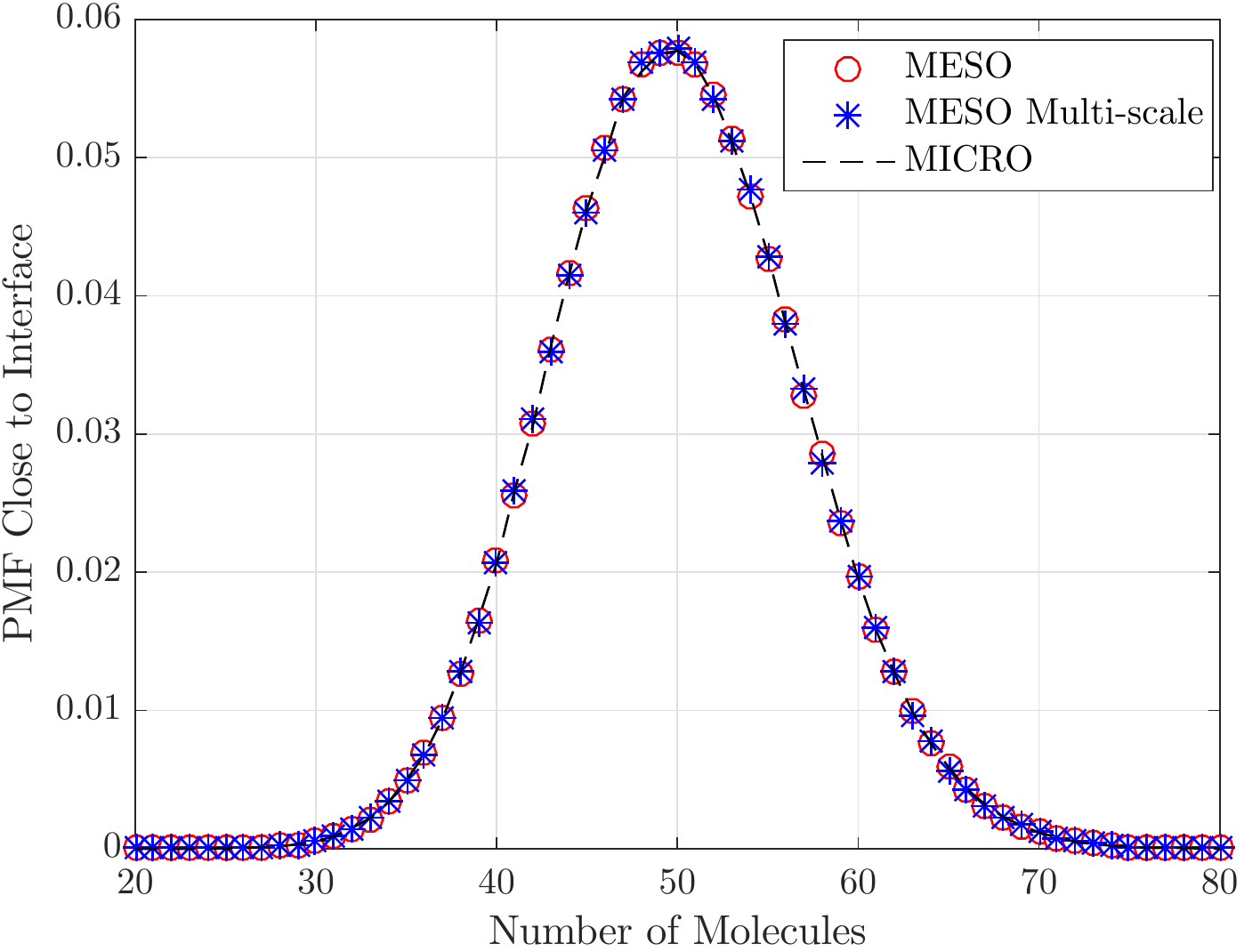}
	\caption{Probability mass functions of number of molecules observed in a subregion in the middle of System 1 ($5\,\%$ of the total size). We model the environment as fully microscopic, fully mesoscopic with subvolumes of size $\subLength{}=0.4\,\mu\meter$, or as fully mesoscopic where half of the subvolumes have size $\subLength{}=0.4\,\mu\meter$ and the other half have size $\subLength{}=1.6\,\mu\meter$ (``MESO Multi-scale''). The corresponding probability mass functions are practically indistinguishable.}
	\label{fig_multiscale_pmf}
\end{figure}

It is also of interest to consider variations of System 1 that are entirely mesoscopic, and also to use subvolumes of different sizes. We consider using mesoscopic subvolumes of size $\subLength{}=0.4\,\mu\meter$ throughout the entire system, as well as a variation where half of the environment is replaced with subvolumes of size $\subLength{}=1.6\,\mu\meter$, and compare the corresponding observation PMFs in the middle of the system with that of the microscopic benchmark in Fig.~\ref{fig_multiscale_pmf}. All three of these PMFs are practically indistinguishable, which also verifies the transition rule that we derived for mesoscopic subvolumes of different sizes in (\ref{eq_meso_diff_3d}). Comparable results can be observed using mesoscopic subvolumes of other sizes.

\begin{figure}[!tb]
	\centering
	\includegraphics[width=3.45in]{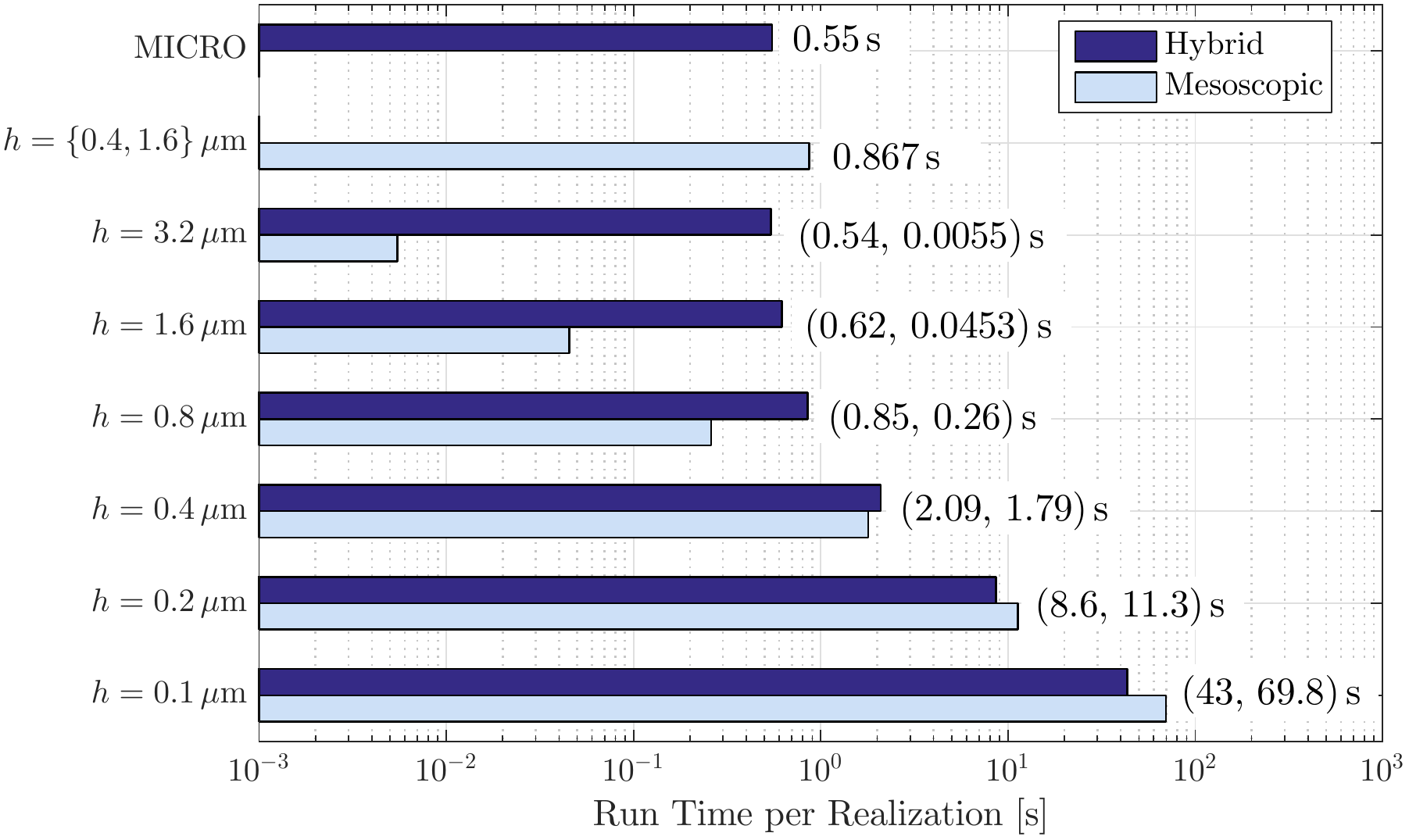}
	\caption{Simulation run times for the variations of System 1. The hybrid variations measured were those assuming that the subvolume size $\subLength{}$ was large. Run time was averaged over at least 100 realizations of each variation on an Intel i7 desktop PC. The run time of the multi-scale mesoscopic variation, with subvolumes of size $\subLength{}=\{0.4,1.6\}\,\mu\meter$, is between that of the corresponding mesoscopic variations with a uniform subvolume size.}
	\label{fig_meso_runtime}
\end{figure}

Now that we have considered the accuracy of microscopic, mesoscopic, and hybrid diffusion, we briefly consider the simulation run times in Fig.~\ref{fig_meso_runtime}. We measure the average run time per realization as determined by running at least 100 realizations on an Intel i7 desktop PC. The run time of the fully mesoscopic system is clearly proportional to the number of subvolumes, as expected from the run time complexity expression in (\ref{eq_meso_complexity}), and is many orders of magnitude faster than the microscopic system when the subvolumes are very large (e.g., in the limiting case of $\subLength{}=3.2\,\mu\meter$, there are only two subvolumes). However, the microscopic system is faster than the mesoscopic system when the subvolumes are $0.4\,\mu\meter$ in size or smaller. As expected, the run time of the multi-scale mesoscopic case, which divides each half of the environment into subvolumes of size $\subLength{}=\{0.4,1.6\}\,\mu\meter$, is between those when the environment has subvolumes of only one of those sizes. The run times in the hybrid variations are generally no faster than when the entire system is microscopic. They become faster than the mesoscopic system when the subvolumes are smaller than $0.4\,\mu\meter$. The run times are not an average of the underlying microscopic and mesoscopic run times, since the hybrid system also introduces a computational overhead to manage the transitions between the two simulation models, as discussed in~\ref{sec_alg_micro} (especially with such a large value of $\hybridDist{\mathrm{max}}=10\,\mu\meter$). We will consider the potential benefits of the hybrid model in further detail when we study System 2.

\begin{figure}[!tb]
	\centering
	\includegraphics[width=3in]{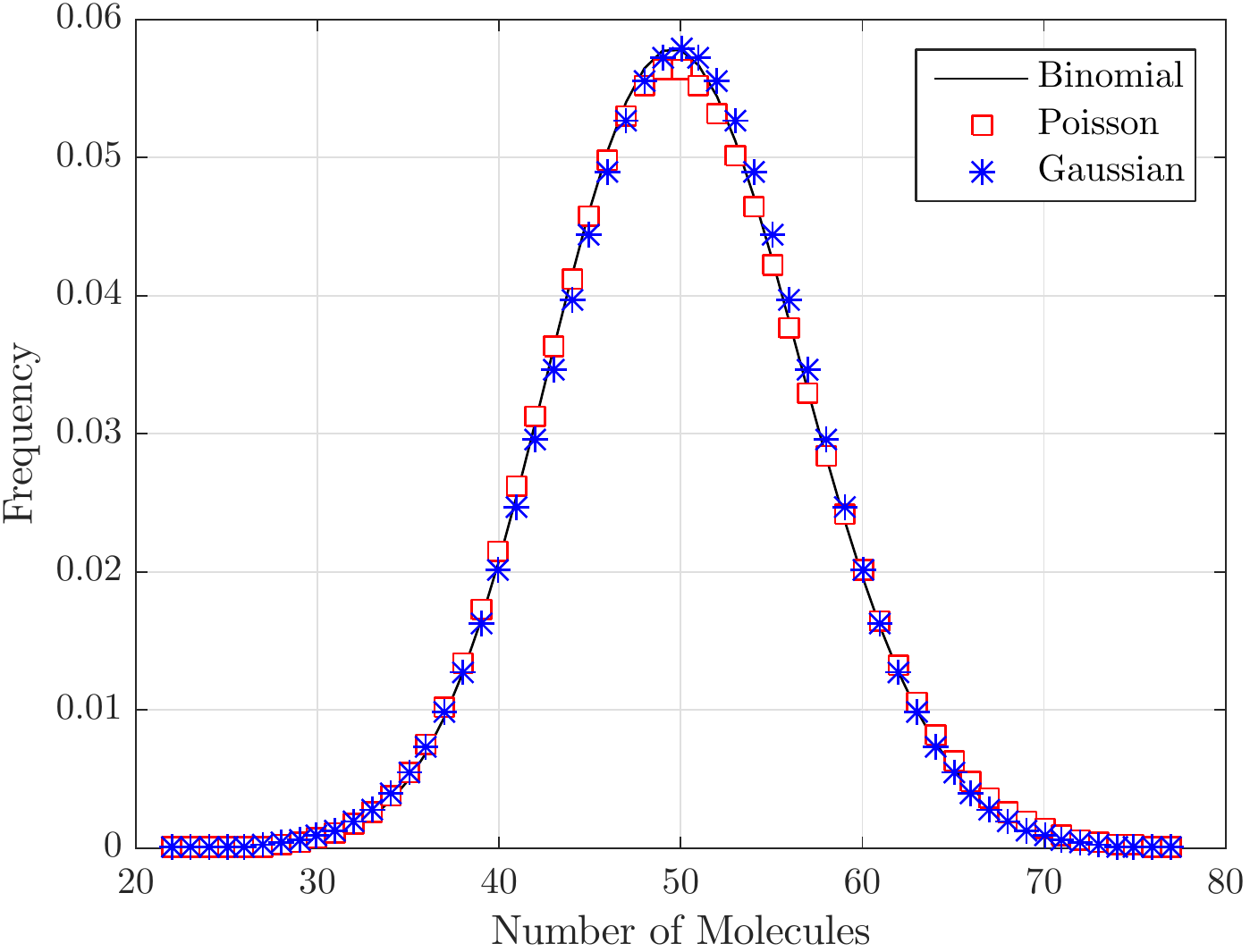}
	\caption{Probability mass function of a Binomial random variable with 1000 trials and a trial success probability of $5\,\%$. This distribution, which matches the expected behavior of the observations in System 1, is also compared with the Poisson and Gaussian approximations of the Binomial. Both approximations show slight deviations from the Binomial, particularly near the peak.}
	\label{fig_expected_pmf}
\end{figure}

Finally, for completeness, we consider the PMFs of the common statistical distributions using the parameters of System 1. Specifically, in Fig.~\ref{fig_expected_pmf} we plot the binomial PMF for 1000 trials that each have a success probability of $5\,\%$ (since there are 1000 molecules that have an unconditional probability of $5\,\%$ of being observed by a given passive actor), and the Poisson and Gaussian approximations of that PMF. The Binomial PMF is identical to that of the microscopic system (which we do not plot again in Fig.~\ref{fig_expected_pmf}; refer to Figs.~\ref{fig_hybrid_pmf} and~\ref{fig_multiscale_pmf}). Both approximations are close to the Binomial PMF in this case, but neither is indistinguishable; the Poisson approximation underestimates the likelihood of the peak observation, whereas the Gaussian approximation underestimates and then overestimates the likelihood of the observations with values less than and greater than 50, respectively. We discussed the general accuracy of these approximations further in \cite{Noel2015}.

\subsection{System 2: Hybrid Reaction-Diffusion}
\label{sec_results_hybrid_rxn_diff}

In System 2, a portion of which is shown in Fig.~\ref{fig_hybrid2_env} on page~\pageref{fig_hybrid2_env}, we demonstrate the potential gain in computation speed by defining an environment with both microscopic and mesoscopic regions. As with System 1, we consider a simple hybrid environment, but this one has a surface reaction in the microscopic regime and (when applicable) a much larger mesoscopic regime. System 2 is a $0.5\,\meter\meter$ long and $5\,\mu\meter$ wide rectangular rod with one \emph{perfectly}-absorbing end; i.e., every molecule that crosses that end is absorbed. 1000 molecules with diffusion coefficient $\Dx{}=10^{-10}\frac{\meter^2}{\second}$ are initially placed uniformly over the entire rod, and we observe the number of molecules over three $5\,\mu\meter$ sections of the rod. The sections are at distances of 0, 5, and $20\,\mu\meter$ from the absorbing end.

The area closest to the absorbing surface is modeled as a microscopic region with a time step of $\dtMicro=1\,\mu\second$, which we found to be small enough to accurately model the perfect absorption reaction at the surface. We first consider the rod modeled as a single microscopic region, and then variations where we model most of the rod as a mesoscopic region with subvolumes of width $\subLength{}=5\,\mu\meter$ (such that the condition $\subLength{}\gg\sqrt{\Dx{}\dtMicro}$ is easily satisfied). In the hybrid variations, the hybrid interface is placed at $\dist_i=\{5,10,25,50\}\,\mu\meter$ from the absorbing surface, such that the environment is either $1\,\%$, $2\,\%$, $5\,\%$, or $10\,\%$ microscopic, respectively. A fully mesoscopic variation was not possible, since AcCoRD cannot yet simulate surface reactions in the mesoscopic regime. Due to the very small microscopic time step, we set $\hybridDist{\mathrm{max}}=1\,\mu\meter$ to limit the proximity of microscopic molecules to the hybrid interface that are tested for entering and exiting the mesoscopic region within a diffusion step.

\begin{figure}[!tb]
	\centering
	\includegraphics[width=3.45in]{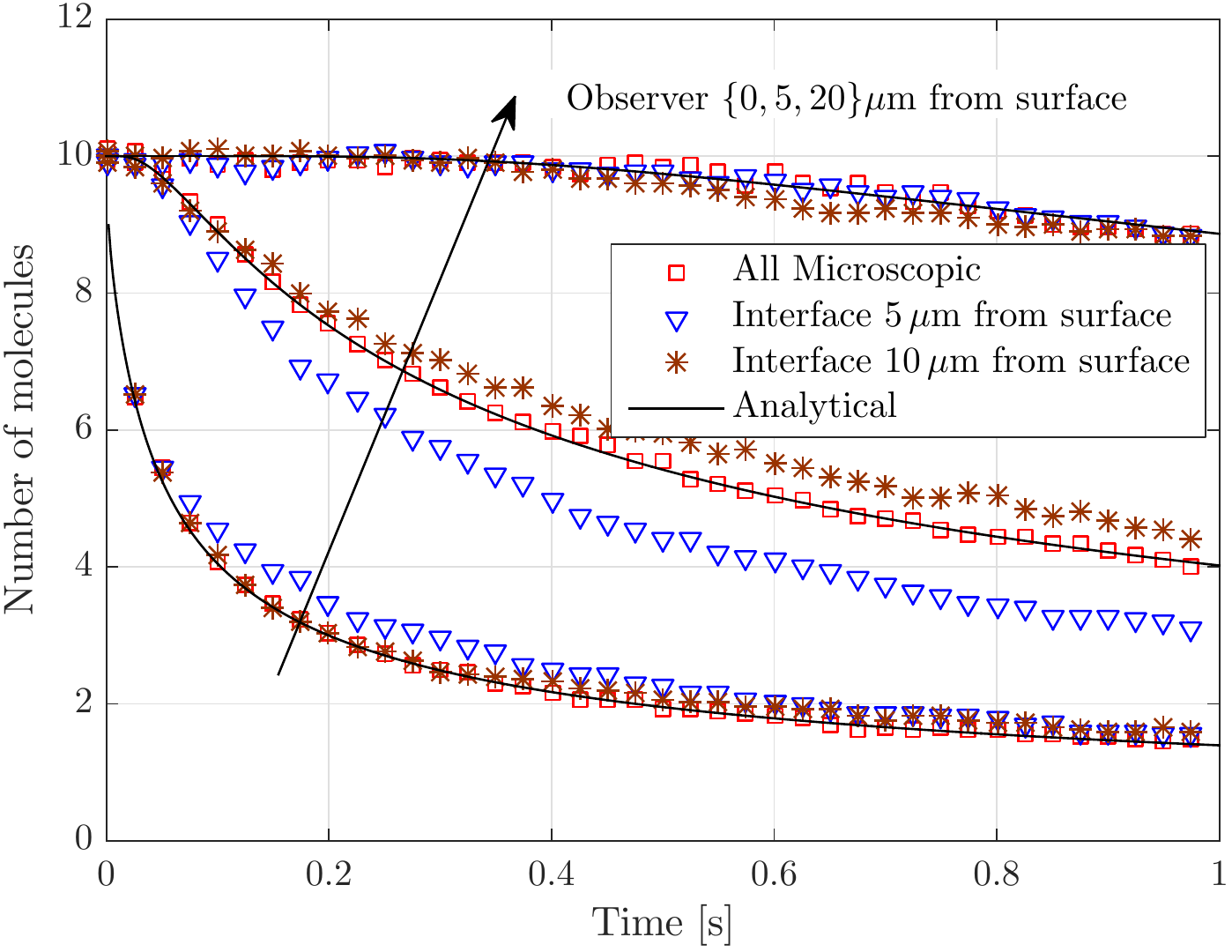}
	\caption{Average time-varying number of molecules observed by each observer in System 2. The hybrid variations place the interface either $\dist_i=\{5,10\}\,\mu\meter$ from the absorbing surface.}
	\label{fig_hybrid_rod_near}
\end{figure}

\begin{figure}[!tb]
	\centering
	\includegraphics[width=3.45in]{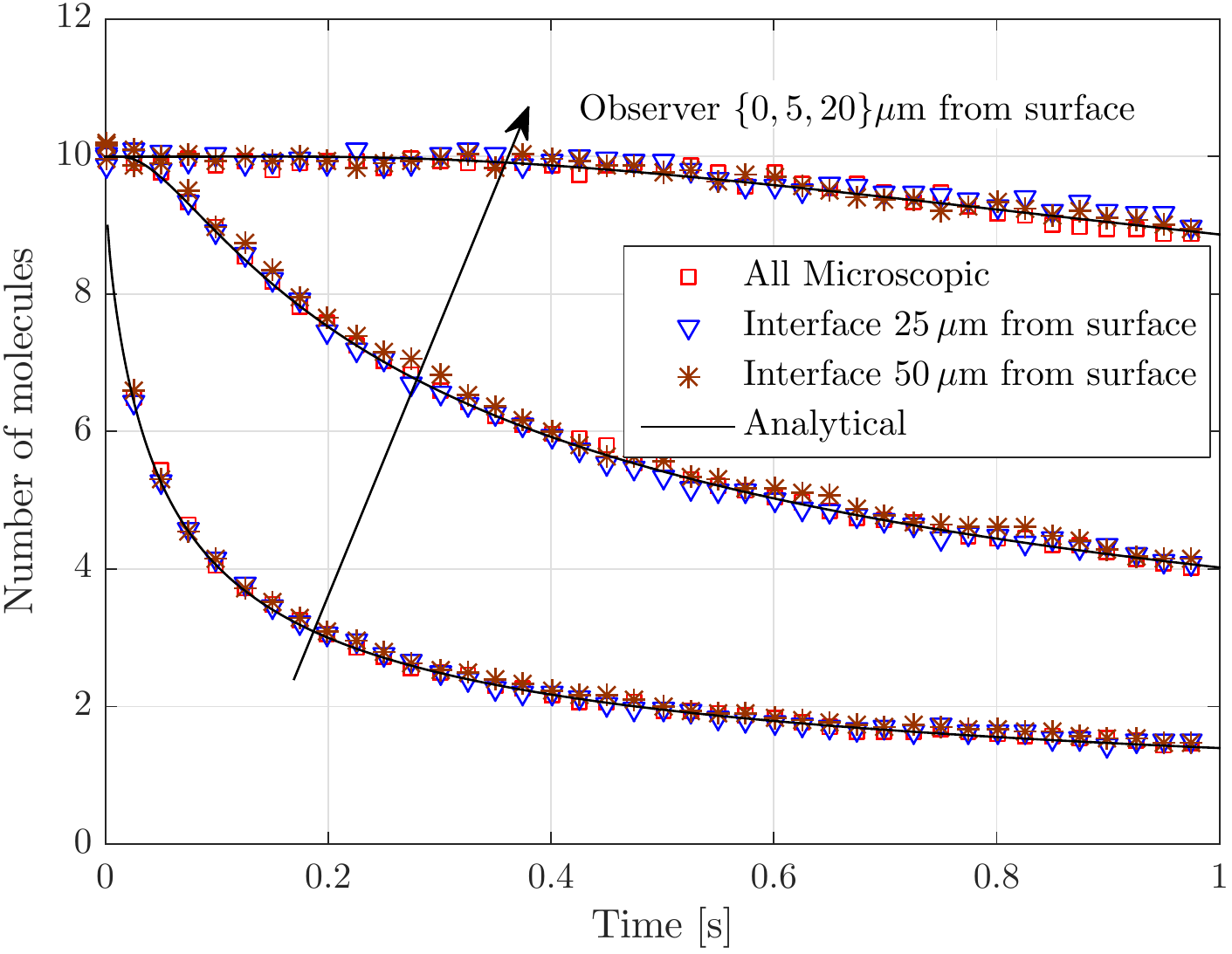}
	\caption{Average time-varying number of molecules observed by each observer in System 2. The hybrid variations place the interface either $\dist_i=\{25,50\}\,\mu\meter$ from the absorbing surface.}
	\label{fig_hybrid_rod_far}
\end{figure}

\begin{figure}[!tb]
	\centering
	\includegraphics[width=3.45in]{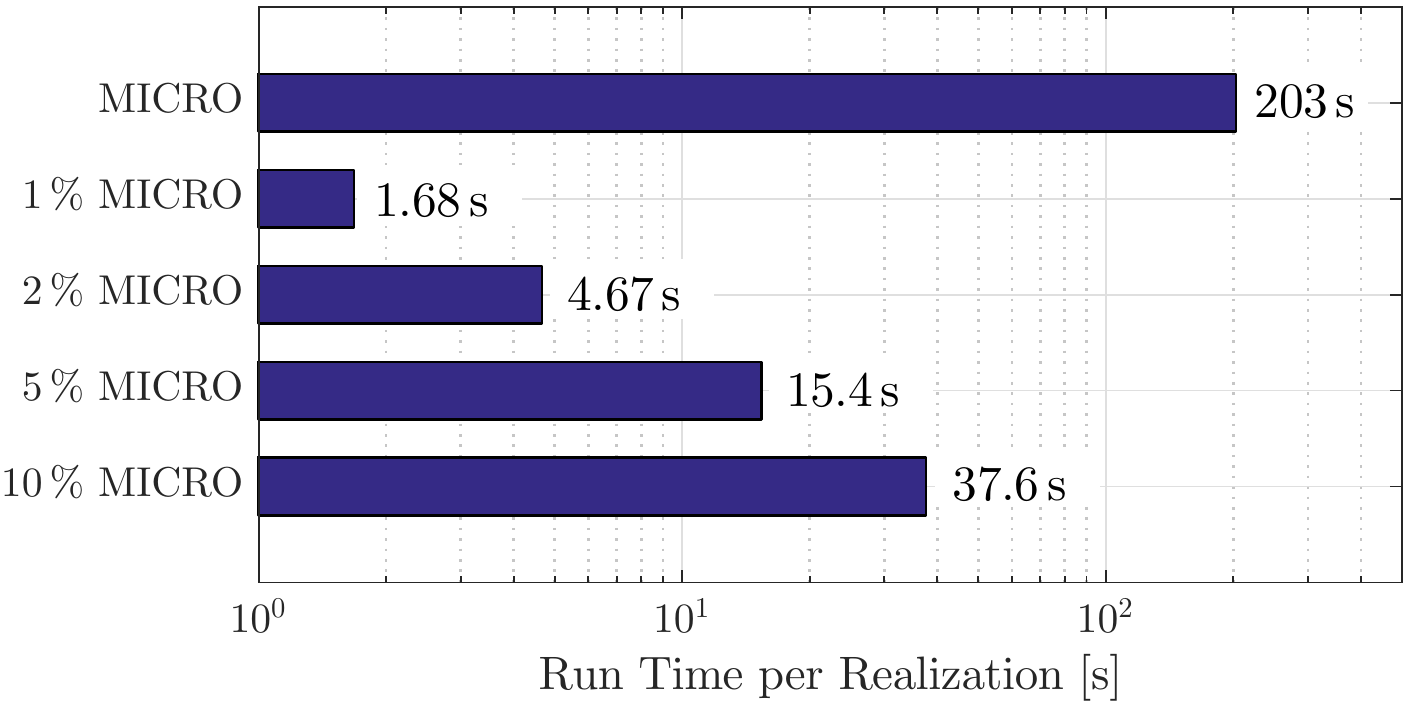}
	\caption{Simulation run times for the variations of System 2. The distances to the hybrid interface are represented via the relative size of the microscopic region, such that the rod is $\{1,2,5,10\}\,\%$ microscopic when the interface is $\dist_i=\{5,10,25,50\}\,\mu\meter$ from the absorbing surface, respectively. Run time was averaged over at least 100 realizations of each variation on an Intel i7 desktop PC.}
	\label{fig_hybrid_rod_runtime}
\end{figure}

We simulated every variation 1000 times and plot the average time-varying signal at each observer in Figs.~\ref{fig_hybrid_rod_near} and \ref{fig_hybrid_rod_far}, where we focus on the hybrid interface being close to and far from the absorbing end, respectively. We measure the average realization run time for all variations in Fig.~\ref{fig_hybrid_rod_runtime}. Video 3 in \cite{Noel2016c} shows a sample realization when $\dist_i=25\,\mu\meter$.

The expected time-varying signal in this environment can be derived in closed form, \emph{if} we assume that the rod has infinite length. Using the point concentration defined in \cite[Eq.~(3.13)]{Crank1979}, and the integral of the error function in \cite[Eq.~(4.1.1)]{Ng1968}
\begin{equation}
\int \ERF{\y}\mathrm{d}\y = \y\ERF{\y} + \frac{1}{\sqrt{\pi}}\EXP{-\y^2},
\end{equation}
then it can be shown that the expected time-varying \emph{concentration} is
\begin{align}
\Cx{}(t) = \,& \Cx{0}\bigg[\rx{2}\ERF{\frac{\rx{2}}{2\sqrt{\Dx{}t}}}
- \rx{1}\ERF{\frac{\rx{1}}{2\sqrt{\Dx{}t}}} \nonumber \\
&+ 2\sqrt{\frac{\Dx{}t}{\pi}}\left(\EXP{-\frac{\rx{2}^2}{4\Dx{}t}}
- \EXP{-\frac{\rx{1}^2}{4\Dx{}t}}\right)\bigg],
\label{eqn_diff_surf}
\end{align}
where $\Cx{0}$ is the initial concentration and $\rx{1}$ and $\rx{2}$ are the distances from the absorbing surface to the start and end of the observer, respectively. In System 2, the initial linear distribution of molecules is $1000$ molecules per $500\,\mu\meter$, so we multiply the time-varying concentration in (\ref{eqn_diff_surf}) by the observer length of $5\,\mu\meter$ to get the time-varying number of molecules at the observer. We apply this scaling of (\ref{eqn_diff_surf}) for the ``Analytical'' curves in Figs.~\ref{fig_hybrid_rod_near} and \ref{fig_hybrid_rod_far}.

Both Figs.~\ref{fig_hybrid_rod_near} and \ref{fig_hybrid_rod_far} demonstrate that the observations in the fully microscopic system are consistent with the analytical curves. Thus, System 2 is sufficiently long to assume that it has infinite length for the time scale considered. In Fig.~\ref{fig_hybrid_rod_near}, we see that the hybrid model with the interface $5\,\mu\meter$ from the surface (i.e., $\dist_i=5\,\mu\meter$) overestimates the number of molecules counted by the observer adjacent to the absorbing surface, and significantly underestimates the number of molecules counted by the observer $5\,\mu\meter$ from the surface. In this environment, since $\dist_i=5\,\mu\meter$, these two observers are adjacent to the hybrid interface. The interface's proximity to the absorbing surface is most likely the cause of these deviations, since the hybrid interface transition rules were originally derived in \cite{Flegg2014} under the assumption that there were no molecule ``sinks'' in the vicinity. There is much less deviation in this variation at the observer $20\,\mu\meter$ from the surface, which is sufficiently far from the interface to not be affected by the inaccuracies that it introduces over the time scale that we simulated.

When the hybrid interface is $\dist_i=10\,\mu\meter$ from the surface, then the accuracy of the molecules observed at $0\,\mu\meter$ and $5\,\mu\meter$ from the interface are more accurate than when $\dist_i=5\,\mu\meter$. However, this variation is actually less accurate for the observer at $20\,\mu\meter$ from the surface, particularly after time $t = 0.5\,\second$. This time is how long it takes for a molecule to travel an average diffusion distance of $10\,\mu\meter$ along one dimension, which is the distance from the hybrid interface to this observer, so the accuracy at the interface is a factor. Overall, placing the interface at $\dist_i=10\,\mu\meter$ gives us better (albeit still imperfect) accuracy near the absorbing surface, at the cost of worse accuracy further from the surface over the time scale shown (i.e., where the average diffusion distance for one molecule is about $14\,\mu\meter$).

In Fig.~\ref{fig_hybrid_rod_far}, we confirm whether the trade-offs in local accuracy are dominated by proximity to the hybrid interface. For the time scale simulated, this is apparently not the case. When the hybrid interface is either $\dist_i=\{25,50\}\,\mu\meter$, the accuracy is comparable to that of the fully microscopic variation. This is even the case when $\dist_i=25\,\mu\meter$ and the observer is $20\,\mu\meter$ from the surface, i.e., when the observer is adjacent to the interface. So, we maintain our claim that the dominant factor affecting accuracy in the hybrid system is the interface's proximity to the absorbing surface, although further analysis would be needed to confirm this analytically.

We complement our study of the accuracy of the System 2 variations by considering the simulation run times in Fig.~\ref{fig_hybrid_rod_runtime}. We observe that placing the hybrid interface $\dist_i=5\,\mu\meter$ from the surface, which made the system only $1\,\%$ microscopic, improved the simulation run time by over 2 orders of magnitude. Even making the system $5\,\%$ microscopic, which corresponded to a negligible loss in accuracy, still improved the run time by over an order of magnitude. From these results, and those observed for System 1 in Fig.~\ref{fig_meso_runtime}, we gain some insight into when it is appropriate to use a hybrid model. We make the following general observations:
\begin{itemize}
	\item The hybrid interface introduces a computational overhead, as discussed in~\ref{sec_alg_micro}, since we need to check microscopic molecules for entering and leaving the microscopic regime within a single simulation time step. Thus, it is beneficial if the size of the interface is small relative to the larger of the adjacent regions. This was true for System 2 but not for System 1.
	\item The hybrid interface is beneficial if the simulation can take advantage of the benefits of both simulation models. In System 2, the microscopic model needed a very small simulation step to accurately simulate the surface reaction (which AcCoRD cannot currently simulate in the mesoscopic regime). However, the mesoscopic model simulates events at the frequency in which they locally occur. Diffusion ``far'' from a reactive surface does not need to be modeled with a small time step, and in System 2 the hybrid model took advantage of this. We also note that, in future work, a microscopic model that uses different time steps in different regions may provide a similar speed benefit, although the hybrid model could still have better memory usage by not needing to model each molecule in the mesoscopic regime.
	\item A user placing a hybrid interface should consider its proximity to reactive surfaces or other environmental features (e.g., unusual boundary shapes) that are not accounted for in the hybrid interface transition rules.
\end{itemize}

\subsection{System 3: Reactive Spherical Surfaces}
\label{sec_results_surface}

In System 3, we consider an approximation of the environment considered in \cite[Fig.~6a]{Andrews2009} to verify the surface reaction probabilities derived in \cite{Andrews2009} (and which we summarize in~\ref{app_surf}). However, instead of the 1D environment considered in \cite{Andrews2009}, System 3 has two concentric spherical surfaces with radii $120\,\mu\meter$ and $122\,\mu\meter$, as shown in Fig.~\ref{fig_surface_env} on page~\ref{fig_surface_env}. This system is locally 1D, and we can also use it to verify our implementation of reflections off of spherical surfaces. This environment is fully microscopic and we use a time step of $\dtMicro=1\,\meter\second$.

For consistency with \cite{Andrews2009}, we consider 3 types of molecule: 1) Type I can be irreversibly absorbed by the outer surface and reflects off of the inner surface; 2) Type II can reversibly adsorb to the outer surface and reflects off of the inner surface; and 3) Type III can reversibly transition across the inner surface and reflects off of the outer surface. We separately consider both ``slow'' and ``fast'' reaction kinetics for each type of molecule, using the reaction coefficients listed in Table~\ref{table_surface_rates}, and we apply the ``steady state'' surface reaction probabilities defined in~\ref{app_surf}. All molecules have a diffusion coefficient of $5\times10^{-12}\,\frac{\meter^2}{\second}$, and each simulation begins by uniformly placing 20000 molecules of the given type between the spherical surfaces. We observe the number of molecules that are absorbed/adsorbed to the outer surface or that are inside the inner sphere (as appropriate for each type of molecule).

\begin{table}[!tb]
	\centering
	\caption{Surface reaction coefficients used by System 3. ``IRR'' and ``REV'' stand for irreversible and reversible, respectively.}
	
	{\renewcommand{\arraystretch}{1.4}\footnotesize
		\begin{tabular}{l||l|l}
			\hline
			\bfseries Reaction & \bfseries Slow Reaction & \bfseries Fast Reaction \\ \hline \hline
			IRR Absorption & $4.23\,\frac{\mu\meter}{\second}$ & $85.9\,\frac{\mu\meter}{\second}$ \\ \hline
			REV Adsorption & $4.23\,\frac{\mu\meter}{\second}$ & $85.9\,\frac{\mu\meter}{\second}$ \\
			Desorption & $28\,\second^{-1}$ & $276\,\second^{-1}$ \\ \hline
			Membrane & $4.36\,\frac{\mu\meter}{\second}$ & $5045\,\frac{\mu\meter}{\second}$ \\ \hline
		\end{tabular}
	}
	\label{table_surface_rates}
\end{table}

The average time-varying results from $100$ realizations of System 3 are shown in Fig.~\ref{fig_surface} and compared with the corresponding analytical equations derived in the appendices of \cite{Andrews2009}\footnote{For clarity of exposition and in consideration of space, we do not write out these analytical expressions in this work. However, we note that their general form is similar to the reaction probabilities that we describe in \ref{app_surf}.}. All of the simulations agree very well with their corresponding analytical curves, despite the system not being truly 1D. Thus, we have confidence in AcCoRD's algorithms for placing molecules within non-trivial spherical environments (since the initialization space was that between the two spheres), detecting imperfect surface reactions, and implementing surface reflections. Videos 4 and 5 in \cite{Noel2016c} show sample realizations with reversible adsorption and membrane transitions, respectively.

\begin{figure}[!tb]
	\centering
	\includegraphics[width=3.45in]{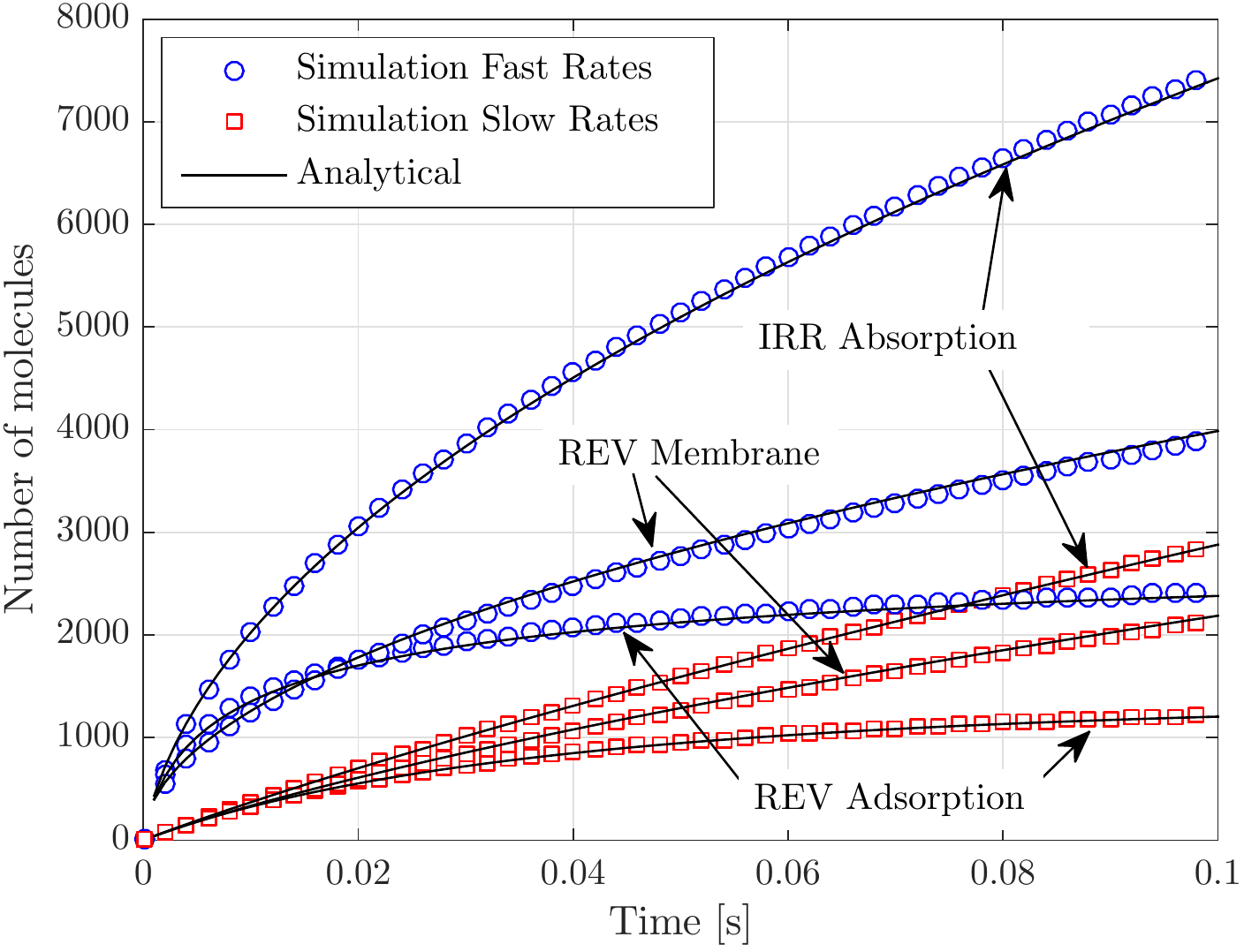}
	\caption{Average time-varying number of molecules absorbed/adsorbed to the outer surface (``Absorption/Adsorption'') or within the inner surface (``Membrane'') in System 3. This environment is a spherical analog to the 1D system studied in \cite[Fig.~6a]{Andrews2009}. ``IRR'' and ``REV'' stand for irreversible and reversible, respectively.}
	\label{fig_surface}
\end{figure}

\subsection{System 4: Simple Communication System}
\label{sec_results_comm}

In System 4, we consider variations of the extensively-studied molecular communication scenario with a point transmitter and a spherical receiver, as shown in Fig.~\ref{fig_comm_env} on page~\pageref{fig_comm_env}. The receiver has a radius of $\rrx=1\,\mu\meter$, and the transmitter is placed $\dist=5\,\mu\meter$ from the center of the receiver. Due to the shape of the receiver, and the many variations that we consider, we consistently use a microscopic model. Unless otherwise noted, the transmitter releases an impulse of $\Nx{\TX}=10^4$ molecules at time $t=0$, molecules have a diffusion coefficient of $\Dx{}=10^{-10}\,\frac{\meter^2}{\second}$, and every simulation is repeated at least 1000 times.

\begin{figure}[!tb]
	\centering
	\includegraphics[width=3.45in]{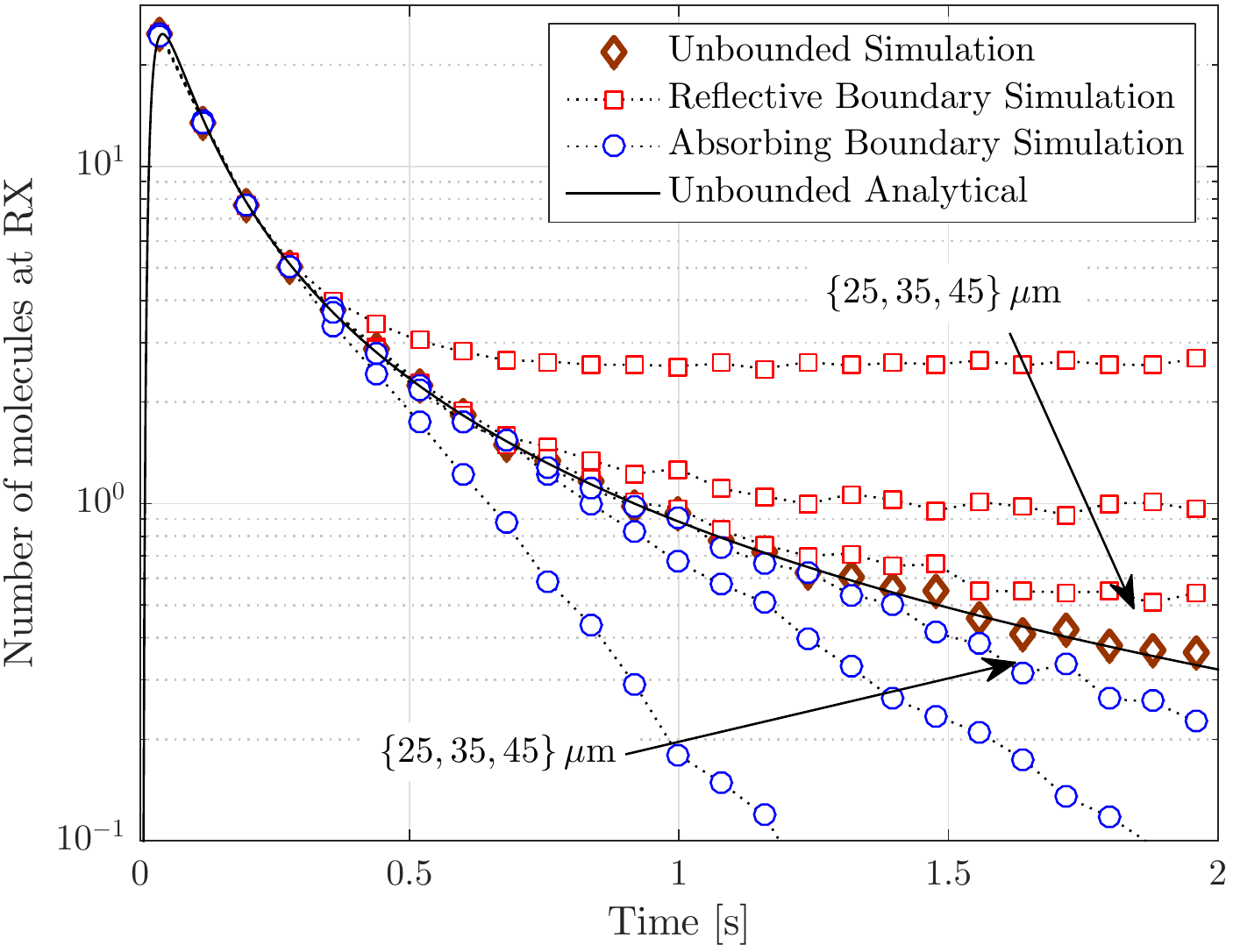}
	\caption{Average time-varying number of molecules inside a passive spherical observer (of radius $1\,\mu\meter$) after release by a point source located $5\,\mu\meter$ from the center of the sphere. The unbounded environment is compared with bounded environments that have either a reflective or absorbing outer surface. The bounded environments are cubes with width $\{25,35,45\}\,\mu\meter$ with the transmitter and observer in the middle.}
	\label{fig_unbounded_vs_bounded}
\end{figure}

First, we consider a passive receiver, i.e., the receiver is a passive spherical actor through which the molecules can freely diffuse\footnote{We clarify that a ``passive'' actor is not synonymous with a passive receiver. In AcCoRD, \emph{all} observers are passive actors. For example, a receiver with a reactive surface would still need a passive actor to count the number of molecules on the surface.}. In Fig.~\ref{fig_unbounded_vs_bounded}, we test three ``outer'' boundary conditions, namely: 1) \emph{unbounded}, where molecules are unimpeded everywhere; 2) \emph{reflective}, where the transmitter and receiver are in the middle of a box with a reflective outer boundary; and 3) \emph{absorbing}, where the transmitter and receiver are in the middle of box with an absorbing outer boundary. Each box is a cube and we consider different box widths. The aim of this test is to assess the accuracy of assuming that a bounded environment is unbounded. For ease of analysis, it is commonly assumed that environments are unbounded. Fully bounded environments can lead to time-varying impulse responses that include infinite sums; see examples in \cite{Crank1979}. However, many envisioned application environments, such as the human body or microfludic devices, are clearly bounded. We expect that the accuracy of the unbounded model will depend on the time scale considered and (in this case) on the size of the box. The expected time-varying number of molecules at the receiver, $\Nx{\RX}(t)$, in the unbounded case is a classical result that can be described as a ``diffusion wave'' and can be expressed from \cite[Eq.~(3.5)]{Crank1979} as
\begin{equation}
\Nx{\RX}(t) = \frac{\Nx{\TX}\Vol{\RX}}{\left(4\pi\Dx{}t\right)^{\frac{3}{2}}}\EXP{-\frac{\dist^2}{4\Dx{}t}},
\label{eqn_3D_diff}
\end{equation}
where $\Vol{\RX}$ is the volume of the receiver, and we assume that the receiver is sufficiently far from the transmitter to assume that the concentration throughout the passive receiver is uniform; see \cite{Noel2013b}.

We plot the average time-varying number of molecules at the receiver for the unbounded and box cases in Fig.~\ref{fig_unbounded_vs_bounded}, where we consider box widths of $\{25,35,45\}\,\mu\meter$ and use a time step of $\dtMicro=2\,\meter\second$. The simulation of the unbounded case agrees with the analytical result, and the slight deviation at higher values of $t$ can be improved with more realizations. We expected that the unbounded equation would overestimate the number of molecules inside the absorbing box and underestimate the number of molecules in the reflective box. This is indeed the case; the unbounded equation is accurate for early time, and for smaller boxes the deviation from this equation occurs sooner. Interestingly, for the same box size, the reflective and absorbing box models begin to deviate at the same time, i.e., at $\{0.35,0.75,1.1\}\,\second$ when the box is $\{25,35,45\}\,\mu\meter$ wide, respectively. In each case, this deviation occurs in about one third of the time it would take for the ``peak'' of the diffusion wave to reach the edge of the box and return to the source\footnote{It can be shown that the expected time of the peak value of (\ref{eqn_3D_diff}) at distance $\dist$ is $\dist^2/(6\Dx{})$, such that the average diffusion distance for time $t$ is $\sqrt{6\Dx{}t}$. For $\dist=\{25,35,45\}\,\mu\meter$, the expected time for molecules to diffuse from the center of the box to the edge and back is $\approx\{1.04,2.04,3.38\}\,\second$, respectively.}. So, for a bounded environment to accurately model a diffusion-only unbounded environment, the distance from the area of interest to the nearest environment boundary should be at least three times the average 3D diffusion distance of $\sqrt{6\Dx{}t}$ for the time scale $t$ of interest.

Next, we study molecule degradation in the propagation environment, which has been considered to address the significant intersymbol interference in a diffusion-only MC system; see \cite{Heren2015} and our previous work in \cite{Noel2014f,Noel2014d}. Specifically, we compare molecule degradation through enzymatic action (using Michaelis-Menten kinetics) with passive, first order molecule degradation. With enzyme kinetics, the diffusing information molecules might reversibly bind to diffusing enzyme molecules. Binding to an enzyme also catalyzes the degradation of the information molecule, but this degradation leaves the enzyme intact to bind with other molecules. The Michaelis-Menten reaction mechanism is commonly used to describe enzyme kinetics and can be written as \cite[Sec.~10.2]{Chang2005}
\begin{align}
	E + S \xrightarrow{\kth{\mathrm{f}}} ES \nonumber \\
	ES \xrightarrow{\kth{\mathrm{r}}} E + S \nonumber \\
	ES \xrightarrow{\kth{\mathrm{d}}} P + E,
	\label{eqn_rxn_mm}
\end{align}
where $E$ is the enzyme molecule, $S$ is the substrate molecule (i.e., a molecule released by the transmitter in this context), $ES$ is the intermediate formed by the binding of the enzyme molecule with its substrate, and $P$ is the product molecule (which we will assume is ignored by the receiver, i.e., $P \equiv \emptyset$). $\kth{\mathrm{f}}$, $\kth{\mathrm{r}}$, and $\kth{\mathrm{d}}$ are the corresponding reaction rate constants. If the enzyme concentration is uniform, and we assume that both $\kth{\mathrm{r}} \to 0$ and $\kth{\mathrm{d}} \to \infty$, then we can approximate Michaelis-Menten kinetics with a first order mechanism that is
\begin{equation}
S \xrightarrow{\kth{1}} \emptyset,
\label{eqn_rxn_first}
\end{equation}
where $\kth{1}$ is $\kth{\mathrm{f}}$ in (\ref{eqn_rxn_mm}) scaled by the average concentration of enzyme molecules. It is much faster to simulate (\ref{eqn_rxn_first}) than (\ref{eqn_rxn_mm}), due to the extra computational load to compare distances between candidate $E$ and $S$ molecules with the binding radius. Furthermore, analytical solutions for signals with first-order degradation are more readily available. For example, it can be shown that first order degradation changes the diffusion wave in (\ref{eqn_3D_diff}) to
\begin{equation}
\Nx{\RX}(t) = \frac{\Nx{\TX}\Vol{\RX}}{\left(4\pi\Dx{}t\right)^{\frac{3}{2}}}\EXP{-\frac{\dist^2}{4\Dx{}t} - \kth{1}t},
\label{eqn_3D_diff_first}
\end{equation}
i.e., an additional decaying exponential factor is added.

We compare Michaelis-Menten enzyme kinetics with first order molecule degradation as follows. We seek parameter values that are convenient to simulate. We begin with a target first order degradation rate of $\kth{1} = 8\,\second^{-1}$, a target bimolecular binding radius of $\rBind=0.1\,\mu\meter$, and a microscopic time step of $\dtMicro=2\,\meter\second$. The value of $\kth{1}$ is sufficient to observe measurable degradation over the time scale of the diffusion wave's peak at the passive receiver. The value of $\rBind$ is smaller than the size of the average 1D diffusion distance in one microscopic time step $\sqrt{2\Dx{}\dtMicro}\approx0.63\,\mu\meter$, which we need to satisfy in order to relate $\rBind$ to $\kth{\mathrm{f}}$ with the closed-form expression
\begin{equation}
\rBind = \left(\frac{3\kth{\mathrm{f}}\dtMicro}{4\pi}\right)^\frac{1}{3},
\label{eqn_enz_rbind}
\end{equation}
as derived in \cite[Eq.~(27)]{Andrews2004}. From (\ref{eqn_enz_rbind}), we calculate that $\rBind$ corresponds to $\kth{\mathrm{f}} = 2.094\times10^{-17}\mol^{-1}\meter^3\second^{-1}$, which (fortunately) is almost two orders of magnitude smaller than the largest value of $\kth{\mathrm{f}}$ physically possible; see \cite[Ch.~10]{Chang2005}. The corresponding uniform enzyme concentration needed to achieve an overall binding rate that matches $\kth{1}$ is $3.82\times10^{18}\frac{\mol}{\meter^3}$. To limit the actual number of enzyme molecules needed, we place the transmitter and receiver in the middle of a reflective box of width $15\,\mu\meter$. This is significantly smaller than the smallest box considered in Fig.~\ref{fig_unbounded_vs_bounded}, but only requires 12891 enzyme molecules to meet the target concentration. In the interest of computational speed, we proceed with these system parameters, while emphasizing that AcCoRD can also accommodate much larger environments. Most enzymes are proteins that tend to be larger than their substrates (see \cite[Ch.~4]{Alberts}), and larger molecules diffuse more slowly, so we set the enzyme $E$ diffusion coefficient to $2\times10^{-11}\,\frac{\meter^2}{\second}$ and that of the intermediate $ES$ to $1.8\times10^{-11}\,\frac{\meter^2}{\second}$.

\begin{figure}[!tb]
	\centering
	\includegraphics[width=3.45in]{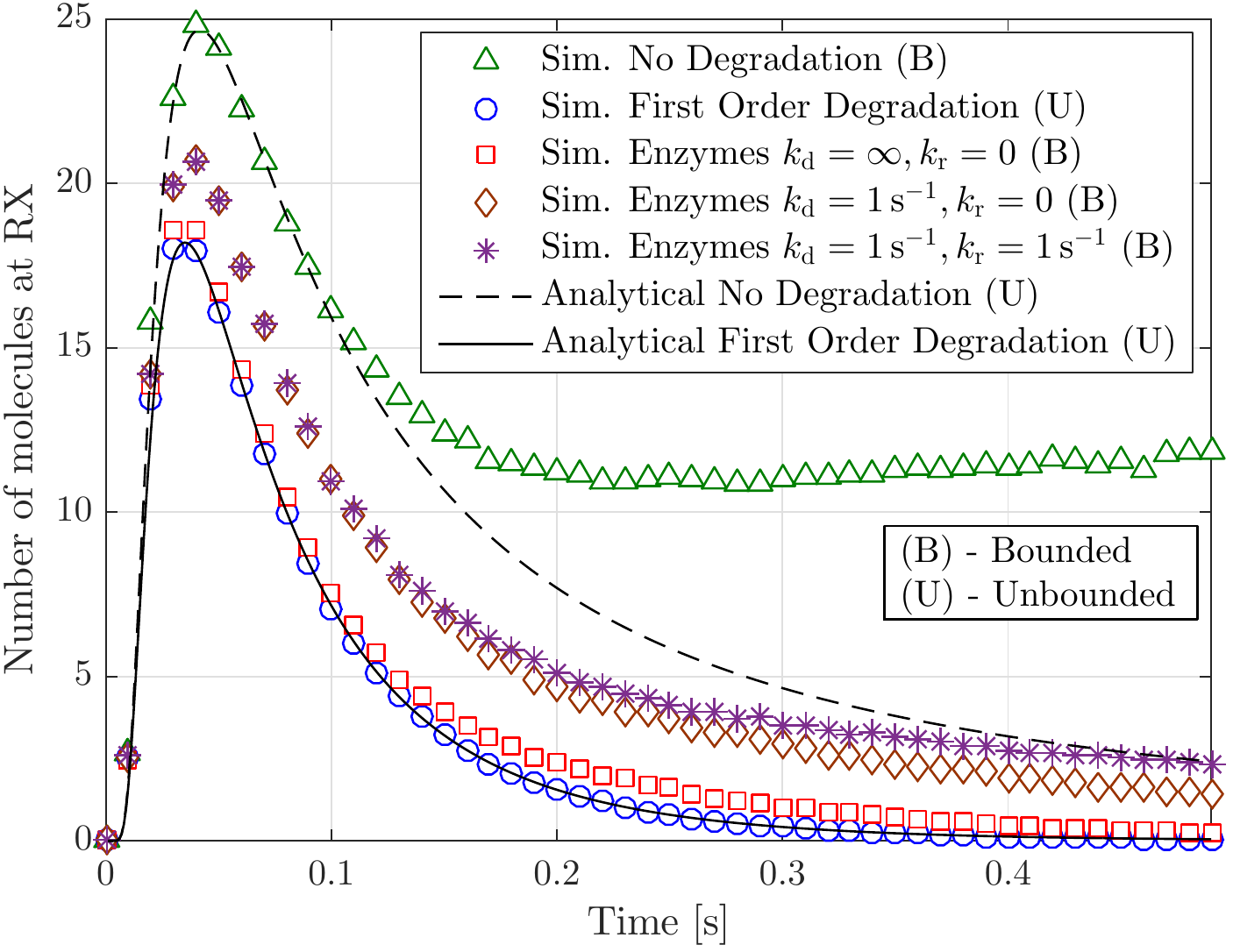}
	\caption{Average time-varying number of molecules inside a passive spherical observer (of radius $1\,\mu\meter$) after release by a point source located $5\,\mu\meter$ from the center of the sphere. An unbounded environment where the molecules undergo first order degradation is compared with 1) a bounded, reflective environment ($15\,\mu\meter$ in width) where the molecules can degrade by binding with diffusing enzymes, and 2) the same reflective environment without degradation. Analytical curves are included for first order degradation and no degradation in an unbounded environment.}
	\label{fig_enzyme}
\end{figure}

In Fig.~\ref{fig_enzyme}, we observe the average time-varying number of molecules inside the passive receiver when the released $S$ molecules can degrade in the propagation environment, and for additional comparison we include the case where there is no degradation but the transmitter and receiver are still in a reflective box of width $15\,\mu\meter$. We see that the simulation of first order degradation agrees well with the analytical expression in (\ref{eqn_3D_diff_first}) and shows significant deviation from the signal without degradation given by (\ref{eqn_3D_diff}) and the simulation of the bounded environment. Interestingly, the curve for the simulation of the bounded environment appears to reach a minimum after about $0.25\,\second$ and then increase. This is because molecules are returning to the center of the box after reflecting off of the boundary. The receiver occupies about $1.24\%$ of the box, so the expected number of molecules should converge to about $12.4$.

For the enzyme kinetics in Fig.~\ref{fig_enzyme}, we simulate three variations in the values of $\kth{\mathrm{r}}$ and $\kth{\mathrm{d}}$, i.e., in the reverse (unbinding) and degradation reactions, respectively. In the ideal limit, where $\kth{\mathrm{r}} \to 0$ and $\kth{\mathrm{d}} \to \infty$, the first order degradation curve is still a clear lower bound. Near the peak of the curve, this difference is because a given enzyme molecule can only react with one observable molecule in a given time slot. For time $t>0.1\,\second$, the finite size of the environment also becomes a factor, although not to the degree observed in the absence of degradation. We also consider very slow but practical values of $\kth{\mathrm{r}}$ and $\kth{\mathrm{d}}$\footnote{$\kth{\mathrm{d}}$ and $\kth{\mathrm{r}}$ typically have values between 1 and	$10^5\,\second^{-1}$; see \cite[Ch.~10]{Chang2005}.}, by first setting $\kth{\mathrm{d}} = 1\,\second^{-1}$ and then adding $\kth{\mathrm{r}} = 1\,\second^{-1}$. When $\kth{\mathrm{d}}$ is finite, an intermediate molecule can exist for multiple time slots. This slows down an enzyme molecule's capacity to bind with multiple substrates, so overall fewer molecules degrade and more are observed at the receiver. When we include a non-zero value of $\kth{\mathrm{r}}$, intermediates can also unbind with no change to the substrate, which further reduces the amount of degradation. From Fig.~\ref{fig_enzyme}, we clearly see that the first order degradation model has limited accuracy in modeling practical enzyme kinetics. Video 6 in \cite{Noel2016c} shows a sample realization with non-zero $\kth{\mathrm{r}}$ and $\kth{\mathrm{d}}$.

\begin{figure}[!tb]
	\centering
	\includegraphics[width=3.45in]{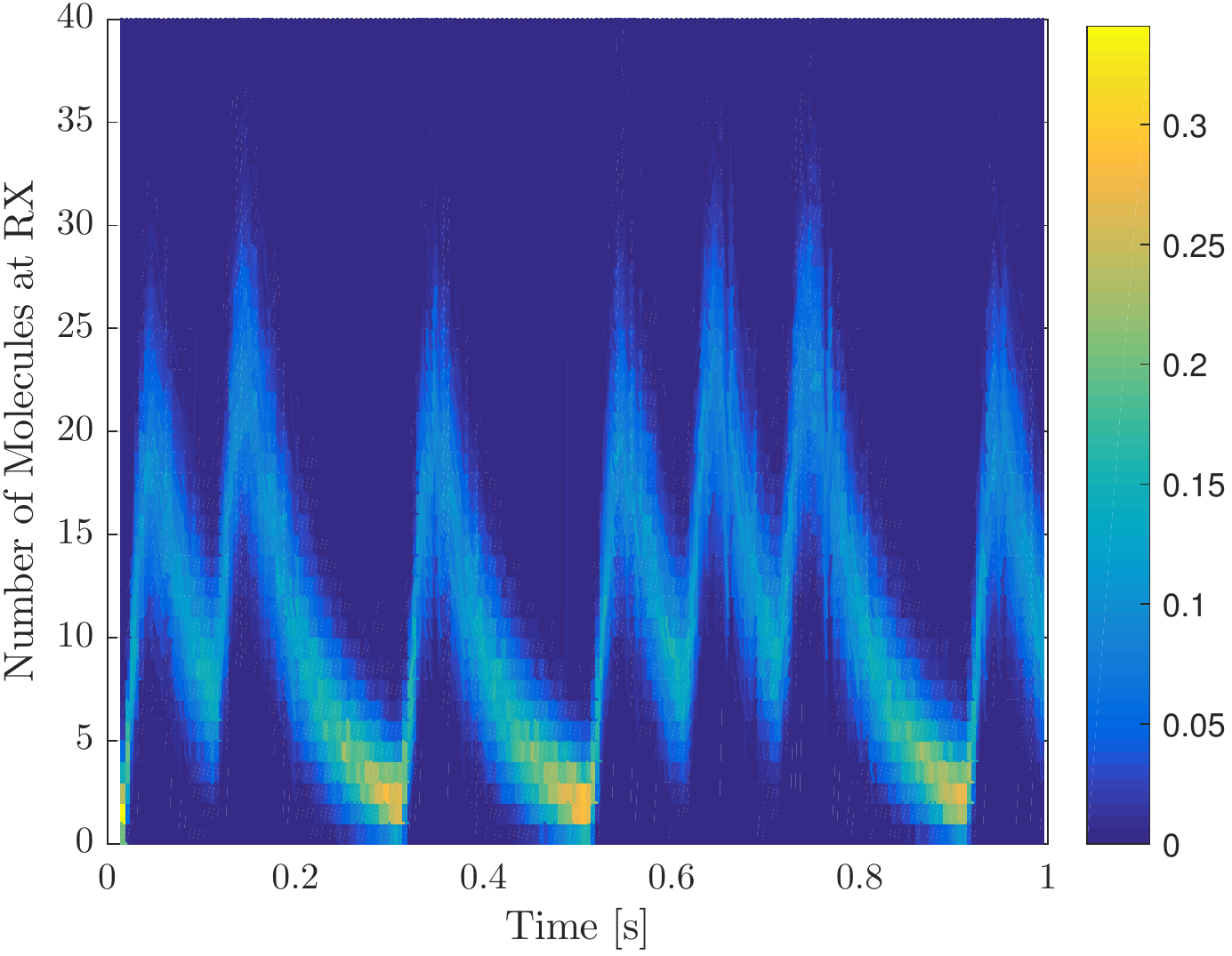}
	\caption{Time-varying probability mass function of the number of molecules inside a passive spherical observer when a point source releases rectangular pulses of molecules according to the binary sequence $[1, 1, 0, 1, 0, 1, 1, 1, 0, 1]$ with a symbol interval of $0.1\,\second$. The pulses have a length of $0.02\,\second$ and a strength of $5\times10^5\,\mol/\second$. Released molecules degrade at rate $\kth{} = 8\,\second^{-1}$. The bar to the right is the legend for observation probabilities.}
	\label{fig_pmf}
\end{figure}

For our final variation of a passive receiver, we study the time-varying statistics of the receiver's observations when the transmitter releases a series of molecule pulses with finite width. Specifically, the transmitter encodes the 10-bit sequence $[1, 1, 0, 1, 0, 1, 1, 1, 0, 1]$ with a bit interval of $0.1\,\second$. For every bit 0, no molecules are released. For every bit 1, molecules are released as a zeroth order reaction for $0.02\,\second$ at a rate of $5\times10^5\,\mol/\second$, i.e., on average there are $10^4$ molecules released for every bit 1. The released molecules also degrade at a rate of $\kth{} = 8\,\second^{-1}$. The time-varying PMF for the observations at the receiver is plotted in Fig.~\ref{fig_pmf}. We can distinguish the seven peaks for the bit 1s, but also notice the accumulation of intersymbol interference since repeated peaks get higher (and this would be even more noticeable in the absence of molecule degradation). The information used to generate Fig.~\ref{fig_pmf} can also be used to analyze the detection performance at the receiver, but that is not a feature currently native to AcCoRD and so is outside the scope of this paper. Video 7 in \cite{Noel2016c} shows the first five symbol intervals of a sample realization of this multi-symbol variation.

Finally, we consider variations of System 4 with reactions at the surface of the receiver. Such a receiver more accurately models the chemical detection of molecules. We focus on the fully-absorbing (i.e., perfectly-absorbing) receiver and the partially-absorbing receiver, which have well known analytical results for the geometry of System 4. When the receiver is fully absorbing, the expected time-varying number of molecules absorbed is given by \cite[Eq.~(3.116)]{Schulten2000}
\begin{equation}
\Nx{\RX}(t) = \frac{\Nx{\TX}\rrx}{\dist}\ERFC{\frac{\dist-\rrx}{\sqrt{4\Dx{}t}}}.
\label{eqn_absorbing_full}
\end{equation}

When the receiver is partially absorbing, the absorption rate $\kth{}$ is finite. The expected time-varying number of molecules absorbed is given by \cite[Eq.~(3.114)]{Schulten2000}
\begin{align}
\Nx{\RX}(t) = & \,\Nx{\TX}\frac{\rrx\beta-1}{\dist\beta}
\bigg[1 + \ERF{\frac{\rrx-\dist}{2\sqrt{\Dx{}t}}} \nonumber \\
& -\EXP{(\dist-\rrx)\beta + \Dx{}t\beta^2}\ERFC{\frac{\dist-\!\rrx + 2\Dx{}\beta t}{2\sqrt{\Dx{}t}}}\!\!\bigg],
\label{eqn_absorbing_partial}
\end{align}
where
\begin{equation}
\beta = \frac{\kth{}}{\Dx{}} + \frac{1}{\rrx}.
\end{equation}

\begin{figure}[!tb]
	\centering
	\includegraphics[width=3.45in]{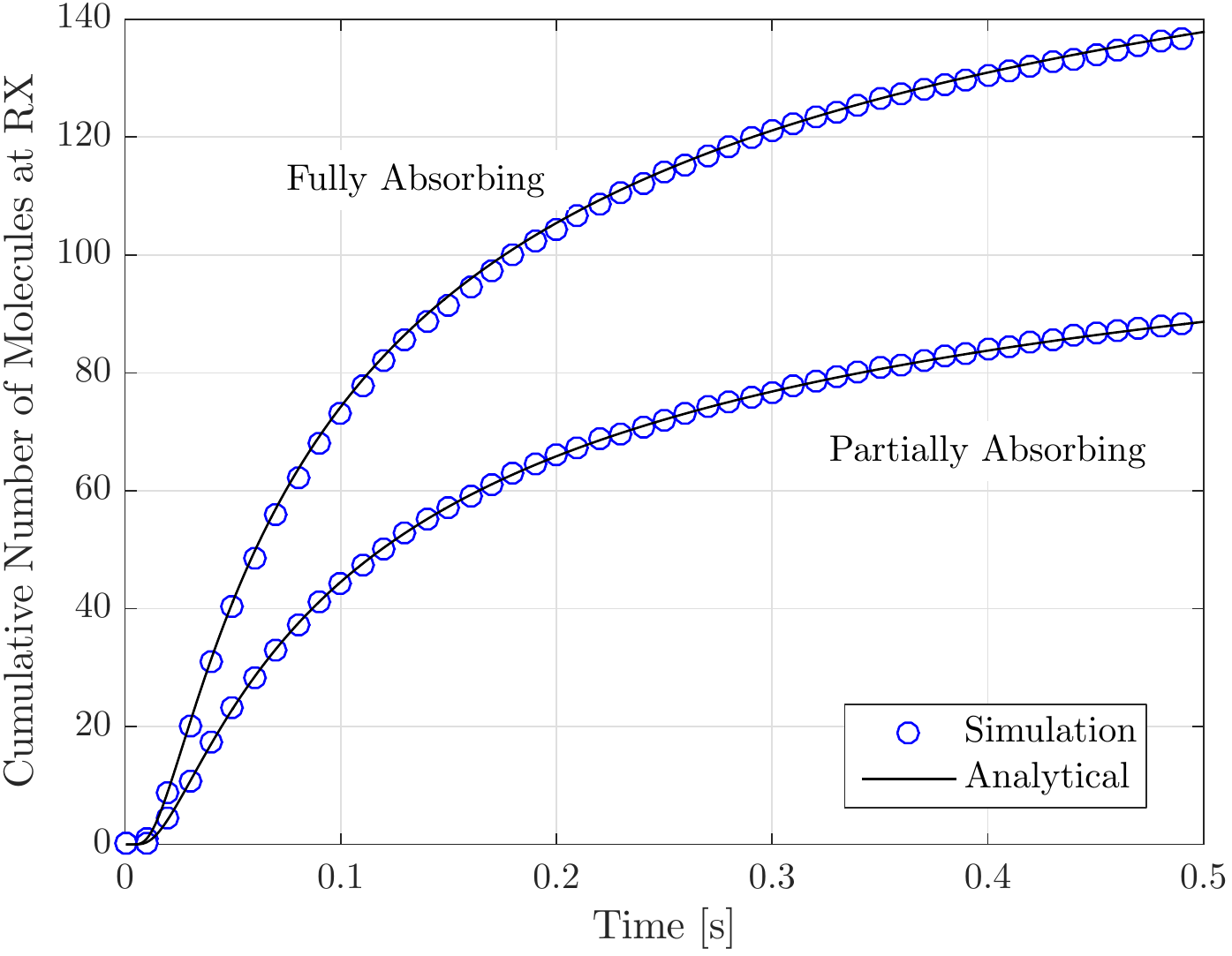}
	\caption{Average time-varying number of molecules absorbed by a spherical surface (of radius $1\,\mu\meter$) after release by a point source located $5\,\mu\meter$ from the center of the sphere. Infinite ($\kth{}=\infty$) and finite ($\kth{}=200\mu\meter/\second$) absorption rates are considered for the fully and partially absorbing cases, respectively.}
	\label{fig_absorbing_rx}
\end{figure}

We plot the average time-varying number of molecules absorbed by the receiver in Fig.~\ref{fig_absorbing_rx}, where we consider full absorption and partial absorption with a rate of $\kth{}=200\,\mu\meter/\second$. A time step of $\dtMicro=1\,\mu\second$ was used to accurately simulate the receiver reactions. Both simulations agree very well with their corresponding analytical curves. Video 8 in \cite{Noel2016c} shows a sample realization of the partially-absorbing receiver.

\section{Limitations and Future Development}
\label{sec_future}

As we have previously noted, the development of AcCoRD is active and ongoing. In this section, we discuss some of AcCoRD's current limitations and how they may be addressed in the future to improve either the accuracy or the applicability of the software. We focus on high level limitations that apply to the implementation of reaction-diffusion phenomena or the design of molecular communication systems. A more detailed description of technical constraints is included in~\ref{app_constraints}.

\subsection{Adding Molecule Flow and Polarity}

We have limited molecule behavior to pure diffusion and selected chemical reactions. We can potentially make minor improvements to the implementations of these phenomena, such as by accounting for environment boundaries in the hybrid transition rules and for surface reactions occurring within a microscopic time step. Furthermore, there are other phenomena that can influence a molecule's behavior. Molecules could be carried by a bulk fluid flow that can vary in both space and time. For example, laminar flow, where layers of fluid slide over one another, is predominant in small blood vessels; see \cite{Nelson2008}. Also, molecular interactions could be introduced by the presence of ionic charges, such that the molecules exert attractive or repulsive forces on each other. These interactions were first considered in a molecular communication context in \cite{Farsad2016b}.

\subsection{Expanding Actor Placement and Behavior}

Actors in AcCoRD must currently be fixed in place and they have independent, predetermined behavior. This is sufficient for unidirectional communication between two fixed devices, but cannot model more complex systems that may require bidirectional communication, relaying, mobility, or other behaviors that rely on real-time signal processing. In both practical biological environments and communication networks, the detection of a signal should trigger a response at the receiver. Under the framework of AcCoRD, we should couple active actors to passive actors so that molecules can be released when specified observation thresholds are met (i.e., detectors are needed to trigger the active actor). We can also consider introducing actor mobility, which can be more accurate for modeling transceivers that are able to move (e.g., bacteria).

Active actors can only release molecules according to the modulation of a concentration shift keying signal, where the number (or rate) of molecules released is linearly proportional to the current information symbol. Other types of modulation can be added, such as those described in \cite{Kuran2011}. Furthermore, adaptive transmission schemes can be considered, where the number of molecules released for a given symbol depends on the interference expected from previous symbols, as we considered in \cite{Ahmadzadeh2015a}.

\subsection{Improving Scalability}

In the simulations of System 2, we observed conditions where the hybrid simulation demonstrated considerable improvement in computation speed over the microscopic model. The potential trade offs in simulation accuracy versus computational complexity can be further improved by adding ``tau-leaping'' to the mesoscopic model; see \cite{Gillespie2001,Iyengar2010,Jedrzejewski-Szmek2016}. In tau-leaping, mesoscopic time steps are used to execute multiple reaction and diffusion events simultaneously. If mesoscopic subvolumes have a sufficient number of molecules, then tau-leaping can improve the simulation run time considerably with minimal loss of accuracy. Finally, a possibility for scalability in the microscopic regime is to enable a different time step in each region.

\section{Conclusions}
\label{sec_concl}

In this paper, we introduced the AcCoRD simulator (Actor-based Communication via Reaction-Diffusion). AcCoRD is a generic 3D reaction-diffusion solver that is designed for MC analysis. It can use a combination of microscopic and mesoscopic simulation models to track time-varying molecule behavior over independent realizations. Simulation environments are defined by the flexible placement of regions. Actors are used to release or observe molecules in the regions. Utilities are provided to visualize the simulation environment and to plot the time-varying behavior or observation statistics. We described all of the simulation components, including the underlying reaction-diffusion theory. We discussed the overall AcCoRD algorithm, including its run time complexity, and we also described the work flow for running and processing a simulation. We demonstrated AcCoRD's consistency with analytical results via a detailed series of simulations.

We hope that the development of AcCoRD will encourage a wider use of simulations within the MC community to support analytical modeling and predict experimental results. MC is a multi-disciplinary field and the availability of a relevant reaction-diffusion ``sandbox'' should improve its accessibility to new researchers while providing a flexible yet convenient platform for exploration and verification.

\section*{Acknowledgments}

This research was funded in part by the Natural Sciences and Engineering Research Council of Canada via a Postdoctoral Fellowship, by the University of British Columbia via a Four-Year-Fellowship, and by the Universit\'{e} de Montr\'{e}al. Computing resources were provided by the Centre for Advanced Computing and Compute/Calcul Canada. The authors would like to thank Modar Halimeh and Tobias Schwering for their help and feedback with software testing.

\appendix

\section{Constraints and Limitations}
\label{app_constraints}

In this appendix, we provide additional detailed notes on AcCoRD's current technical constraints and limitations.

The following constraints exist for the placement of regions:
\begin{itemize}
	\item Two regions can overlap only if one region is entirely inside the other. In that case the inner (i.e., \emph{child}) region must identify the outer region as its \emph{parent} (to let AcCoRD know that the nesting is intended). Multiple levels of nesting can be used, i.e., a child's parent can also have its own parent region.
	\item If a spherical region (which must always be microscopic) is the parent of a box region (or vice versa), then their surfaces must be separated by at least a distance of the box region's subvolume length $\subLength{}$. If the box region is the parent of the spherical region, then the box must also be microscopic. However, a spherical region can have a mesoscopic child region.
	\item If a box region is the parent of a box region, then the placement of the child must be flush with subvolumes of the parent (i.e., every subvolume in the parent region is either fully within the child or not). This applies whether the parent is microscopic or mesoscopic.
	\item Regions can be placed adjacent to each other, including as children of the same parent region.
	\item A square surface cannot have a normal 3D region as its parent, but they can be placed adjacent to each other.
	\item Surface regions must be microscopic and can only be adjacent to other microscopic regions. This could be relaxed in a future release.
	\item Regions that are supposed to be neighbors but which are separated by multiple levels of nesting (e.g., grandparents and grandchildren regions) are not properly detected as neighbors. This occurs when a region shares part of its outer boundary with its parent but the parent does not share part of its outer boundary with the grandparent. This could be corrected in a future release.
\end{itemize}

The following constraints exist for the placement of actors:
\begin{itemize}
	\item Spherical actors must be placed entirely in the microscopic regime.
	\item If two actors are configured to act at the same time, then they will act in a random order (according to whichever actor happens to be at the top of the timer queue). For example, if a user intends to observe molecules immediately after they are placed, then they should add a small offset to the start time of the passive actor. This could be improved in a future release by assigning an actor priority to define which actor should go ``first'' in the case of simultaneous behavior.
\end{itemize}

The following constraints and limitations exist for chemical reactions:
\begin{itemize}
	\item Membrane surface reactions can only be defined at membrane surface regions.
	\item Absorbing and desorbing surface reactions can only be defined at non-membrane surface regions.
	\item There is currently no mechanism to enable a surface-bound molecule to diffuse along its surface, which should be possible if the surface is fluid (e.g., lipid bilayers in cells are 2D fluids whose molecules can move in the plane of the bilayer; see \cite[Ch.~11]{Alberts}).
	\item Surface reactions only occur by testing a candidate molecule's diffusion trajectory for crossing a reactive surface. It is more accurate to also consider the possibility that a molecule reacted with a surface even if its diffusion trajectory does not intersect the surface. We ignored this more general case, as in \cite{Andrews2009}, but this might be added in a future release.
	\item Unbinding radii are only applied to \emph{second order} microscopic reactions that have multiple products. This could be expanded in a future release to apply to all microscopic reactions with multiple products, including the unbinding of enzyme intermediates.
	\item Bimolecular reactions in the microscopic regime are tested at the \emph{region} level, which can result in many unnecessary distance comparisons if the region is much larger than the binding radius. One workaround is to define regions on the order of the size of the binding radius. A future release could instead test bimolecular reactions at the level of microscopic \emph{subvolumes}, although this would introduce overhead to keep track of which subvolume every candidate reactant is in.
\end{itemize}

The following limitations exist for hybrid diffusion between mesoscopic and microscopic regions:
\begin{itemize}
	\item The assumption on subvolume size at a hybrid interface (i.e., assuming that subvolume length $\subLength{}$ is $\gg \sqrt{\Dx{}\dtMicro}$ or $\sim \sqrt{\Dx{}\dtMicro}$) is applied uniformly to the entire environment and does not consider local values of diffusion coefficient or subvolume size. A future release could enable local definition of this assumption based on the specific interface and diffusing molecule.
	\item Mesoscopic molecules added to the microscopic regime via transition rule (\ref{eq_hybrid_tangential}), i.e., under the assumption that the subvolume is small, could be placed beyond the environment boundary. The validity of the molecule's location is not corrected until the following microscopic time step. This can be corrected in a future release.
	\item A mesoscopic subvolume that is adjacent to a microscopic region along some face of the subvolume should be ``exposed'' to that region along the \emph{entire} face. The equation for the diffusion propensity in (\ref{eq_prop_hybrid}) assumes that this is the case.
\end{itemize}

\section{Surface Reaction Probabilities}
\label{app_surf}

In this appendix, we list and describe the surface reaction probabilities. We also discuss the placement of the corresponding product molecules. Most equations in this appendix were derived in \cite{Andrews2009} and are included here because they are implemented directly in AcCoRD. All probabilities are described for arbitrary reaction $\indChemRxn$ with reactant $\indMolType$ and reaction rate\footnote{Strictly speaking, absorption and membrane reactions are defined by a reaction \emph{coefficient}, but we use the term ``rate'' here for consistency.} constant $\kth{\indChemRxn}$. Where applicable, $\kth{\indChemRxnRev}$ is the reaction rate of reaction $\indChemRxn$'s reverse reaction. A direct comparison of the accuracy of the different equations for surface reactions is outside the scope of this work.

Generally, \cite{Andrews2009} derives ``steady state'' reaction probabilities which imply that the interactions between the molecules and the surface are in a state of equilibrium. However, these probabilities were shown to also be accurate over transient time scales. In some cases there is also a ``well-mixed'' reaction probability which assumes that the molecule distribution in the vicinity of the surface is homogeneous. The well-mixed probability, originally derived in \cite{Erban2007}, is in general not as accurate as the steady state expressions but it is simpler to calculate.

\subsection{Absorption}

First, we consider absorption, where a molecule binds to or is consumed by the surface. If absorption occurs, then the product molecule(s) is (are) placed at the surface where the surface intersected the trajectory of the diffusing reactant. If the reaction is unsuccessful, then the molecule has a perfectly-elastic collision with the surface and is reflected off of the surface at the intersection point.

The well-mixed absorption probability is \cite[Eq.~(1)]{Andrews2009}
\begin{equation}
\label{eq_adsorb_irr_mixed}
\prob{abs}|_\mathrm{mixed} = \kth{\indChemRxn}\sqrt{\frac{\pi\dtInf}{\Dx{\indMolType}}},
\end{equation}
where $\dtInf$ is the corresponding time step (which can be less than or equal to the simulation's microscopic time step $\dtMicro$, depending on when the molecule was created), and $\Dx{\indMolType}$ is the molecule's diffusion coefficient. Eq.~(\ref{eq_adsorb_irr_mixed}) is accurate if
\begin{equation}
\label{eq_adsorb_irr_mixed_acc}
\kth{\indChemRxn} \ll \sqrt{\frac{\Dx{\indMolType}}{\pi\dtInf}},
\end{equation}
i.e., if $\prob{abs}$ is sufficiently small.

The steady state reaction probability depends on whether the absorption is reversible. In the irreversible case, a polynomial fit was used to find the reaction probability as \cite[Eq.~(21)]{Andrews2009}
\begin{equation}
\label{eq_adsorb_irr_ss}
\prob{abs,irr}|_\mathrm{ss} = \kth{\indChemRxn}'\sqrt{2\pi} - 3.33321\kth{\indChemRxn}'^2 + 3.35669\kth{\indChemRxn}'^3 - 1.52092\kth{\indChemRxn}'^4,
\end{equation}
where \cite[Eq.~(12)]{Andrews2009}
\begin{equation}
\label{eq_adsorb_rate_norm}
\kth{\indChemRxn}' = \kth{\indChemRxn}\sqrt{\frac{\dtInf}{2\Dx{\indMolType}}},
\end{equation}
is the reduced absorption coefficient.

The steady state reaction probability for reversible absorption (i.e., \emph{adsorption}) was derived in closed form as \cite[Eq.~(37)]{Andrews2009}
\begin{equation}
\label{eq_adsorb_rev_ss}
\prob{abs,rev}|_\mathrm{ss} = \frac{\kth{\indChemRxn}'\sqrt{2\pi}
	\left(\varX{2}-\varX{1}\! - \varX{2}\ERFCX{\varX{1}} + \varX{1}\ERFCX{\varX{2}}\right)}
{\varX{1}\varX{2}(\varX{2}-\varX{1})},
\end{equation}
where
\begin{align}
\label{eq_adsorb_rev_ss_terms}
\varX{1} = & \frac{\kth{\indChemRxn}' - \sqrt{\kth{\indChemRxn}'^2 -2\kth{\indChemRxnRev}\dtInf}}{\sqrt{2}}, \\
\varX{2} = & \frac{\kth{\indChemRxn}' + \sqrt{\kth{\indChemRxn}'^2 -2\kth{\indChemRxnRev}\dtInf}}{\sqrt{2}},
\end{align}
$\kth{\indChemRxnRev}$ is the rate of the reverse desorption reaction, $\ERFCX{\cdot}$ is the scaled complementary error function
\begin{equation}
\label{eq_erfcx}
\ERFCX{\varX{}} = \EXP{\varX{}^2}\left(1-\ERF{\varX{}}\right),
\end{equation}
and the error function $\ERF{\cdot}$ is defined in (\ref{eq_erf}).

\subsection{Desorption}
\label{app_surf_desorption}

Desorption releases a molecule that was bound to the surface. In the irreversible case, the probability of reaction is the same as that for a non-surface first order reaction, i.e., as in (\ref{eq_prob_first}). However, we must still determine where to place the product. We place the product along the surface normal from the point where the reactant was bound. A steady state distance $\x$ along the normal can be generated via \cite[Eq.~(27)]{Andrews2009}
\begin{equation}
\label{eq_desorb_dist_irr_ss}
\x = \frac{0.571825\uniRV{} - 0.552246\uniRV{}^2}{1 - 1.53908\uniRV{} + 0.546424\uniRV{}^2}\sqrt{2\Dx{\indMolType}\dtInf},
\end{equation}
where $\Dx{\indMolType}$ is the diffusion coefficient of the molecule once it is unbound from the surface and $\uniRV{}$ is a single uniform random variable between 0 and 1. Eq.~(\ref{eq_desorb_dist_irr_ss}) was determined numerically with the assumption that the precise reaction time was unknown. In AcCoRD, we can determine the time of irreversible desorption using the method described in Section~\ref{sec_theory_rxn}. Given desorption time $\timeX{1}$, the molecule will have $\dtInf-\timeX{1}$ seconds to diffuse from the surface, and the distance $\x$ along the normal can be generated via
\begin{equation}
\label{eq_desorb_dist_irr_forced}
\x = \sqrt{8\Dx{\indMolType}(\dtInf-\timeX{1})}\normRV{},
\end{equation}
where $\normRV{}$ is a normal random variable with mean 0 and variance 1. Eq.~(\ref{eq_desorb_dist_irr_forced}) is twice the unbounded diffusion distance for this time interval because here we have a reflective surface (since the desorption is irreversible), and in AcCoRD we use (\ref{eq_desorb_dist_irr_forced}) by specifying the ``full diffusion'' option for placing desorbed molecules.

The reaction probability for reversible desorption can be written as a function of the corresponding absorption (i.e., adsorption) reaction in (\ref{eq_adsorb_rev_ss}), i.e., \cite[Eq.~(32)]{Andrews2009}
\begin{equation}
\label{eq_desorb_rev_ss}
\prob{des,rev}|_\mathrm{ss} = \frac{\kth{\indChemRxn}}{\kth{\indChemRxnRev}}
\sqrt{\frac{\Dx{\indMolType}\dtInf}{\pi}}\prob{abs,rev}|_\mathrm{ss},
\end{equation}
where here $\kth{\indChemRxn}$ is the desorption rate and $\kth{\indChemRxnRev}$ is the rate of the reverse adsorbing reaction.

Since the exact desorption times are unknown in the reversible case, we should only consider molecule placement via the corresponding steady state equation, which was determined numerically as \cite[Eq.~(35)]{Andrews2009}
\begin{equation}
\label{eq_desorb_dist_rev_ss}
\x = \frac{0.729614\uniRV{} - 0.70252\uniRV{}^2}{1 - 1.47494\uniRV{} + 0.484371\uniRV{}^2}\sqrt{2\Dx{\indMolType}\dtInf}.
\end{equation}

\subsection{Membrane Transitions}

Membrane reactions enable selective transitions across a surface. When these reactions occur successfully, the diffusing molecule continues along its original trajectory. Otherwise, it is reflected off of the surface at the point where the trajectory intersected the surface.

In the irreversible case, a molecule can only transition in one direction. Thus, the transition probabilities in this case are the same as for an absorption reaction with the same reaction rate, i.e., the well-mixed probability is given by (\ref{eq_adsorb_irr_mixed}) and the steady state probability is given by (\ref{eq_adsorb_irr_ss}). 

The forward and reverse reactions in reversible membrane transitions are analogous to each other. Thus, unlike the irreversible case, reversible membrane reactions are not analogous to reversible adsorption. The transition probability for reversible membrane transitions was derived as \cite[Eq.~(47)]{Andrews2009}
\ifOneCol
	\begin{equation}
	\label{eq_mem_rev_ss}
	\prob{mem,rev}|_\mathrm{ss} = 
	\frac{\kth{\indChemRxn}}{\left(\kth{\indChemRxn}+\kth{\indChemRxnRev}\right)^2}
	\bigg[2\left(\kth{\indChemRxn}+\kth{\indChemRxnRev}\right) + \sqrt{\frac{\pi}{2}}\left(\ERFCX{\sqrt{2}\left(\kth{\indChemRxn}+\kth{\indChemRxnRev}\right)}-1\right)\bigg],
	\end{equation}
\else
	\begin{align}
	\label{eq_mem_rev_ss}
	\prob{mem,rev}|_\mathrm{ss} = &
	\frac{\kth{\indChemRxn}}{\left(\kth{\indChemRxn}+\kth{\indChemRxnRev}\right)^2}
	\bigg[2\left(\kth{\indChemRxn}+\kth{\indChemRxnRev}\right) \nonumber \\
	&+\sqrt{\frac{\pi}{2}}\left(\ERFCX{\sqrt{2}\left(\kth{\indChemRxn}+\kth{\indChemRxnRev}\right)}-1\right)\bigg],
	\end{align}
\fi
where $\kth{\indChemRxn}$ is the reaction rate for the transition in the direction being considered.

\section{Detailed AcCoRD Algorithms}
\label{app_algorithms}

In this appendix, we describe each of the four system evolution steps of the overall AcCoRD algorithm that was presented in Section~\ref{sec_top_algorithm}. Specifically, we describe active actor behavior, passive actor behavior, evolution of the microscopic regime, and evolution of the mesoscopic regime. We also include comments as appropriate on the computational speed of the microscopic and mesoscopic algorithms using big O notation ($\O{\cdot}$).

\subsection{Active Actors}
\label{sec_active}

The algorithm for executing an active actor simulation step is presented in Algorithm~\ref{alg_active}. Each step takes one of two actions: either a new symbol is \emph{created} for some actor $\indActor$, or molecules from an existing symbol are released by actor $\indActor$. We separate these two actions in order to accommodate the release of molecules over a \emph{release interval}. The release interval can be instantaneous or any finite interval independent of the actor's \emph{action interval}, i.e., symbol interval. By defining the release interval to be longer than the action interval, it is technically possible for an actor to be releasing molecules for multiple symbols simultaneously (although this is most likely impractical for a molecular communication system).

\begin{algorithm}[!tb]
	\caption{Active Actor Step}
	\label{alg_active}
	\begin{algorithmic}[1]
		\Procedure{Active\_Action}{$\indActor$}
		\If{Next actor $\indActor$ action is new symbol}
		\If{Bits are random}
		\State Generate random symbol bits
		\Else
		\State Read defined symbol bits
		\EndIf
		\State Determine modulation parameters
		\State Update list of current symbols
		\State Update time of next new symbol
		\Else  \Comment{Releasing molecule(s) for a current symbol}
		\If{Release times random}
		\State $\numNewMol \gets 1$ \Comment{Only one new molecule}
		\State Generate next release time
		\Else
		\State Calculate $\numNewMol$
		\State Calculate next release time
		\EndIf
		\For{All $\numNewMol$}
		\State Place new molecule within actor $\indActor$
		\EndFor
		\State Update time of symbol's next molecule release
		\EndIf
		\State Determine actor $\indActor$'s next action
		\State Update actor $\indActor$'s timer
		\EndProcedure
	\end{algorithmic}	
\end{algorithm}

When a new symbol is needed, we generate the symbol bits randomly (with some independent probability) or read them from a sequence provided by the configuration file. The modulation properties and the new symbol bits are used to determine how many molecules to release and how often. For example, concentration shift keying (CSK) releases molecules at a uniform rate over the release interval, where the rate is linearly proportional to the symbol value.

When the current actor action is the actual release of molecules, we use the modulation properties to determine how many molecules to release at the current time. If the release times within the interval are random, then only one molecule will be released and we must randomly generate the next molecule release time as a zeroth order reaction process (see Section~\ref{sec_theory_rxn} for the simulation of zeroth order reactions). Otherwise, the number of molecules will depend on the modulation strength, and we calculate the next molecule release time using a \emph{slot interval}, which is a subset of the release interval. In either case, if the next molecule release time is beyond the current release interval, then it is ignored and transmission for the corresponding symbol is complete.

Every molecule released by actor $\indActor$ is initialized at a uniformly-generated random location within the actor volume. If the location is within the microscopic regime, then the molecule is added to the corresponding region's linked list for that molecule type. If the location is within the mesoscopic regime, then the molecule count in the corresponding subvolume is incremented and the subvolume's reaction propensities are updated as described in~\ref{sec_alg_meso}.

At the end of the active actor simulation step, we determine actor $\indActor$'s next action by comparing the time of the next new symbol with that of the next molecule release. The corresponding time is used to update the actor's timer.

\subsection{Passive Actors}
\label{sec_passive}

The algorithm for executing a passive actor simulation step is presented in Algorithm~\ref{alg_passive} for passive actor $\indActor$. We test the location of all candidate molecules against the boundary of the actor. This test is trivial for molecules whose region is known to be entirely inside the actor. Any molecule that is within the actor is recorded. In general, we only count the \emph{number} of each type of molecule that the actor is configured to observe. Optionally, the observation can also include the molecule location coordinates. These coordinates are necessary if making a video from the simulation output.

\begin{algorithm}[!tb]
	\caption{Passive Actor Step}
	\label{alg_passive}
	\begin{algorithmic}[1]
		\Procedure{Passive\_Action}{$\indActor$}
		\For{Each type of molecule being observed}
		\For{Every region within actor $\indActor$}
		\If{Region microscopic}
		\For{Every molecule in ``recent'' list} \label{step_read_recent}
		\If{Molecule within actor $\indActor$}
		\State Record molecule
		\EndIf
		\EndFor \label{step_read_recent_end}
		\State Repeat lines~\ref{step_read_recent} to \ref{step_read_recent_end} for ``normal'' list
		\Else \Comment{Region mesoscopic}
		\For{Every subvolume within actor $\indActor$}
		\If{Entire subvolume inside actor}
		\State Record all molecules
		\Else \Comment{Partial overlap}
		\For{Every molecule}
		\State Flip biased coin for location
		\If{Molecule within actor $\indActor$}
		\State Record molecule
		\EndIf									
		\EndFor								
		\EndIf							
		\EndFor \Comment{Subvolumes in region}
		\EndIf \Comment{Microscopic or mesoscopic?}
		\EndFor \Comment{Regions in actor $\indActor$}
		\EndFor \Comment{Molecule types}
		\State Append observations to realization list
		\State Update actor $\indActor$'s timer
		\EndProcedure
	\end{algorithmic}	
\end{algorithm}

In microscopic regions, the precise coordinates of every molecule are already known and so the test for being inside actor $\indActor$ is a purely geometric problem. It is also straightforward to record molecule locations if needed. In mesoscopic regions, we consider the individual subvolumes. If a subvolume is known to be inside the actor, then all corresponding molecules are recorded. Otherwise, we flip a biased coin for each molecule based on the relative fraction of the subvolume that is within actor $\indActor$. If molecule locations are needed, then they are randomly generated within the subvolume (or a fraction thereof when the entire subvolume is not within the actor).

Once all molecules have been recorded, we append this information to the master list of observations for the current realization. Finally, we update actor $\indActor$'s timer by incrementing it by the actor's \emph{action interval}.

\subsection{Microscopic Steps}
\label{sec_alg_micro}

The algorithm for executing a microscopic simulation step is presented in Algorithm~\ref{alg_micro}. The primary stages in this algorithm are (in order) the execution of zeroth order reactions, first order reactions, diffusion, and second order reactions. For clarity of presentation in Algorithm~\ref{alg_micro}, we omit showing that each stage is completed for all regions before proceeding to the next stage.

\begin{algorithm}[!tb]
	\caption{Microscopic Regime Step}
	\label{alg_micro}
	\begin{algorithmic}[1]
		\Procedure{Microscopic\_Step}{}
		\For{Every zeroth order reaction type}
		\While{Next reaction time within time step}
		\State Execute reaction
		\State Generate next reaction time
		\EndWhile
		\EndFor \Comment{Zeroth order reactions}
		
		\For{Every molecule}
		\While{First order reaction is possible}
		\State Flip coin for first order reaction
		\If{Reaction occurs}
		\State Generate reaction time
		\State Execute reaction
		\EndIf
		\EndWhile
		\EndFor \Comment{First order reactions}
		
		\For{Every molecule}
		\State Generate new location via diffusion
		\State Validate new location
		\If{Molecule entered mesoscopic regime}
		\State Place molecule in subvolume
		\ElsIf{Molecule reacted with surface}
		\State Execute reaction
		\Else
		\State Update molecule in ``normal'' list
		\EndIf
		\EndFor \Comment{Diffusion}
		
		\For{Every second order reaction}
		\For{Every valid first reactant}
		\For{Every valid second reactant}
		\If{Reactants within binding radius}
		\If{Reactants reach reaction site}
		\State Execute reaction
		\State Separate with unbinding radius
		\EndIf
		\EndIf
		\EndFor \Comment{Second reactants}
		\EndFor \Comment{First reactants}
		\EndFor \Comment{Second order reactions}
		
		\State Update subvolume propensities as needed
		
		\State Increment microscopic timer by $\dtMicro$
		\EndProcedure
	\end{algorithmic}	
\end{algorithm}

\subsubsection{Zeroth Order Reactions}

Zeroth order reactions in a microscopic region each have their own internal timer to track the time of the most recent reaction event. Within a microscopic time step, we generate the delay between reaction times as exponential random variables according to (\ref{eq_t_zeroth}) in Section~\ref{sec_theory_rxn}. We continue until the reaction time exceeds the time remaining in the step and is saved for the next time step. Random reaction locations are generated uniformly within the region volume. Every product molecule is added to the region's corresponding ``recent'' molecule list with the reaction location and the molecule's time step.

The generation of zeroth order reaction times is effectively analogous to how zeroth order reaction events can occur in the mesoscopic regime. Thus, this process for zeroth order reaction $\indChemRxn$ in region $\indRegion$ takes $\O{\kth{\indChemRxn}\Vol{\indRegion}}$ time, where $\Vol{\indRegion}$ is the region's volume.

\subsubsection{First Order Reactions}

The stage for \emph{non-surface} first order reactions is as follows. For each molecule that can be the reactant in a first order reaction, we generate a uniform random variable $\uniRV{}$ between 0 and 1 and compare it with the probability of the reaction occurring in (\ref{eq_prob_first}) (or with the probabilities generated by (\ref{eq_prob_first_mult}) if the molecule is a potential reactant in more than one non-surface first order reaction). If a reaction is successful, then we generate the precise reaction time as a constrained exponential random variable via (\ref{eq_t_first}). Product molecules are initialized at the same location as the reactant. This process is iterative, so a product molecule that can also be the reactant in a first order reaction is tested for reacting in the time still remaining within the time step. As with zeroth order reactions, the generation of reaction times is effectively analogous to how it is done in the mesoscopic regime, so the number of instances of first order reaction $\indChemRxn$ grows with $\O{\kth{\indChemRxn}\Nx{\indChemRxn, \indRegion}}$, where $\Nx{\indChemRxn,\indRegion}$ is the number of corresponding reactant molecules in the region.

\subsubsection{Diffusion}

The diffusion stage generates a potential diffusion trajectory for every molecule with a nonzero diffusion coefficient. The time to generate and test the trajectories grows with $\O{\numMolMicroRegion}$, where $\numMolMicroRegion$ is the number of molecules in microscopic region $\indRegion$. We follow each proposed trajectory from the initial location to the proposed new location in a process that we approximate as taking $\O{|\setRegionNeigh|}$ time, where $|\setRegionNeigh|$ is the number of regions that are neighbors of region $\indRegion$. In practice, $\O{|\setRegionNeigh|}$ time can be considerable, since we need to check every neighbor for a trajectory intersection and implement the corresponding behavior at the closest intersecting neighbor. This process must then be repeated for every boundary reflection or transition to a normal (i.e., non-surface) neighbor.

Molecules are permitted to transition between adjacent normal regions without impedance. If the trajectory crosses a region boundary where there is no adjacent region, then we reflect the trajectory from the point of intersection with the boundary. If the trajectory crosses a surface region, then we test the corresponding surface reaction (either absorption or a membrane transition, if applicable) with its probability calculated in \ref{app_surf} and we reflect the trajectory if the surface reaction fails. If absorption occurs, then we execute the corresponding reaction with a reaction time that is at the \emph{end} of the time step. If a membrane transition occurs, then the molecule can pass through the surface and we continue to test the proposed trajectory.

If the diffusion trajectory crosses a mesoscopic region, then we place the molecule in the first mesoscopic subvolume crossed by the trajectory. Even if a molecule's trajectory does not cross a mesoscopic region, it can be placed in a mesoscopic region if the molecule's final location is in a region that neighbors a mesoscopic region, which we test by comparing a uniform random variable $\uniRV{}$ with the probability in (\ref{eq_micro_to_meso}). This probability is a function of the initial and final distances from the hybrid interface. Performing this test takes considerable time, since we need to calculate the corresponding distances and generate a random variable, so we ignore this test if either the initial or final distance is greater than the defined maximum distance $\hybridDist{\mathrm{max}}$. Every molecule that is still in the microscopic regime at the end of the diffusion stage is in the corresponding region's ``normal'' list.

\subsubsection{Second Order Reactions}

In the stage for second order reactions, we compare the distances between pairs of candidate reactants for every possible reaction. This takes up to $\O{\numMolMicroRegion\left(\numMolMicroRegion+\sum_{\setRegionNeigh}\numMolMicroRegionX{\indRegionNeigh}\right)}$ time for each reaction in region $\indRegion$, where the summation is over the set of regions that are neighbors to region $\indRegion$ and $\numMolMicroRegionX{\indRegionNeigh}$ is the number of microscopic molecules in neighboring region $\indRegionNeigh$. To be a candidate pair, the two molecules must satisfy all of the following criteria:
\begin{itemize}
	\item Be the specific reactants for the current second order reaction.
	\item Be located in the same or neighboring microscopic regions.
	\item Have not yet participated in a second order reaction within the current time step (either as a reactant or product).
\end{itemize}

If a candidate pair of reactants are separated by less than the reaction binding radius $\rBind$, then the candidate reaction site is determined from the reactants' diffusion coefficients $\Dx{\reactant{}}$ and $\Dx{\product{}}$. If the current locations (i.e., after diffusing) of reactants $1$ and $2$ are $\rad{1} = \{\x_{1}, \y_{1}, \z_{1}\}$ and $\rad{2} = \{\x_{2}, \y_{2}, \z_{2}\}$, respectively, then the candidate reaction site $\rad{\rxn}$ is
\begin{equation}
\label{eq_second_order_rxn_location}
\rad{\rxn} = \frac{\Dx{1}}{\Dx{1} + \Dx{2}}\left(\rad{2} - \rad{1}\right) + \rad{1}.
\end{equation}

To validate $\rad{\rxn}$, we follow the trajectory of each reactant from their current location. Both reactants must be able to reach $\rad{\rxn}$. We deem $\rad{\rxn}$ to be invalid if \emph{any} of the following are true:
\begin{itemize}
	\item Either reactant reflects off of a surface.
	\item Either reactant crosses a mesoscopic region.
	\item $\rad{\rxn}$ is in a region where the current second order reaction was not defined.
\end{itemize}

We assume that all potential surface reactions at this stage would lead to reflection (which invalidates the candidate second order reaction site) because we force the time step for reactions to $\dtInf = 0$. This is an approximation since we assume that the translation of reactants to the reaction site occurs spontaneously at the end of the current time step (since surface reactions were already assessed for each reactant in the diffusion stage). If $\rad{\rxn}$ is valid, then we remove the reactants and add product molecules that are centered around $\rad{\rxn}$ according to the unbinding radius $\rUnbind$. Otherwise, the reaction does not occur and the reactants remain at their current locations. If there is one product, then it is placed at $\rad{\rxn}$. If there are at least two products, then the displacement of each product is $\rUnbind$ scaled by the relative diffusion coefficient (similar to how we determined $\rad{\rxn}$ in (\ref{eq_second_order_rxn_location})). If there are two products, then they are placed in the same directions as the locations of the two reactants. If there are more than two products, then the directions are determined randomly. We track the trajectory of each product molecule from the reaction site in case the molecule reflects off of a surface or enters a mesoscopic region, and correct the product molecule's location as needed.

At the end of the microscopic simulation step, we update the propensities of mesoscopic subvolumes that had molecules added via diffusion (as described in \ref{sec_alg_meso}), and then we increment the microscopic timer by the microscopic time step $\dtMicro$.

\subsection{Mesoscopic Steps}
\label{sec_alg_meso}

The algorithm for executing a mesoscopic simulation step is presented in Algorithm~\ref{alg_meso}. Each potential reaction event in every mesoscopic subvolume, including chemical reactions and diffusion between adjacent subvolumes, is assigned a propensity, $\prop{}$. We described the calculation of event propensities in Section~\ref{sec_theory} as functions of the type of event, the subvolume size, and the number of each type of molecule in the subvolume.

\begin{algorithm}[!tb]
	\caption{Mesoscopic Regime Step}
	\label{alg_meso}
	\begin{algorithmic}[1]
		\Procedure{Mesoscopic\_Step}{}
		\State Determine subvolume of next reaction (via NRM)
		\State Determine next reaction in subvolume (via DM)
		\If{Next event is diffusion}
		\State Determine molecule type and destination
		\State Decrement molecule count in source subvolume
		\If{Destination subvolume is microscopic}
		\State Create microscopic molecule
		\Else \Comment{Destination is also mesoscopic}
		\State Increment molecule count in destination
		\State Update destination propensity (via NRM)
		\EndIf
		\Else \Comment{Next event is chemical reaction}
		\State Adjust molecule counts
		\EndIf
		\State Update source propensity (via NRM)
		\State Update mesoscopic timer
		\EndProcedure
	\end{algorithmic}	
\end{algorithm}

The primary feature in a mesoscopic algorithm is how the reaction propensities are used to determine the order of reaction events, when the events occur, and how the order is updated. There are several mathematically equivalent methods for doing so, and they can be integrated as desired for a given implementation. In AcCoRD we adopt the Next Subvolume Method (NSM), which was originally proposed in \cite{Elf2004} and uses both Gillespie's original Direct Method (DM; see \cite{Gillespie1976}) and Gibson and Bruck's Next Reaction Method (NRM; see \cite{Gibson2000}). Specifically, we use the NRM to determine the subvolume where the next event will occur and the DM to determine the next event in a given subvolume. Even though the NRM is efficient when there is a very large number of events, the NSM was shown in \cite{Elf2004} to be faster than applying the NRM to all events.

The primary difference between the original NSM and the implementation in AcCoRD is that AcCoRD accounts for the hybrid interface between microscopic and mesoscopic regimes. The event propensity for a subvolume molecule to enter a microscopic region, and the algorithm for determining a new microscopic molecule's location, were described in Section~\ref{sec_diff_hybrid}.

\subsubsection{Next Reaction Method in NSM}

The NSM uses the NRM to determine the subvolume where the next reaction occurs. We sort all subvolumes in an indexed priority queue. Subvolume $\indSub$ has an associated subvolume propensity $\prop{\indSub}$, which is the sum of all propensities of all possible reaction events in that subvolume. Given the subvolume's initial propensity $\prop{\indSub,i}$, the time $\timeX{i}$ until the first reaction event can be generated as an exponential random variable via
\begin{equation}
\label{eq_t_dm}
\timeX{i} = -\frac{\log\uniRV{}}{\prop{\indSub,i}},
\end{equation}
where $\uniRV{}$ is a uniform random number between 0 and 1. We sort the subvolumes in the indexed priority queue according to their reaction times, such that the subvolume with the lowest time is at the front of the queue (and can be immediately found, i.e., in $\O{1}$ time). When a reaction occurs in subvolume $\indSub$, then the propensity $\prop{\indSub,i}$ is updated (if the number of molecules of any type changed) and the time until the next subvolume event is obtained via (\ref{eq_t_dm}). Sometimes, the propensity can change \emph{without} a reaction occurring in the subvolume, i.e., if molecules are added to the subvolume by an active actor or via diffusion. If this occurs, then the reaction time can be updated \emph{without} generating a new random number. Given the initial propensity $\prop{\indSub,i}$, the updated propensity $\prop{\indSub,f}$, the original reaction time $\timeX{i}$, and the current simulation time $\timeX{c}$, then the updated subvolume event time $\timeX{f}$ is \cite[p.~1881]{Gibson2000}
\begin{equation}
\label{eq_t_nrm}
\timeX{f} = \timeX{c} + \frac{\prop{\indSub,i}}{\prop{\indSub,f}}\left(\timeX{i} - \timeX{c}\right),
\end{equation}
and this update is much faster than using (\ref{eq_t_dm}). When the subvolume reaction time is updated (either via (\ref{eq_t_dm}) or (\ref{eq_t_nrm})), we also update the subvolume's place in the indexed priority queue, which takes $\O{\log\numSub}$ time, where $\numSub$ is the total number of subvolumes. However, as noted in \cite{Gibson2000}, updating the queue often takes much less than $\O{\log\numSub}$ time if a small number of subvolumes are much more likely to have an event than the others. At the end of the mesoscopic simulation step, we read the time associated with the subvolume at the front of the queue to update the mesoscopic timer.

\subsubsection{Direct Method in NSM}

The NSM uses the DM to determine which subvolume reaction takes place. This process is effectively a biased die roll where the probability of each event is proportional to the propensity of that event. If reaction $\indChemRxn$ has propensity $\prop{\indChemRxn}$, then it has probability $\prop{\indChemRxn}/\prop{\indSub}$ of being the next reaction in subvolume $\indSub$. We perform the die roll by generating a uniform random number $\uniRV{}$ and comparing it with the cumulative sum of $\prop{\indChemRxn}/\prop{\indSub}$ terms for subvolume $\indSub$; see \cite[Eq.~(27b)]{Gillespie1976}.

Using the DM takes $\O{\numEvent}$ time, where $\numEvent$ is the total number of possible events in that subvolume, i.e., all chemical reactions and diffusion of each type of molecule to every neighboring subvolume. Every distinguishable reaction is a distinct event. For example, consider a subvolume with two types of molecules. If the molecules can participate in one of three chemical reactions (either as a reactant or product), and both types of molecule can diffuse into any of the subvolume's six neighbors, then there are a total of $3 + 2 \times 6 = 15$ possible events.

\subsubsection{Executing Events}

Once the NSM has determined the current event, it is executed by adjusting the corresponding molecule counts. This takes $\O{1}$ time for a diffusion event and $\O{|\setMolTypes|}$ time for a chemical reaction event, where $|\setMolTypes|$ is the number of molecule types. It also takes $\O{\numEvent}$ time to update the event propensities within the subvolume (or subvolumes in the case of diffusion). As mentioned previously, the individual event propensities are added to get the total subvolume propensity $\prop{\indSub}$ and then we can recalculate the subvolume's next event time.

Given the time to execute the NSM and the time to execute the current event, one mesoscopic step takes $\O{\log\numSub + \numEvent + |\setMolTypes|} = \O{\log\numSub + \numEvent}$ time, since the number of events is a linear function of the number of types of molecules.

\section{Review of Statistics}
\label{app_prob}

In this appendix, we review expressions for the probability distributions and mutual information used in AcCoRD's post-processing utilities.

\subsection{Binomial Distribution}

If we assume that all molecules behave independently and have the same probability of being observed at some time by a passive actor, then it is natural to describe the number of molecules observed by the actor at some time as a Binomial random variable; see \cite[Ch.~5]{Ross2009}. Suppose that $\varOne$ is a Binomial random variable and corresponds to the number of molecules observed. If $\numTrials$ is the total number of molecules, and $\trialProb$ is probability of a given molecule being observed, then the probability mass function (PMF) of $\varOne$ is \cite[Eq.~(5.1.2)]{Ross2009}
\begin{equation}
\label{eq_pmf_binomial}
\Pr\{\varOne = \numSuccess\} = \binom{\numTrials}{\numSuccess}\trialProb^\numSuccess
\left(1-\trialProb\right)^{\numTrials-\numSuccess},
\end{equation}
where $\numSuccess \in \{0,1,\ldots,\numTrials\}$. The PMF can be more convenient to calculate via  \cite[Eq.~(26.5.25)]{Abramowitz1964}
\begin{equation}
\label{eq_pmf_binomial_beta}
\Pr\{\varOne = \numSuccess\} = I_{\trialProb}\left(\numSuccessThresh,\numTrials-\numSuccessThresh+1\right) - I_{\trialProb}\left(\numSuccessThresh+1,\numTrials-\numSuccessThresh\right),
\end{equation}
where \cite[Eq.~(6.6.2)]{Abramowitz1964}
\begin{equation}
\label{eq_beta_incomplete_reg}
I_{x}\left(a,b\right) = \frac{B_{x}\left(a,b\right)}{B\left(a,b\right)}
\end{equation}
is the regularized incomplete Beta function, \cite[Eq.~(6.6.1)]{Abramowitz1964}
\begin{equation}
\label{eq_beta_incomplete}
B_{x}\left(a,b\right) = \int_{0}^{x}t^{a-1}\left(1-t\right)^{b-1}\mathrm{d}t
\end{equation}
is the incomplete Beta function, and \cite[Eq.~(6.2.1)]{Abramowitz1964}
\begin{equation}
\label{eq_beta}
B\left(a,b\right) = B_{x}\left(a,b\right)|_{x=1},
\end{equation}
is the Beta function. The cumulative distribution function (CDF) of $\varOne$ is \cite[Eq.~(5.1.4)]{Ross2009}
\begin{equation}
\label{eq_cdf_binomial}
\Pr\{\varOne \le \numSuccessThresh\} = \sum_{\numSuccess=0}^{\numSuccessThresh}\binom{\numTrials}{\numSuccess}\trialProb^\numSuccess
\left(1-\trialProb\right)^{\numTrials-\numSuccess},
\end{equation}
which can be more convenient to calculate via \cite[Eq.~(26.5.24)]{Abramowitz1964}
\begin{equation}
\label{eq_cdf_binomial_beta}
\Pr\{\varOne \le \numSuccessThresh\} = I_{1-\trialProb}\left(\numTrials-\numSuccessThresh,\numSuccessThresh+1\right).
\end{equation}

\subsection{Poisson Approximation}

One approximation of the Binomial distribution uses the Poisson distribution with mean $\numTrials\trialProb$, which is valid when $\numTrials$ is sufficiently large and $\trialProb$ is sufficiently small; see \cite[Ch.~5]{Ross2009}. If $\varOne$ is a Poisson random variable with mean $\numTrials\trialProb$, then its PMF is \cite[Eq.~(5.2.1)]{Ross2009}
\begin{equation}
\label{eq_pmf_poisson}
\Pr\{\varOne = \numSuccess\} = \frac{\left(\numTrials\trialProb\right)^\numSuccess\EXP{-\numTrials\trialProb}}{\numSuccess!},
\end{equation}
and its CDF is
\begin{equation}
\label{eq_cdf_poisson}
\Pr\{\varOne \le \numSuccessThresh\} = \sum_{\numSuccess=0}^{\numSuccessThresh}\frac{\left(\numTrials\trialProb\right)^\numSuccess\EXP{-\numTrials\trialProb}}{\numSuccess!}.
\end{equation}

Using \cite[Eqs.~(26.4.19), (26.4.21)]{Abramowitz1964}, it can be shown that the CDF can also be calculated as
\begin{equation}
\label{eq_cdf_poisson_gamma}
\Pr\{\varOne \le \numSuccessThresh\} = \frac{\Gamma(\lfloor\numSuccessThresh+1\rfloor,\numTrials\trialProb)}{\Gamma(\lfloor\numSuccessThresh+1\rfloor)},
\end{equation}
where \cite[Eq.~(6.5.3)]{Abramowitz1964}
\begin{equation}
\label{eq_gamma_incomplete}
\Gamma\left(a,x\right) = \int_{x}^{\infty}t^{a-1}\EXP{-t}\mathrm{d}t
\end{equation}
is the incomplete Gamma function and \cite[Eq.~(6.1.1)]{Abramowitz1964}
\begin{equation}
\label{eq_gamma}
\Gamma\left(a\right) = \Gamma\left(a,x\right)|_{x=0}
\end{equation}
is the Gamma function.

\subsection{Gaussian Approximation}

By the central limit theorem, a Binomial random variable can also be approximated as a Gaussian random variable with mean $\numTrials\trialProb$ and variance $\numTrials\trialProb(1-\trialProb)$; see \cite[Ch.~6]{Ross2009}. However, since a Gaussian distribution has real support, we should include a continuity correction to approximate the discrete Binomial distribution. If Binomial random variable $\varOne$ is approximated by Gaussian random variable $\varTwo$, then the PMF of $\varOne$, $\Pr\{\varOne = \numSuccess\}$, should be approximated using $\Pr\{\numSuccess-0.5 < \varTwo < \numSuccess + 0.5\}$. By adding the continuity correction to \cite[Eq.~(2.1-93)]{Proakis2000}, we approximate the CDF of Binomial random variable $\varOne$ as
\begin{equation}
\label{eq_cdf_gaussian}
\Pr\{\varOne \le \numSuccess\} = \frac{1}{2}\left[1 + \ERF{\frac{\numSuccess+0.5-\numTrials\trialProb}{\sqrt{2\numTrials\trialProb(1-\trialProb)}}}\right],
\end{equation}
and we obtain the PMF $\Pr\{\varOne = \numSuccess\}$ by subtracting $\Pr\{\varOne \le \numSuccess-1\}$ from $\Pr\{\varOne \le \numSuccess\}$.

\subsection{Mutual Information}

The mutual information between two variables $\varOne$ and $\varTwo$, $I\left(\varOne;\varTwo\right)$, measures how much the knowledge of one variable aids in the prediction of a second variable. It is defined as \cite[Eq.~(2.28)]{Cover2006}
\ifOneCol
	\begin{equation}
	\label{eq_mutual_information}
	I\left(\varOne;\varTwo\right) = \sum_{\varOneVal}\sum_{\varTwoVal}
	\Pr\left(\varOne=\varOneVal,\varTwo=\varTwoVal\right)
	\log\frac{\Pr\left(\varOne=\varOneVal,\varTwo=\varTwoVal\right)}
	{\Pr\left(\varOne=\varOneVal\right)\Pr\left(\varTwo=\varTwoVal\right)},
	\end{equation}
\else
	\begin{multline}
	\label{eq_mutual_information}
	I\left(\varOne;\varTwo\right) = \sum_{\varOneVal}\sum_{\varTwoVal}
	\Pr\left(\varOne=\varOneVal,\varTwo=\varTwoVal\right) \\
	\times \log\frac{\Pr\left(\varOne=\varOneVal,\varTwo=\varTwoVal\right)}
	{\Pr\left(\varOne=\varOneVal\right)\Pr\left(\varTwo=\varTwoVal\right)},
	\end{multline}
\fi
where $\Pr\left(\varOne=\varOneVal,\varTwo=\varTwoVal\right)$ is the joint probability distribution of $\varOne$ and $\varTwo$, and $\Pr\left(\varOne=\varOneVal\right)$ and $\Pr\left(\varTwo=\varTwoVal\right)$ are the marginal distributions of $\varOne$ and $\varTwo$, respectively. Given realizations of the two variables $\varOne$ and $\varTwo$, we can calculate these distributions empirically. If $\varOne$ and $\varTwo$ are independent, then knowing $\varOne$ does not provide any greater certainty in predicting the value of $\varTwo$ (or vice versa). Theoretically, independent variables have $I\left(\varOne;\varTwo\right) = 0$, though calculating mutual information from a finite number of realizations will generally result in a non-zero value; see \cite{Goebel2005}.

\section{Video Summary}
\label{app_video}

There are eight videos in \cite{Noel2016c}. Video 1 shows a sample environment described in Section~\ref{sec_components}. The environments in the remaining videos are shown in Section~\ref{sec_intro} and complete descriptions can be found in Section~\ref{sec_results}. For clarity, each video generally shows fewer molecules than the corresponding simulation. Summaries of the video files are as follows:
\begin{enumerate}
	\item \texttt{Video 1}: This video shows a simulation of the environment in Fig.~\ref{fig_fancy_env} on page~\pageref{fig_fancy_env}. Molecules are released by the blue surface. They can diffuse (in one direction) through the green membrane surface and be absorbed by the sphere. All other boundaries are reflective. The trajectory of one molecule is emphasized by showing it in red (until it is absorbed and turns blue).
	\item \texttt{Video 2}: This video shows a simulation of System 1, whose environment is shown in Fig.~\ref{fig_hybrid_env_mol} on page~\pageref{fig_hybrid_env_mol} and Fig.~\ref{fig_hybrid_env} on page~\pageref{fig_hybrid_env}. The system has adjacent microscopic and mesoscopic regions of equal size. Molecules are initialized uniformly throughout the environment and freely diffuse throughout. All boundaries are reflective. The trajectory of one molecule is emphasized by showing it in blue in the mesoscopic regime and in red in the microscopic regime.
	\item \texttt{Video 3}: This video shows a simulation of a subset of System 2, whose environment is shown in Fig.~\ref{fig_hybrid2_env} on page~\pageref{fig_hybrid2_env}. The system is a rectangular rod that is mostly a mesoscopic region but the end with an absorbing surface is microscopic. Molecules are initialized uniformly throughout the environment and freely diffuse until they are absorbed. The molecules are dark grey in the mesoscopic region, light grey in the microscopic region, and turn red when absorbed.
	\item \texttt{Video 4}: This video shows a simulation of a small subset of System 3, whose full environment is shown in Fig.~\ref{fig_surface_env} on page~\pageref{fig_surface_env}. Molecules are initialized in the narrow space between two spheres. These molecules can irreversibly absorb to the outer sphere (and turn blue) and are reflected off of the inner sphere.
	\item \texttt{Video 5}: This video shows a simulation of a small subset of System 3, whose full environment is shown in Fig.~\ref{fig_surface_env} on page~\pageref{fig_surface_env}. Molecules are initialized in the narrow space between two spheres. These molecules can probabilistically diffuse through the inner sphere in either direction (i.e., the inner sphere is a bidirectional membrane) and are reflected off of the outer sphere. The molecules turn yellow while they are inside the inner sphere.
	\item \texttt{Video 6}: This video shows a simulation of System 4, whose environment is shown in Fig.~\ref{fig_comm_env} on page~\pageref{fig_comm_env}. A point transmitter releases molecules that are observed at a passive sphere. The released molecules are blue while they diffuse freely, yellow while they are bound to enzyme molecules, and red while they are inside the sphere. The unbound enzymes are white.
	\item \texttt{Video 7}: This video shows a simulation of System 4, whose environment is shown in Fig.~\ref{fig_comm_env} on page~\pageref{fig_comm_env}. A point transmitter releases multiple pulses of molecules that are observed at a passive sphere. The molecules turn red while they are inside the sphere.
	\item \texttt{Video 8}: This video shows a simulation of System 4, whose environment is shown in Fig.~\ref{fig_comm_env} on page~\pageref{fig_comm_env}. A point transmitter releases molecules that are observed at an absorbing sphere. The molecules turn red when they are absorbed.
\end{enumerate}

\section*{References}

\bibliography{../references/library_fixed}

\begin{thebibliography}{10}
\expandafter\ifx\csname url\endcsname\relax
  \def\url#1{\texttt{#1}}\fi
\expandafter\ifx\csname urlprefix\endcsname\relax\def\urlprefix{URL }\fi
\expandafter\ifx\csname href\endcsname\relax
  \def\href#1#2{#2} \def\path#1{#1}\fi

\bibitem{Hiyama2005}
S.~Hiyama, Y.~Moritani, T.~Suda, R.~Egashira, A.~Enomoto, M.~J. Moore,
  T.~Nakano, Molecular communication, in: Proc. NSTI Nanotech, 2005, pp.
  391--394.

\bibitem{Antunes2009}
L.~C.~M. Antunes, R.~B.~R. Ferreira, Intercellular communication in bacteria.,
  Crit. Rev. Microbiol. 35~(2) (2009) 69--80.
\newblock \href {http://dx.doi.org/10.1080/10408410902733946}
  {\path{doi:10.1080/10408410902733946}}.

\bibitem{Sadava2014}
D.~E. Sadava, D.~M. Hillis, H.~C. Heller, M.~Berenbaum, Life: The Science of
  Biology, 10th Edition, Sinauer Associates, 2014.

\bibitem{Alberts}
B.~Alberts, D.~Bray, K.~Hopkin, A.~D. Johnson, J.~Lewis, M.~Raff, K.~Roberts,
  P.~Walter, Essential Cell Biology, 3rd Edition, Garland Science, 2009.

\bibitem{Nakano2013c}
T.~Nakano, A.~W. Eckford, T.~Haraguchi, Molecular Communication, Cambridge
  University Press, 2013.

\bibitem{Farsad2016}
N.~Farsad, H.~B. Yilmaz, A.~Eckford, C.-B. Chae, W.~Guo, A comprehensive survey
  of recent advancements in molecular communication, IEEE Commun. Surv.
  Tutorials 18~(3) (2016) 1887--1919.
\newblock \href {http://dx.doi.org/10.1109/COMST.2016.2527741}
  {\path{doi:10.1109/COMST.2016.2527741}}.

\bibitem{Crank1979}
J.~Crank, The Mathematics of Diffusion, 2nd Edition, Oxford University Press,
  1979.

\bibitem{Cussler1984}
E.~L. Cussler, Diffusion: Mass Transfer in Fluid Systems, Cambridge University
  Press, 1984.

\bibitem{Berg1993}
H.~C. Berg, Random Walks in Biology, Princeton University Press, 1993.

\bibitem{Truskey2009}
G.~A. Truskey, F.~Yuan, D.~F. Katz, Transport Phenomena in Biological Systems,
  2nd Edition, Pearson Prentice Hall, 2009.

\bibitem{Chang2005}
R.~Chang, Physical Chemistry for the Biosciences, University Science Books,
  2005.

\bibitem{Farsad2013a}
N.~Farsad, W.~Guo, A.~W. Eckford, Tabletop molecular communication: Text
  messages through chemical signals, PLoS One 8~(12) (2013) e82935.
\newblock \href {http://dx.doi.org/10.1371/journal.pone.0082935}
  {\path{doi:10.1371/journal.pone.0082935}}.

\bibitem{Noel2014f}
A.~Noel, K.~C. Cheung, R.~Schober, Improving receiver performance of diffusive
  molecular communication with enzymes, IEEE Trans. Nanobiosci. 13~(1) (2014)
  31--43.
\newblock \href {http://dx.doi.org/10.1109/TNB.2013.2295546}
  {\path{doi:10.1109/TNB.2013.2295546}}.

\bibitem{Farsad2013}
N.~Farsad, N.-R. Kim, A.~W. Eckford, C.-B. Chae, Channel and noise models for
  nonlinear molecular communication systems, IEEE J. Sel. Areas Commun. 32~(12)
  (2014) 2392--2401.
\newblock \href {http://dx.doi.org/10.1109/JSAC.2014.2367662}
  {\path{doi:10.1109/JSAC.2014.2367662}}.

\bibitem{Andrews2004}
S.~S. Andrews, D.~Bray, Stochastic simulation of chemical reactions with
  spatial resolution and single molecule detail, Phys. Biol. 1~(3-4) (2004)
  137--151.
\newblock \href {http://dx.doi.org/10.1088/1478-3967/1/3/001}
  {\path{doi:10.1088/1478-3967/1/3/001}}.

\bibitem{Wei2013a}
G.~Wei, P.~Bogdan, R.~Marculescu, Efficient modeling and simulation of
  bacteria-based nanonetworks with {BNSim}, IEEE J. Sel. Areas Commun. 31~(12)
  (2013) 868--878.
\newblock \href {http://dx.doi.org/10.1109/JSAC.2013.SUP2.12130019}
  {\path{doi:10.1109/JSAC.2013.SUP2.12130019}}.

\bibitem{Felicetti2013}
L.~Felicetti, M.~Femminella, G.~Reali, Simulation of molecular signaling in
  blood vessels: Software design and application to atherogenesis, Nano Commun.
  Net. 4~(3) (2013) 98--119.
\newblock \href {http://dx.doi.org/10.1016/j.nancom.2013.06.002}
  {\path{doi:10.1016/j.nancom.2013.06.002}}.

\bibitem{Llatser2014}
I.~Llatser, D.~Demiray, A.~Cabellos-Aparicio, D.~T. Altilar, E.~Alarc{\'{o}}n,
  {N3Sim}: Simulation framework for diffusion-based molecular communication
  nanonetworks, Simul. Model. Pract. Theory 42 (2014) 210--222.
\newblock \href {http://dx.doi.org/10.1016/j.simpat.2013.11.004}
  {\path{doi:10.1016/j.simpat.2013.11.004}}.

\bibitem{Yilmaz2014a}
H.~B. Yilmaz, C.-B. Chae, Simulation study of molecular communication systems
  with an absorbing receiver: Modulation and {ISI} mitigation techniques,
  Simul. Model. Pract. Theory 49 (2014) 136--150.
\newblock \href {http://dx.doi.org/10.1016/j.simpat.2014.09.002}
  {\path{doi:10.1016/j.simpat.2014.09.002}}.

\bibitem{Jian2016}
Y.~Jian, B.~Krishnaswamy, C.~M. Austin, A.~O. Bicen, J.~E. Perdomo, S.~C.
  Patel, I.~F. Akyildiz, C.~R. Forest, R.~Sivakumar, {nanoNS3}: Simulating
  bacterial molecular communication based nanonetworks in {Network Simulator
  3}, in: Proc. ACM NANOCOM, 2016, pp. 1--7.
\newblock \href {http://dx.doi.org/10.1145/2967446.2967464}
  {\path{doi:10.1145/2967446.2967464}}.

\bibitem{Noel2016}
A.~Noel, \href{https://github.com/adamjgnoel/AcCoRD/}{{AcCoRD} ({A}ctor-based
  {C}ommunication via {R}eaction-{D}iffusion)}.
\newline\urlprefix\url{https://github.com/adamjgnoel/AcCoRD/}

\bibitem{Noel2015a}
A.~Noel, K.~C. Cheung, R.~Schober, Multi-scale stochastic simulation for
  diffusive molecular communication, in: Proc. IEEE ICC, 2015, pp. 1109--1115.
\newblock \href {http://dx.doi.org/10.1109/ICC.2015.7248471}
  {\path{doi:10.1109/ICC.2015.7248471}}.

\bibitem{Noel2016c}
A.~Noel,
  \href{https://www.youtube.com/watch?v=7QcN6eGrC4w&list=PLZ7uYXG-7XF8UyhFrIuQIiZig1XA89e3i}{Videos
  for {AcCoRD} journal paper} (2016).
\newline\urlprefix\url{https://www.youtube.com/watch?v=7QcN6eGrC4w&list=PLZ7uYXG-7XF8UyhFrIuQIiZig1XA89e3i}

\bibitem{Andrews2009}
S.~S. Andrews, Accurate particle-based simulation of adsorption, desorption and
  partial transmission, Phys. Biol. 6~(4) (2009) 046015.
\newblock \href {http://dx.doi.org/10.1088/1478-3975/6/4/046015}
  {\path{doi:10.1088/1478-3975/6/4/046015}}.

\bibitem{Noel2016a}
A.~Noel, D.~Makrakis, A.~Hafid, Channel impulse responses in diffusive
  molecular communication with spherical transmitters, in: Proc. CSIT BSC 2016,
  to appear, pp. 1--6.

\bibitem{Noel2015}
A.~Noel, K.~C. Cheung, R.~Schober, On the statistics of reaction-diffusion
  simulations for molecular communication, in: Proc. ACM NANOCOM, 2015, pp.
  1--6.
\newblock \href {http://dx.doi.org/10.1145/2800795.2800821}
  {\path{doi:10.1145/2800795.2800821}}.

\bibitem{Plimpton1995}
S.~Plimpton, Fast parallel algorithms for short-range molecular dynamics, J.
  Comput. Phys. 117~(1) (1995) 1--19.
\newblock \href {http://dx.doi.org/10.1006/jcph.1995.1039}
  {\path{doi:10.1006/jcph.1995.1039}}.

\bibitem{StevePl}
S.~Plimpton, \href{http://lammps.sandia.gov}{{LAMMPS} molecular dynamics
  simulator}.
\newline\urlprefix\url{http://lammps.sandia.gov}

\bibitem{Andrews}
S.~S. Andrews, \href{http://www.smoldyn.org}{Smoldyn: A spatial stochastic
  simulator for chemical reaction networks}.
\newline\urlprefix\url{http://www.smoldyn.org}

\bibitem{VanZon2005}
J.~S. van Zon, P.~R. ten Wolde, Green's-function reaction dynamics: A
  particle-based approach for simulating biochemical networks in time and
  space, J. Chem. Phys. 123~(23) (2005) 234910.
\newblock \href {http://dx.doi.org/10.1063/1.2137716}
  {\path{doi:10.1063/1.2137716}}.

\bibitem{Ramaswamy2011}
R.~Ramaswamy, I.~F. Sbalzarini, Exact on-lattice stochastic reaction-diffusion
  simulations using partial-propensity methods., J. Chem. Phys. 135~(24) (2011)
  244103.
\newblock \href {http://dx.doi.org/10.1063/1.3666988}
  {\path{doi:10.1063/1.3666988}}.

\bibitem{Drawert2012}
B.~Drawert, S.~Engblom, A.~Hellander, {URDME}: A modular framework for
  stochastic simulation of reaction-transport processes in complex geometries,
  BMC Syst. Biol. 6~(1) (2012) 76.
\newblock \href {http://dx.doi.org/10.1186/1752-0509-6-76}
  {\path{doi:10.1186/1752-0509-6-76}}.

\bibitem{Bayati2011}
B.~Bayati, P.~Chatelain, P.~Koumoutsakos, Adaptive mesh refinement for
  stochastic reaction-diffusion processes, J. Comput. Phys. 230~(1) (2011)
  13--26.
\newblock \href {http://dx.doi.org/10.1016/j.jcp.2010.08.035}
  {\path{doi:10.1016/j.jcp.2010.08.035}}.

\bibitem{Gillespie2001}
D.~T. Gillespie, Approximate accelerated stochastic simulation of chemically
  reacting systems, J Chem Phys 115~(4) (2001) 1716 -- 1733.
\newblock \href {http://dx.doi.org/10.1063/1.1378322}
  {\path{doi:10.1063/1.1378322}}.

\bibitem{Iyengar2010}
K.~A. Iyengar, L.~A. Harris, P.~Clancy, Accurate implementation of leaping in
  space: The spatial partitioned-leaping algorithm, J. Chem. Phys. 132~(9)
  (2010) 094101.
\newblock \href {http://dx.doi.org/10.1063/1.3310808}
  {\path{doi:10.1063/1.3310808}}.

\bibitem{COMSOL}
{COMSOL Inc.}, \href{http://www.comsol.com}{{COMSOL Multiphysics}}.
\newline\urlprefix\url{http://www.comsol.com}

\bibitem{ANSYS}
{ANSYS Inc.}, \href{http://www.ansys.com}{{ANSYS}}.
\newline\urlprefix\url{http://www.ansys.com}

\bibitem{Klann2012}
M.~Klann, A.~Ganguly, H.~Koeppl, Hybrid spatial {Gillespie} and particle
  tracking simulation, Bioinformatics 28~(18) (2012) i549--i555.
\newblock \href {http://dx.doi.org/10.1093/bioinformatics/bts384}
  {\path{doi:10.1093/bioinformatics/bts384}}.

\bibitem{Flegg2014}
M.~B. Flegg, S.~J. Chapman, L.~Zheng, R.~Erban, Analysis of the two-regime
  method on square meshes, SIAM J. Sci. Comput. 36~(3) (2014) 561--588.
\newblock \href {http://dx.doi.org/10.1137/130915844}
  {\path{doi:10.1137/130915844}}.

\bibitem{Hellander2012}
A.~Hellander, S.~Hellander, P.~L{\"{o}}tstedt, Coupled mesoscopic and
  microscopic simulation of stochastic reaction-diffusion processes in mixed
  dimensions, Multiscale Model. Simul. 10~(2) (2012) 585--611.
\newblock \href {http://dx.doi.org/10.1137/110832148}
  {\path{doi:10.1137/110832148}}.

\bibitem{Robinson2015}
M.~Robinson, S.~S. Andrews, R.~Erban, Multiscale reaction-diffusion simulations
  with smoldyn, Bioinformatics 31~(14) (2015) 2406--2408.
\newblock \href {http://dx.doi.org/10.1093/bioinformatics/btv149}
  {\path{doi:10.1093/bioinformatics/btv149}}.

\bibitem{Wagner2008}
G.~Wagner, R.~Jones, J.~Templeton, M.~Parks, An atomistic-to-continuum coupling
  method for heat transfer in solids, Comput. Methods Appl. Mech. Eng. 197~(41)
  (2008) 3351--3365.
\newblock \href {http://dx.doi.org/10.1016/j.cma.2008.02.004}
  {\path{doi:10.1016/j.cma.2008.02.004}}.

\bibitem{Resasco2012}
D.~C. Resasco, F.~Gao, F.~Morgan, I.~L. Novak, J.~C. Schaff, B.~M. Slepchenko,
  {Virtual Cell}: Computational tools for modeling in cell biology, Wiley
  Interdiscip. Rev. Syst. Biol. Med. 4~(2) (2012) 129--140.
\newblock \href {http://dx.doi.org/10.1002/wsbm.165}
  {\path{doi:10.1002/wsbm.165}}.

\bibitem{Gul2010}
E.~Gul, B.~Atakan, O.~B. Akan, {NanoNS}: A nanoscale network simulator
  framework for molecular communications, Nano Commun. Net. 1~(2) (2010)
  138--156.
\newblock \href {http://dx.doi.org/10.1016/j.nancom.2010.08.003}
  {\path{doi:10.1016/j.nancom.2010.08.003}}.

\bibitem{Akkaya2014}
A.~Akkaya, G.~Genc, T.~Tugcu, {HLA} based architecture for molecular
  communication simulation, Simul. Model. Pract. Theory 42 (2014) 163--177.
\newblock \href {http://dx.doi.org/10.1016/j.simpat.2013.12.012}
  {\path{doi:10.1016/j.simpat.2013.12.012}}.

\bibitem{Gillespie1976}
D.~T. Gillespie, A general method for numerically simulating the stochastic
  time evolution of coupled chemical reactions, J. Comput. Phys. 22~(4) (1976)
  403--434.
\newblock \href {http://dx.doi.org/10.1016/0021-9991(76)90041-3}
  {\path{doi:10.1016/0021-9991(76)90041-3}}.

\bibitem{Gillespie2007}
D.~T. Gillespie, Stochastic simulation of chemical kinetics., Annu. Rev. Phys.
  Chem. 58 (2007) 35--55.
\newblock \href {http://dx.doi.org/10.1146/annurev.physchem.58.032806.104637}
  {\path{doi:10.1146/annurev.physchem.58.032806.104637}}.

\bibitem{Llatser2011}
I.~Llatser, I.~Pascual, N.~Garralda, A.~Cabellos-Aparicio, M.~Pierobon,
  E.~Alarc{\'{o}}n, J.~{Sol{\'{e}} Pareta}, Exploring the physical channel of
  diffusion-based molecular communication by simulation, in: Proc. IEEE
  GLOBECOM, 2011, pp. 1--5.
\newblock \href {http://dx.doi.org/10.1109/GLOCOM.2011.6134028}
  {\path{doi:10.1109/GLOCOM.2011.6134028}}.

\bibitem{Llatser}
I.~Llatser, \href{http://www.n3cat.upc.edu/n3sim}{{N3Sim}: A simulation
  framework for diffusion-based molecular communication}.
\newline\urlprefix\url{http://www.n3cat.upc.edu/n3sim}

\bibitem{Bush2015}
S.~F. Bush, J.~L. Paluh, G.~Piro, V.~Rao, R.~V. Prasad, A.~W. Eckford, Defining
  communication at the bottom, IEEE Trans. Mol. Biol. Multi-Scale Commun. 1~(1)
  (2015) 90--96.
\newblock \href {http://dx.doi.org/10.1109/TMBMC.2015.2465513}
  {\path{doi:10.1109/TMBMC.2015.2465513}}.

\bibitem{Matsumoto1998}
M.~Matsumoto, T.~Nishimura, Mersenne twister: A 623-dimensionally
  equidistributed uniform pseudo-random number generator, ACM Trans. Model.
  Comput. Simul. 8~(1) (1998) 3--30.
\newblock \href {http://dx.doi.org/10.1145/272991.272995}
  {\path{doi:10.1145/272991.272995}}.

\bibitem{ONeill}
M.~E. O'Neill, {PCG}: A family of simple fast space-efficient statistically
  good algorithms for random number generation, ACM Trans. Math. Softw.
  (submitted) 1--46.

\bibitem{Kuran2011}
M.~S. Kuran, H.~B. Yilmaz, T.~Tugcu, I.~F. Akyildiz, Modulation techniques for
  communication via diffusion in nanonetworks, in: IEEE Int. Conf. Commun.,
  2011, pp. 1--5.
\newblock \href {http://dx.doi.org/10.1109/icc.2011.5962989}
  {\path{doi:10.1109/icc.2011.5962989}}.

\bibitem{Petrucci2016}
R.~H. Petrucci, F.~G. Herring, J.~D. Madura, C.~Bissonnette, General Chemistry:
  Principles and Modern Applications, 11th Edition, Pearson Prentice Hall,
  2016.

\bibitem{Crockford}
D.~Crockford, \href{http://www.json.org}{{JSON}: {JavaScript Object Notation}}.
\newline\urlprefix\url{http://www.json.org}

\bibitem{Bernstein2005}
D.~Bernstein, Simulating mesoscopic reaction-diffusion systems using the
  gillespie algorithm, Phys. Rev. E - Stat. Nonlinear, Soft Matter Phys. 71~(4)
  (2005) 1--13.
\newblock \href {http://dx.doi.org/10.1103/PhysRevE.71.041103}
  {\path{doi:10.1103/PhysRevE.71.041103}}.

\bibitem{Ng1968}
E.~W. Ng, M.~Geller, A table of integrals of the error functions, J. Res. Natl.
  Bur. Stand. Sect. B Math. Sci. 73B~(1) (1969) 1--20.
\newblock \href {http://dx.doi.org/10.6028/jres.073B.001}
  {\path{doi:10.6028/jres.073B.001}}.

\bibitem{Ross2009}
S.~M. Ross, Introduction to Probability and Statistics for Engineers and
  Scientists, 4th Edition, Academic Press, 2009.

\bibitem{Hellander2016}
S.~Hellander, L.~Petzold, Reaction rates for a generalized reaction-diffusion
  master equation, Phys. Rev. E 93~(1) (2016) 13307.
\newblock \href {http://dx.doi.org/10.1103/PhysRevE.93.013307}
  {\path{doi:10.1103/PhysRevE.93.013307}}.

\bibitem{Huang2011}
R.~Huang, I.~Chavez, K.~M. Taute, B.~Luki{\'{c}}, S.~Jeney, M.~G. Raizen, E.-L.
  Florin, Direct observation of the full transition from ballistic to diffusive
  brownian motion in a liquid, Nat. Phys. 7~(7) (2011) 576--580.
\newblock \href {http://dx.doi.org/10.1038/nphys1953}
  {\path{doi:10.1038/nphys1953}}.

\bibitem{Noel2014d}
A.~Noel, K.~C. Cheung, R.~Schober, Optimal receiver design for diffusive
  molecular communication with flow and additive noise, IEEE Trans. Nanobiosci.
  13~(3) (2014) 350--362.
\newblock \href {http://dx.doi.org/10.1109/TNB.2014.2337239}
  {\path{doi:10.1109/TNB.2014.2337239}}.

\bibitem{Goebel2005}
B.~Goebel, Z.~Dawy, J.~Hagenauer, J.~Mueller, An approximation to the
  distribution of finite sample size mutual information estimates, in: Proc.
  IEEE ICC, 2005, pp. 1102--1106.
\newblock \href {http://dx.doi.org/10.1109/ICC.2005.1494518}
  {\path{doi:10.1109/ICC.2005.1494518}}.

\bibitem{Deng2016b}
Y.~Deng, A.~Noel, W.~Guo, A.~Nallanathan, M.~Elkashlan, Stochastic geometry
  model for large-scale molecular communication systems, in: Proc. IEEE
  GLOBECOM 2016, to appear, pp. 1--7.

\bibitem{Noel2016b}
A.~Noel, Y.~Deng, D.~Makrakis, A.~Hafid, Active versus passive: Receiver model
  transforms for diffusive molecular communication, in: Proc. IEEE GLOBECOM
  2016, to appear, pp. 1--6.

\bibitem{Heren2015}
A.~C. Heren, H.~B. Yilmaz, C.-B. Chae, T.~Tugcu, Effect of degradation in
  molecular communication: Impairment or enhancement?, IEEE Trans. Mol. Biol.
  Multi-Scale Commun. 1~(2) (2015) 217--229.
\newblock \href {http://dx.doi.org/10.1109/TMBMC.2015.2502859}
  {\path{doi:10.1109/TMBMC.2015.2502859}}.

\bibitem{Meng2014}
L.-S. Meng, P.-C. Yeh, K.-C. Chen, I.~F. Akyildiz, On receiver design for
  diffusion-based molecular communication, IEEE Trans. Signal Process. 62~(22)
  (2014) 6032--6044.
\newblock \href {http://dx.doi.org/10.1109/TSP.2014.2359644}
  {\path{doi:10.1109/TSP.2014.2359644}}.

\bibitem{Mahfuz2015}
M.~U. Mahfuz, D.~Makrakis, H.~T. Mouftah, A comprehensive analysis of
  strength-based optimum signal detection in concentration-encoded molecular
  communication with spike transmission, IEEE Trans. Nanobiosci. 14~(1) (2015)
  67--83.
\newblock \href {http://dx.doi.org/10.1109/TNB.2014.2368593}
  {\path{doi:10.1109/TNB.2014.2368593}}.

\bibitem{ShahMohammadian2012}
H.~ShahMohammadian, G.~G. Messier, S.~Magierowski, Optimum receiver for
  molecule shift keying modulation in diffusion-based molecular communication
  channels, Nano Commun. Net. 3~(3) (2012) 183--195.
\newblock \href {http://dx.doi.org/10.1016/j.nancom.2012.09.006}
  {\path{doi:10.1016/j.nancom.2012.09.006}}.

\bibitem{Noel2013b}
A.~Noel, K.~C. Cheung, R.~Schober, Using dimensional analysis to assess
  scalability and accuracy in molecular communication, in: Proc. IEEE ICC
  MoNaCom, 2013, pp. 818--823.
\newblock \href {http://dx.doi.org/10.1109/ICCW.2013.6649346}
  {\path{doi:10.1109/ICCW.2013.6649346}}.

\bibitem{Schulten2000}
K.~Schulten, I.~Kosztin, Lectures in Theoretical Biophysics, University of
  Illinois, 2000.

\bibitem{Nelson2008}
P.~Nelson, Biological Physics: Energy, Information, Life, Updated 1st Edition,
  W. H. Freeman and Company, 2008.

\bibitem{Farsad2016b}
N.~Farsad, A.~Goldsmith, A molecular communication system using acids, bases
  and hydrogen ions, in: Proc. IEEE SPAWC, 2016, pp. 1--6.
\newblock \href {http://dx.doi.org/10.1109/SPAWC.2016.7536834}
  {\path{doi:10.1109/SPAWC.2016.7536834}}.

\bibitem{Ahmadzadeh2015a}
A.~Ahmadzadeh, A.~Noel, R.~Schober, Analysis and design of multi-hop
  diffusion-based molecular communication networks, IEEE Trans. Mol. Biol.
  Multi-Scale Commun. 1~(2) (2015) 144--157.
\newblock \href {http://dx.doi.org/10.1109/TMBMC.2015.2501741}
  {\path{doi:10.1109/TMBMC.2015.2501741}}.

\bibitem{Jedrzejewski-Szmek2016}
Z.~Jedrzejewski-Szmek, K.~T. Blackwell, Asynchronous tau-leaping, J. Chem.
  Phys. 144~(12) (2016) 125104.
\newblock \href {http://dx.doi.org/10.1063/1.4944575}
  {\path{doi:10.1063/1.4944575}}.

\bibitem{Erban2007}
R.~Erban, S.~J. Chapman, Reactive boundary conditions for stochastic
  simulations of reaction-diffusion processes., Phys. Biol. 4~(1) (2007)
  16--28.
\newblock \href {http://dx.doi.org/10.1088/1478-3975/4/1/003}
  {\path{doi:10.1088/1478-3975/4/1/003}}.

\bibitem{Elf2004}
J.~Elf, M.~Ehrenberg, Spontaneous separation of bi-stable biochemical systems
  into spatial domains of opposite phases, Syst. Biol. (Stevenage). 1~(2)
  (2004) 230--236.
\newblock \href {http://dx.doi.org/10.1049/sb:20045021}
  {\path{doi:10.1049/sb:20045021}}.

\bibitem{Gibson2000}
M.~A. Gibson, J.~Bruck, Efficient exact stochastic simulation of chemical
  systems with many species and many channels, J. Phys. Chem. A 104~(9) (2000)
  1876--1889.
\newblock \href {http://dx.doi.org/10.1021/jp993732q}
  {\path{doi:10.1021/jp993732q}}.

\bibitem{Abramowitz1964}
M.~Abramowitz, I.~A. Stegun, Handbook of Mathematical Functions with Formulas,
  Graphs, and Mathematical Tables, 1st Edition, United States Department of
  Commerce, National Bureau of Standards, 1964.

\bibitem{Proakis2000}
J.~G. Proakis, Digital Communications, 4th Edition, McGraw-Hill, 2000.

\bibitem{Cover2006}
T.~M. Cover, J.~A. Thomas, Elements of Information Theory, 2nd Edition,
  Wiley-Interscience, 2006.

\end{thebibliography}

\vspace*{5mm}

\noindent\includegraphics[width=1in,height=1.25in,
	clip,keepaspectratio]{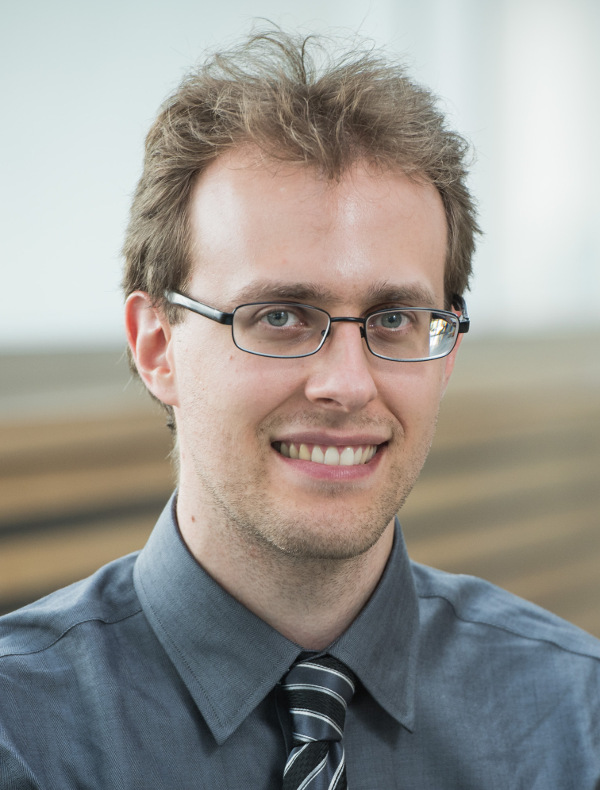} Adam Noel received the B.Eng. degree in Electrical Engineering in 2009 from Memorial University in St. John’s, Canada and the M.A.Sc. and Ph.D. degrees in Electrical and Computer Engineering from the University of British Columbia in Vancouver, Canada, in 2011 and 2015, respectively. In 2013, he was a Visiting Scientist at the Institute for Digital Communication at Friedrich-Alexander-University in Erlangen, Germany. He is currently a Postdoctoral Fellow at the University of Ottawa, Canada. Dr. Noel's current research interests include channel modeling, system design, and simulation methods for molecular communication networks.\newline

\noindent\includegraphics[width=1in,height=1.25in,
clip,keepaspectratio]{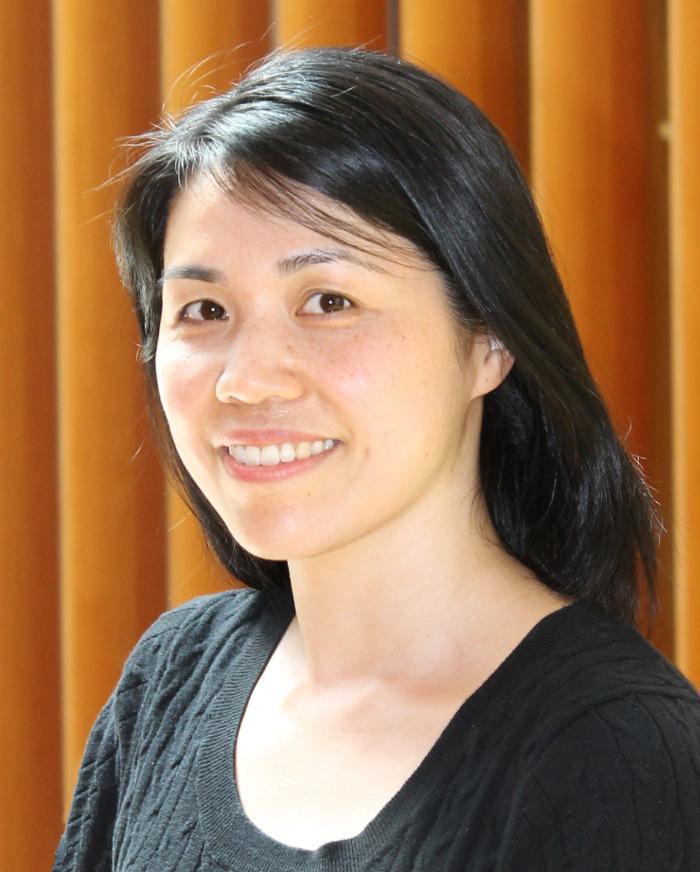}
Karen C. Cheung received the B.S. and Ph.D. degrees in bioengineering from the University of California, Berkeley, in 1998 and 2002, respectively. From 2002 to 2005, she was	a postdoctoral researcher at the Ecole Polytechnique F\'{e}d\'{e}rale de Lausanne, Lausanne, Switzerland. She is now at the University of British Columbia, Vancouver, BC, Canada. Her research interests include lab-on-a-chip systems for	cell culture and characterization, inkjet printing for tissue engineering, and implantable neural interfaces.\newline

\noindent\includegraphics[width=1in,height=1.25in,
clip,keepaspectratio]{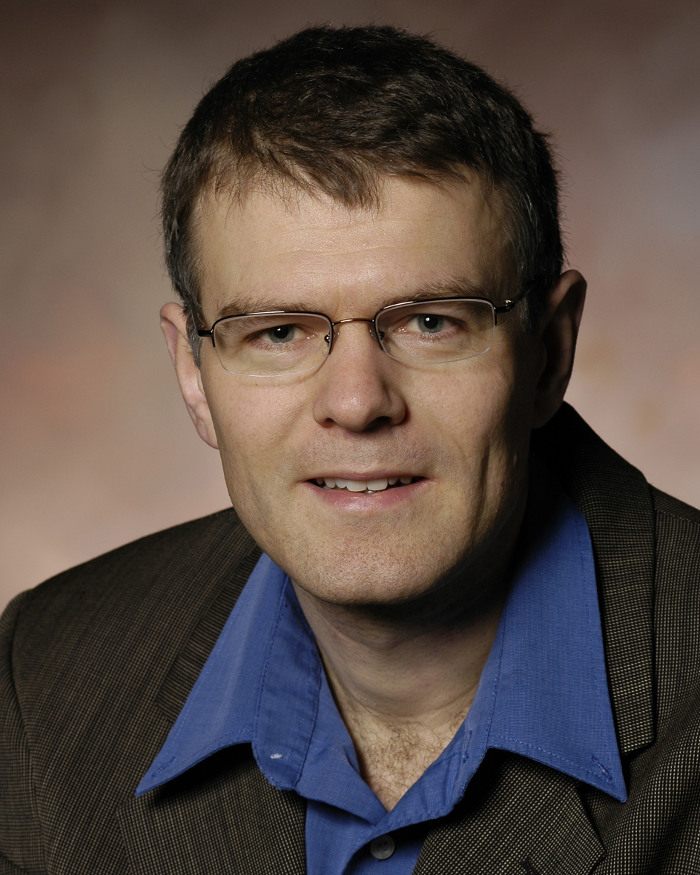}
Robert Schober received the Diplom and the Ph.D. degrees in electrical engineering from Friedrich-Alexander-University (FAU), Erlangen, Germany, in 1997 and 2000, respectively. Since May 2002 he has been with the University of British Columbia, Vancouver, Canada, where he is now a Full Professor. Since January 2012 he is an Alexander von Humboldt Professor and the Chair for Digital Communication at FAU. His research interests fall into the broad areas of Communication Theory, Wireless Communications, and Statistical Signal Processing. From 2012 to 2015, he was the Editor-in-Chief of the IEEE Transactions on Communications.\newline

\noindent Professor Dimitrios Makrakis is currently with the School of Electrical Engineering and Computer Science (EECS) of the University of Ottawa, and is the founding Director of the Broadband and Wireless Research Laboratory (BroadWIRLab).  Prior to joining the University of Ottawa, Dr. Makrakis was a faculty member at the Department of Electrical Engineering of Western University (prior the University of Western Ontario) and Adjunct Professor at the Computer Science Department of Western U.  His research evolves around the subject of information transfer, computing and cyber security and is both theoretical and experimental.\newline

\noindent\includegraphics[width=1in,height=1.25in,
clip,keepaspectratio]{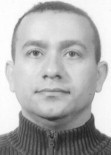}
Abdelhakim Hafid is Full Professor at the University of Montreal, where he founded the Network Research Lab (NRL) in 2005. He is also a research fellow at CIRRELT (Interuniversity Research Center on Enterprise Networks, Logistics and Transportation). He supervised to graduation more than 30 graduate students. He published over 200 journal and conference papers, and he also holds 3 US patents. Dr. A. Hafid has extensive academic and industrial research experience in the area of the management of next generation networks, QoS management, distributed systems, and communication protocols.

\end{document}